\global\def\draftcontrol{0}

   \def\versionno{ Hubble cascade }
\catcode`\@=11

\expandafter\ifx\csname draftcontrol\endcsname\relax\global\def\draftcontrol{0}
\fi

{\count255=\time\divide\count255 by 60
\xdef\hourmin{\number\count255}
\multiply\count255 by-60\advance\count255 by\time
\xdef\hourmin{\hourmin:\ifnum\count255<10 0\fi\the\count255}}
\def\draftdate{\number\month/\number\day/\number\year\ \ \ \hourmin }

\newcommand\makepapertitle{\par
  \begingroup
    \renewcommand\thefootnote{\@fnsymbol\c@footnote}%
    \def\@makefnmark{\rlap{\@textsuperscript{\normalfont\@thefnmark}}}%
    \long\def\@makefntext##1{\parindent 1em\noindent
            \hb@xt@1.8em{%
                \hss\@textsuperscript{\normalfont\@thefnmark}}##1}%
     \newpage
     \global\@topnum\z@   % Prevents figures from going at top of page.
     \@makepapertitle
     \thispagestyle{empty}\@thanks
  \endgroup
  \setcounter{footnote}{0}%
  \global\let\thanks\relax
  \global\let\makepapertitle\relax
  \global\let\@makepapertitle\relax
  \global\let\@thanks\@empty
  \global\let\@author\@empty
  \global\let\@date\@empty
  \global\let\@title\@empty
  \global\let\title\relax
  \global\let\author\relax
  \global\let\date\relax
  \global\let\and\relax
  \def\version{\let\version\@version\@gobble}
}
\def\@makepapertitle{%
  \newpage
   \ifnum\draftcontrol=1 {}
   \version\versionno
   \vskip 3em%
   \else
   \hfill\hbox to 3cm {\parbox{4cm}{\@pubnum}\hss}%
   \vskip 3em%
   \fi
   \begin{center}%
   \let \footnote \thanks
     {\LARGE {\@title}}%
     \vskip 1.5em%
     {\normalsize%\large
       \lineskip .5em%
       \begin{tabular}[t]{c}%
         \@author
       \end{tabular}\par}%
     \vskip 1.5em%
     {\@bstract}%
     \end{center}%
     \vskip 1.5em
     \@date%
   \par
}

\gdef\@pubnum{}
\def\pubnum#1{%
  \gdef\@pubnum{#1}}

\gdef\@bstract{}
\def\Abstract#1{%
  \gdef\@bstract{%
   \parbox{\textwidth-0pc}{%
   \centerline{\bf Abstract}\penalty1000%
\kern.2cm%
\noindent%\abstractfont \baselineskip=12pt
\renewcommand\baselinestretch{1.0}%
{#1}}}
}

\def\ps@paper{\let\@mkboth\@gobbletwo%
     \ifnum\draftcontrol=1
    \def\@oddfoot{\hbox to \textwidth{\tiny \versionno \hfil\tiny\draftdate}%
    \hskip -\textwidth \hbox to \textwidth{\hfil\rm\thepage\hfil}}%
     \else\def\@oddfoot{\hbox to \textwidth{\hfil\rm\thepage\hfil}}
     \fi
     \let\@evenfoot\@oddfoot
}
\def\body{\clearpage
          \pagestyle{paper}
    }
\def\@version#1{\ifnum\draftcontrol=1
\typeout{}\typeout{#1}\typeout{}
\vskip3mm\centerline{\hbox{\fbox{\normalsize{\tt DRAFT -- #1 -- }
                   {\draftdate}}}}\vskip3mm
\fi}
\let\version\@version
\long\def\eqlabel#1{\ifnum\draftcontrol=1
                    \tag@false  % there are some problems with multline without this
                    \tag*{(\theequation) \hbox to -0.2cm{\hspace{0cm}\small{#1}\hss}}
                    \refstepcounter{equation}
                    \edef\@currentlabel{\theequation}
                    \ltx@label{#1}          % use old LaTeX \label instead of new definition
                    \else
                    \label{#1}
                    \fi
                    }
\let\st@bibitem\@bibitem
\let\st@lbibitem\@lbibitem
\ifnum\draftcontrol=1
  \def\@bibitem#1{%
    \st@bibitem{#1}\a@@label{#1}\ignorespaces}
  \def\@lbibitem[#1]#2{%
    \st@lbibitem[#1]{#2}\a@@label{#2}\ignorespaces}
  \def\a@@label#1{%
    \gdef\a@lab{\smash{\normalfont\small#1}}
    \ifvmode
      \if@inlabel
        \global\setbox\@labels\hbox{%
          \llap{\a@lab\let\a@lab\relax
                \kern\@totalleftmargin\kern\marginparsep}%
          \box\@labels}%
      \fi
    \fi}
\fi
\documentclass[12pt,letterpaper]{article}
\usepackage{amsmath,amssymb,array,calc,epsfig,rotating,bm}
\usepackage[sort]{cite}
\usepackage{graphicx}
\usepackage{psfrag,verbatim}
\ifnum\draftcontrol=1
\fi
\renewcommand\baselinestretch{1.25}
\renewcommand\section{\@startsection {section}{1}{\z@}%
                                   {-3.5ex \@plus -1ex \@minus -.2ex}%
                                   {2.3ex \@plus.2ex}%
                                   {\normalfont\large\bfseries}}
\renewcommand\subsection{\@startsection{subsection}{2}{\z@}%
                                   {-3.25ex\@plus -1ex \@minus -.2ex}%
                                   {1.5ex \@plus .2ex}%
                                   {\normalfont\normalsize\bfseries}}
\renewcommand\subsubsection{\@startsection{subsubsection}{3}{\z@}%
                                   {-3.25ex\@plus -1ex \@minus -.2ex}%
                                   {1.5ex \@plus .2ex}%
                                   {\normalfont\normalsize\it}}
\renewcommand\paragraph{\@startsection{paragraph}{4}{\z@}%
                                   {-3.25ex\@plus -1ex \@minus -.2ex}%
                                   {1.5ex \@plus .2ex}%
                                   {\normalfont\normalsize\bf}}

\numberwithin{equation}{section}

\def\revise#1       {\raisebox{-0em}{\rule{3pt}{1em}}%
                     \marginpar{\raisebox{.5em}{\vrule width3pt\
                     \vrule width0pt height 0pt depth0.5em
                     \hbox to 0cm{\hspace{0cm}{%
                     \parbox[t]{4em}{\raggedright\footnotesize{#1}}}\hss}}}}

\newcommand\nxt[1]  {\\\fnxt#1}
\newcommand{\ie}{{\it i.e.,}\ }

\def\calz {{\cal Z}}

\def\cali         {{\cal I}}
\def\calj         {{\cal J}}

\def\call         {{\cal L}}
\def\calm         {{\cal M}}
\def\caln         {{\cal N}}
\def\calo         {{\cal O}}

\def\calq         {{\cal Q}}

\def\calv         {{\cal V}}

\def\zet          {{\mathbb Z}}

\def\del          {\partial}

\def\tr           {\mathop{\rm Tr}}

\def\sqr#1#2{{\vcenter{\vbox{\hrule height.#2pt
 \hbox{\vrule width.#2pt height#1pt \kern#1pt
 \vrule width.#2pt}\hrule height.#2pt}}}}

\def\a{\alpha}

\def\w{\omega}
\def\r{\rho}
\def\dd{\delta}
\def\e{\epsilon}

\def\g{\gamma}

\def\hh{\hat{h}}
\def\hf{\hat{f}}

\def\hK{\hat{K}}
\def\aa1{\phi}
\def\cc1{\psi}
\def\hh{\hat{h}}

\def\k{\kappa}

\def\l{\lambda}

\def\Om{\Omega}
\def\om{\Omega}

\def\hr{\hat{\r}}
\def\hf{\hat{f}}
\def\hK{\hat{K}}

\def\csb{{\chi\rm{SB}}}

\catcode`\@=12

\begin{document}

%%%
%%%%%% text starts here
%%%%%%%%%

\title{\bf  Cascading gauge theory on $dS_4$ \\
and \\
\vspace{0.4cm}
String Theory Landscape}

\pubnum{UWO-TH-13/15}

\date{October 4, 2013}
%\date\today

\author{
Alex Buchel and Dami\'an A. Galante\\
%
% \begin{small}
% \href{mailto:abuchel@perimeterinstitute.ca}{abuchel@perimeterinstitute.ca}, \href{mailto:dgalante@perimeterinstitute.ca}{dgalante@perimeterinstitute.ca}
% \end{small} \\ [0.2cm]
 %
%
\it Department of Applied Mathematics\\
\it University of Western Ontario\\
\it London, Ontario N6A 5B7, Canada\\[0.2cm]
\it Perimeter Institute for Theoretical Physics\\
\it Waterloo, Ontario N2J 2W9, Canada\\
 }

\Abstract{Placing anti-D3 branes at the tip of the conifold in Klebanov-Strassler geometry provides 
a generic way of constructing meta-stable de Sitter ($dS$) vacua in String Theory. A local 
geometry of such vacua exhibit  gravitational solutions with a D3 charge measured at the tip 
opposite to the asymptotic charge. We discuss a restrictive set of such geometries, where 
anti-D3 branes are smeared at the tip. Such geometries represent holographic dual of cascading 
gauge theory in $dS_4$ with or without chiral symmetry breaking. We find that in the phase 
with unbroken chiral symmetry the D3 charge at the tip is always positive. Furthermore, this charge is
zero in the phase with spontaneously broken chiral symmetry. We show that the effective potential of 
the chirally symmetric phase is lower than that in the symmetry broken phase, \ie there is no 
spontaneous chiral symmetry breaking for cascading gauge theory in $dS_4$. The positivity 
of the D3 brane charge in smooth de-Sitter deformed conifold geometries with fluxes presents
difficulties in uplifting AdS vacua to dS ones in String Theory via smeared anti-D3 branes. 
}

\makepapertitle

\body

\version\versionno
\tableofcontents

\section{Introduction and Summary}
String Theory is expected to have a Landscape of (meta-stable) de-Sitter vacua \cite{bp}. 
A generic way to construct such vacua was presented in \cite{kklt} (KKLT): 
\nxt first, turning on fluxes on Calabi-Yau compactifications of type IIB string theory 
produces highly warped geometry with stabilized complex structure 
(but not K\"ahler) moduli of the compactification \cite{gkp};
\nxt next, including non-perturbative effects (which are under control given the unbroken
supersymmetry), one obtains anti-de Sitter ($AdS_4$) vacua with {\it all} moduli fixed;
\nxt finally, one uses anti-D3 branes of type IIB string theory to uplift $AdS_4$ to
de Sitter ($dS_4$) vacua.

As the last step of the construction completely breaks supersymmetry, it is much less controlled. 
In fact, in \cite{Bena:2009xk,Bena:2011hz,Bena:2011wh,Bena:2012bk} it was argued that putting 
anti-D3 branes at the tip of the Klebanov-Strassler (KS) \cite {ks} geometry (as done in KKLT construction)
leads to a naked singularity. Whether or not the resulting singularity is physical 
is subject to debates\footnote{See \cite{Dymarsky:2011pm} for arguments in favour of this
singularity.}. In \cite{landscape} it was shown that the singularity can not be cloaked 
by a regular event horizon, and thus must be unphysical \cite{Gubser:2000nd}. This conclusion is reached 
analyzing local Klebanov-Tseytlin (KT) \cite{kt} or KS geometry with  regular 
Schwarzschild horizon. Such geometry is dual to strongly coupled cascading gauge theory 
plasma with unbroken \cite{bh0,Buchel:2001gw,Gubser:2001ri,abk,Buchel:2009bh}           
(in KT case) or broken \cite{ksbh} (in KS case) chiral symmetry. It was shown that a D3-brane charge 
measured at the horizon is always positive, and thus can not cloak a physical negative-D3-charge 
singularity. 

The {\it good} versus {\it bad} gravitational singularity criteria of Gubser  \cite{Gubser:2000nd}
is based on a simple principle that singularities in gravitational backgrounds holographically dual to some 
strongly coupled gauge theories arise in the interior of the bulk space-time geometry, corresponding to the
infrared (IR) in the dual gauge theories.  Physical infrared singularities in gauge theories can be removed 
with an infrared cutoff. In the original paper, \cite{Gubser:2000nd}, this cutoff is provided by a temperature. 
However, the role of the cutoff can be served by a curvature scale of a boundary 
compactification manifold \cite{bt}, or by a Hubble scale when the strongly coupled gauge theory is formulated  
in $dS_4$  \cite{bds4}.  
In this paper we extend analysis of \cite{landscape} considering\footnote{The early discussion 
of this problem was presented in \cite{bds4}.} de Sitter deformation of the KT/KS geometries 
(holographically dual to cascading gauge theory in $dS_4$ with unbroken/broken chiral symmetry). 
As in \cite{landscape}, we ask the question whether it is possible to construct smooth geometries 
with a negative D3 charge in the interior of the space.    
 
The analysis presented here closely follow \cite{abs3}. In section
\ref{action} we review dual five-dimensional effective gravitational actions describing 
states of cascading gauge theory on $\calm_4$ with (un-)broken chiral symmetry. 
In section \ref{symmetric} we construct states of cascading gauge theory in $dS_4$ 
with unbroken chiral symmetry. In section \ref{broken} we repeat the exercise for states of the 
theory with spontaneous broken chiral symmetry. In section \ref{transition} we compare effective 
potentials of the cascading gauge theory in $dS_4$ with broken and unbroken chiral symmetry and 
identify the true ground state of the theory. In section \ref{propertiesnew} we compute the 
D3 charge in the interior of the  bulk of de Sitter deformed KT/KS geometries. 
Using results of \cite{abs3}, we compute the 
D3 charge in the interior of the  bulk of $S^3$ deformed KT/KS geometries --- in this last section 
we use the radius of the three-sphere $\ell_3$ as an infrared cutoff to distinguish {\it good} versus
{\it bad} gravitational singularities.    

Our discussion is rather technical; so, for benefits of the readers who are interesting in results only,
we collect them here. Recall that cascading gauge theory is a 
four-dimensional $\caln=1$ supersymmetric 
$SU(K+P)\times SU(K)$ gauge theory with two chiral superfields $A_1, A_2$ in the $(K+P,\overline{K})$
representation, and two fields $B_1, B_2$ in the $(\overline{K+P},K)$.
Perturbatively, this gauge theory has two gauge couplings $g_1, g_2$ 
associated with 
two gauge group factors,  and a quartic 
superpotential
\begin{equation}
W\sim \tr \left(A_i B_j A_kB_\ell\right)\e^{ik}\e^{j\ell}\,.
\end{equation} 
The theory has a global $SU(2)\times SU(2)$ (flavor) symmetry under which $A_i$ and $B_k$ (separately) transform 
as doublets. As this symmetry is always unbroken (both in the field theory and in the gravitational dual) 
all our conclusions concerning uplifting to de Sitter vacua with anti-D3 branes are strictly applicable
when the anti-D3 branes are smeared on the tip of the conifold --- it is only in this case that the 
dual gauge theory flavor symmetry is unbroken. To define a theory, one needs to specify the space-time four-manifold 
$\calm_4$ in  which the theory is formulated. In case when $\calm_4=R^{3,1}$, \ie Minkowski space-time, 
one finds that the sum of the gauge couplings 
does not run
\begin{equation}
\frac{d}{d\ln\mu}\left(\frac{\pi}{g_s}\equiv \frac{4\pi}{g_1^2(\mu)}+\frac{4\pi}{g_2^2(\mu)}\right)=0\,,
\eqlabel{sum}
\end{equation}
while the difference between the two couplings is  
\begin{equation}
\frac{4\pi}{g_2^2(\mu)}-\frac{4\pi}{g_1^2(\mu)}\sim P \ \left[3+2(1-\g_{ij})\right]\ \ln\frac{\mu}{\Lambda}\,,
\eqlabel{diff}
\end{equation}
where $\Lambda$  is the strong coupling scale of the theory and $\g_{ij}$ are anomalous dimensions\footnote{When $K\gg P$, 
$\g_{ij}\approx -\frac 12$, see \cite{ks}.} of 
operators $\tr A_i B_j$. For generic $\calm_4$, the sum of the gauge couplings runs; however, the theory is 
still determined by 2 parameters: the asymptotic value of the dilaton $g_0$, 
\begin{equation}
g_0\equiv \lim_{\mu\to \infty} g_s(\mu) =\lim_{\mu \to \infty} \left( \frac{4}{g_1^2(\mu)}+\frac{4}{g_2^2(\mu)}\right)^{-1}\,,
\eqlabel{defg0}
\end{equation}
and the strong coupling scale $\Lambda$ arising in the renormalization group running of the difference of two couplings 
\eqref{diff}.  To summarize, cascading gauge theory is characterized by $\{P, g_0, \Lambda \}$ and the choice 
of a four-manifold $\calm_4$. Relevant to the discussion here, when $\calm_4=dS_4$ or $R\times S^3$, the manifold provides 
one additional scale to the problem: the Hubble scale $H$ (in case of $dS_4$) or the compactification scale $\ell_3^{-1}$ 
(in case of $S^3$ compactification).  Depending on the ratio of the mass scale supplied by $\calm_4$ and the 
strong coupling scale $\Lambda$, the cascading theory might undergo phase transition in the infrared
associated with spontaneous breaking  of the chiral symmetry\footnote{When $\calm_4$ is 
Minkowski, the chiral symmetry is spontaneously broken, see \cite{ks}.} $\zet_{2P}\to \zet_2$. 
Ideally, we would like to explore the phase structure of the theory for arbitrary values of parameters --- 
in practice, we are restricted to regions of parameter space 
where our numerical code used to generate $\calm_4$ deformed KT/KS throat geometries is stable.   

We now present the summary of our results:
\nxt When $\calm_4=dS_4$ and the chiral symmetry is unbroken, the D3 brane charge at the tip of the 
conifold is always positive, as long as 
\begin{equation}
\ln \frac{H^2}{\Lambda^2 P^2 g_0} \ge -0.4\,.
\eqlabel{hlsymm}
\end{equation}   
\nxt When $\calm_4=dS_4$ and the chiral symmetry is broken, the D3 brane charge at the tip of the 
conifold is always zero; we managed to construct geometries of this type for 
\begin{equation}
\ln  \frac{H^2}{\Lambda^2 P^2 g_0} \ge -0.03 \,.    
\eqlabel{hlbroken}
\end{equation}
\nxt Comparing effective potential of the gauge theory in broken $\calv_{eff}^b$ 
and unbroken $\calv_{eff}^s$  phases we 
establish that in all cases, when we can construct the phase with spontaneously broken chiral symmetry, 
\begin{equation}
\calv_{eff}^b>\calv_{eff}^s\,,\qquad {\rm when}\qquad \ln  \frac{H^2}{\Lambda^2 P^2 g_0} \ge -0.03   \,,
\eqlabel{result3}
\end{equation} 
\ie spontaneous symmetry breaking does not happen for given values of the gauge theory parameters.  
To put these parameters in perspective, note that the (first-order) confinement/deconfinement and chiral symmetry 
breaking  phase transition in 
cascading gauge theory plasma occurs at temperature $T$ such that \cite{abk}
\begin{equation}
\ln  \frac{T_{deconfinement,\csb}^2}{\Lambda^2 P^2 g_0}=0.2571(2)\,,
\eqlabel{compare1}
\end{equation}
and the (first-order) chiral symmetry breaking in cascading gauge theory on $S^3$ occurs 
for compactification scale $\mu_3\equiv \ell_3^{-1}$ such that \cite{abs3}
\begin{equation}
\ln  \frac{\mu_{3,\csb}^2}{\Lambda^2 P^2 g_0}=0.4309(8)\,.
\eqlabel{compare2}
\end{equation}
 \nxt When $\calm_4=R\times S^3$ and the chiral symmetry is unbroken, the D3 brane charge at the tip of the 
conifold is {\bf negative} when 
\begin{equation}
\ln  \frac{\mu_3^2}{\Lambda^2 P^2 g_0}< \ln  \frac{\mu_{3,negative}^2}{\Lambda^2 P^2 g_0}=0.0318(3)\,.
\eqlabel{result4}
\end{equation}
However, since cascading gauge theory undergoes a first order phase transition with spontaneous breaking 
of the chiral symmetry at 
\begin{equation}
\mu_{3,\csb}> \mu_{3,negative}\,,
\eqlabel{compare3}
\end{equation}
and the D3 brane charge at the tip of the conifold in broken phase is zero, the charge in the ground 
state is in fact zero whenever 
\begin{equation}
\mu_3\le \mu_{3,\csb}\,.
\eqlabel{result5}
\end{equation}
Furthermore, chirally symmetric states of cascading gauge theory on $S^3$ develop symmetry breaking tachyonic 
instabilities at $\mu_{3,tachyon}$ (below the first order chiral symmetry breaking scale $\mu_{3,\csb}$)
\begin{equation}
\ln  \frac{\mu_{3,tachyon}^2}{\Lambda^2 P^2 g_0}=0.3297(3)\,.
\eqlabel{compare4}
\end{equation}
which is again above  $\mu_{3,negative}$.

Our results represented here, together with those reported in \cite{landscape}, point that the singularity of 
smeared anti-D3 branes at the tip of the conifold is unphysical: had it been otherwise, we should have been able to implement 
an infrared cutoff in the geometry with a D3 brane charge measured at the cutoff being negative. The role of the cutoff 
is played by the temperature (as discussed in \cite{landscape}), by the compactification scale 
(when $\calm_4=R\times S^3$), or by the Hubble scale (when $\calm_4=dS_4$). Interesting, we  find that 
the D3 brane charge can become negative when the KT throat geometry is $S^3$ deformed; however this occurs 
in the regime where this phase is unstable both via the first order phase transition and the tachyon condensation 
to $S^3$ deformed KS throat geometry --- the latter geometry has zero D3 brane charge at the tip.    
All this raises questions about construction of generic de Sitter vacua in String Theory \cite{kklt}.

We stress, however, that our analysis does not definitely exclude local non-singular supergravity description of de Sitter vacua
in String Theory. The issue stems from the anti-D3 brane "smearing approximation'' used. Early discussion of the relevant
smearing approximation appeared in \cite{Bena:2011wh,Dymarsky:2011pm}. There, the authors carefully analyzed non-supersymmetric 
deformations of KS geometry, invariant under the $SU(2)\times SU(2)$ global symmetry of the latter. They further identified 
a class of perturbations that is being sources by anti-D3 branes, placed at the tip of the conifold, and then computed the leading-order 
backreaction of those perturbations on KS geometry. Insistence on preserving  the $SU(2)\times SU(2)$ global symmetry is a 
smearing approximation 
--- from the brane perspective it implies that  anti-D3 branes are uniformly distributed (uniformly smeared) 
over the transverse compact five-dimensional manifold. Our discussion here shares the same smearing approximation as in 
\cite{Bena:2011wh,Dymarsky:2011pm}, but extends the analysis to the full (rather than leading-order) backreaction.  
Smearing approximation is a practical tool enabling the analysis of the  complicated cascading geometries involved. However, 
it must be questioned: it is not clear that non-supersymmetric uniform distribution along $T^{1,1}$ directions of anti-D3 branes    
is stable against 'clumping'. While it is highly  desirable to lift this approximation, it is very difficult to do this in practice: 
one is forced to analyze  a coupled nonlinear system of partial differential equations, rather than ordinary differential 
equations. We feel that until fully localized anti-D3 brane analysis in cascading geometries are performed, 
the singularity question of local supergravity description of de Sitter vacua
in String Theory will remain open.

\section{Dual effective actions of cascading gauge theory}\label{action}

Consider $SU(2)\times SU(2)\times \zet_2$ invariant states of cascading gauge theory on a 4-dimensional 
manifold $\calm_4\equiv \del\calm_5$. 
Effective gravitational action on a 5-dimensional manifold $\calm_5$ 
describing holographic dual of such states was derived in \cite{ksbh}:
\begin{equation}
\begin{split}
S_5\left[g_{\mu\nu},\Omega_i,h_i,\Phi\right]=& \frac{108}{16\pi G_5} 
\int_{\calm_5} {\rm vol}_{\calm_5}\ \Omega_1 \Omega_2^2\om_3^2\ 
\biggl\lbrace 
 R_{10}-\frac 12 \left(\nabla \Phi\right)^2\\
&-\frac 12 e^{-\Phi}\left(\frac{(h_1-h_3)^2}{2\om_1^2\om_2^2\om_3^2}+\frac{1}{\om_3^4}\left(\nabla h_1\right)^2
+\frac{1}{\om_2^4}\left(\nabla h_3\right)^2\right)
\\
&-\frac 12 e^{\Phi}\left(\frac{2}{\om_2^2\om_3^2}\left(\nabla h_2\right)^2
+\frac{1}{\om_1^2\om_2^4}\left(h_2-\frac P9\right)^2
+\frac{1}{\om_1^2\om_3^4} h_2^2\right)
\\
&-\frac {1}{2\Omega_1^2\Omega_2^4\om_3^4}\left(4{\om}_0+ h_2\left(h_3-h_1\right)+\frac 19 P h_1\right)^2
\biggr\rbrace\,,\\
\end{split}
\eqlabel{5action}
\end{equation}
where $\Omega_0$ is a constant, $R_{10}$ is given by
\begin{equation}
\begin{split}
R_{10}=R_5&+\left(\frac{1}{2\om_1^2}+\frac{2}{\om_2^2}+\frac{2}{\om_3^2}-\frac{\om_2^2}{4\om_1^2\om_3^2}
-\frac{\om_3^2}{4\om_1^2\om_2^2}-\frac{\om_1^2}{\om_2^2\om_3^2}\right)-2\Box \ln\left(\om_1\om_2^2\om_3^2\right)\\
&-\biggl\{\left(\nabla\ln\om_1\right)^2+2\left(\nabla\ln\om_2\right)^2
+2\left(\nabla\ln\om_3\right)^2+\left(\nabla\ln\left(\om_1\om_2^2\om_3^2\right)\right)^2\biggr\}\,,
\end{split}
\eqlabel{ric5}
\end{equation}
and $R_5$ is the five-dimensional Ricci scalar of the metric 
\begin{equation}
ds_{5}^2 =g_{\mu\nu}(y) dy^{\mu}dy^{\nu}\,,
\eqlabel{5met}
\end{equation}
that forms part of the ten dimensional full metric
\begin{equation}
ds_{10}^2 = ds_{5}^2 + ds^2_{T^{1,1}}\,, \qquad ds^2_{T^{1,1}} = \Omega_1^2(y) g_5^2 + \Omega_2^2(y) (g_3^2 + g_4^2) + \Omega_3^2(y) (g_1^2 + g_2^2).
\eqlabel{10dmetric}
\end{equation}

One-forms $\{g_i\}$ (for $i=1,\cdots,5$) are the usual forms defined in the warp-squashed $T^{1,1}$ and are given as in \cite{ksbh}, for coordinates $0 \leq \psi \leq 4 \pi$, $0 \leq \theta_a \leq \pi$ and $0 \leq \phi_a \leq 2 \pi$ ($a=1,2$).

All the covariant derivatives $\nabla_\lambda$  are
with respect to the metric \eqref{5met}. Fluxes (and dilaton $\Phi$) are parametrized in such a way that functions $h_1(y), h_2(y), h_3(y)$ appear as
\begin{equation}
\begin{split}
B_2 & =  h_1(y) g_1 \wedge g_2 + h_3(y) g_3 \wedge g_4, \\
F_3=&\frac 19 P\ g_5\wedge g_3\wedge g_4+h_2(y)\ \left(g_1\wedge g_2-g_3\wedge g_4\right)\wedge g_5
\\
&\qquad +\left(g_1\wedge g_3+g_2\wedge g_4\right)\wedge d\left(h_2(y)\right)\,,\\
\Phi & = \Phi (y),
\end{split}
\eqlabel{fluxes}
\end{equation}
where $P$ corresponds to the number of fractional branes in the conifold.

Finally, $G_5$ is the five dimensional effective gravitational constant  
\begin{equation}
G_5\equiv \frac{729}{4\pi^3}G_{10}\,,
\eqlabel{g5deff}
\end{equation}
where $G_{10}$ is a 10-dimensional gravitational constant of 
type IIB supergravity.

Chirally symmetric 
states of the cascading gauge theory are described by the gravitational configurations of \eqref{5action} 
subject to constraints   
\begin{equation}
h_1=h_3\,,\qquad h_2=\frac{P}{18}\,,\qquad \om_2=\om_3\,.
\eqlabel{cinv}
\end{equation}

In what follows, we find it convenient to introduce 
\begin{equation}
\begin{split}
h_1=&\frac 1P\left(\frac{K_1}{12}-36\Om_0\right)\,,\qquad h_2=\frac{P}{18}\ K_2\,,\qquad 
h_3=\frac 1P\left(\frac{K_3}{12}-36\Om_0\right)\,,\\
\Om_1=&\frac 13 f_c^{1/2} h^{1/4}\,,\qquad \Om_2=\frac {1}{\sqrt{6}} f_a^{1/2} h^{1/4}\,,\qquad 
\Om_3=\frac {1}{\sqrt{6}} f_b^{1/2} h^{1/4}\,,
\end{split}
\eqlabel{redef}
\end{equation}

\section{Chirally symmetric phase of cascading gauge theory on $dS_4$}\label{symmetric}

We consider here $SU(2)\times SU(2)\times U(1)\times SO(4)$ (chirally-symmetric) states of the 
strongly coupled cascading gauge theory. 
We find it convenient to use  a radial coordinate introduced in \cite{aby}: 
\begin{equation}
ds_5^2=g_{\mu\nu}(y)dy^\mu dy^\nu=h^{-1/2}\r^{-2}\ \biggl(-dt^2
+\frac{1}{H^2 }\cosh^2(Ht) \left(dS^3\right)^2\biggr)
+h^{1/2}\r^{-2}\ (d\r)^2\,,
\eqlabel{metricaby}
\end{equation}  
where  $h=h(\r)$.
Furthermore, we use parametrization \eqref{redef} and denote\footnote{Recall that for the 
unbroken chiral symmetry we must set $K_2(\r)\equiv 1$.}
\begin{equation}
f_c=f_2\,,\qquad f_a=f_b=f_3\,,\qquad K_1=K_3=K\,,\qquad \Phi=\ln g\,,
\eqlabel{om12}
\end{equation}
with $f_i=f_i(\r)$, and $K=K(\r)$, $g=g(\r)$. 

Notice that parametrization \eqref{metricaby} is not unique --- the diffeomorphisms 
of the type 
\begin{equation}
\left( \begin{array}{c}
\r  \\
h  \\
f_2\\
f_3\\
K\\
g   \end{array} \right)\
\Longrightarrow \left( \begin{array}{c}
\hr  \\
\hh  \\
\hf_2\\
\hf_3\\
\hK\\
\hat{g}   \end{array} \right)
=
\left( \begin{array}{c}
{\r}/{(1+\a\ \r)}  \\
(1+\a\ \r)^4\ h \\
(1+\a\ \r)^{-2}\ f_2\\
(1+\a\ \r)^{-2}\ f_3\\
K\\
{g}   \end{array} \right)\,,\qquad \a={\rm const}\,,
\eqlabel{leftover}
\end{equation}
preserve the general form of the metric. We can completely fix \eqref{leftover}, \ie
parameter $\a$ in \eqref{leftover},
requiring that for a geodesically complete $\calm_5$  the radial coordinate $\rho$
extends as 
\begin{equation}
\rho\in [0,+\infty)\,.
\eqlabel{extend}
\end{equation}

\subsection{Equations of motion}

For a background ansatz \eqref{metricaby}, \eqref{om12}, 
the  equations of motion obtained from \eqref{5action} take form
\begin{equation}
\begin{split}
0=&f_2''+\frac{f_2 (g')^2}{8g^2}- \frac{3 f_2 (K')^2}{16 h f_3^2 g P^2}+\frac{f_2 (h')^2}{8h^2}
-\frac{3f_2 (f_3')^2}{4f_3^2}- \frac{f_2'^2}{2f_2}+\frac{f_2 h'}{h \r}
+\left( \frac{3f_3'}{2f_3}-\frac3\r\right) f_2'\\
&+ \frac{3g P^2}{4h f_3^2 \r^2}- \frac{K^2}{8 h^2 f_3^4 \r^2}+\frac{f_2 (5 f_3^2-9 f_2+6 f_3)}{f_3^2 \r^2}
-3 h f_2 H^2\,,
\end{split}
\eqlabel{eq2}
\end{equation}
\begin{equation}
\begin{split}
0=&f_3''+\frac{(K')^2}{16 h f_3 g P^2}+ \frac{f_3 (g')^2}{8 g^2}+ \frac{f_3 (h')^2}{8h^2}
+\frac{(f_3')^2}{4f_3}-\frac{3 f_3'}{\r}+\frac{f_3 h'}{h \r}
- \frac{g P^2}{4f_2 h f_3 \r^2}\\
&-\frac{K^2}{8f_2 h^2 f_3^3 \r^2}
+\frac{5 f_3^2-6 f_3+3 f_2}{f_3 \r^2}-3 h f_3 H^2\,,
\end{split}
\eqlabel{eq3}
\end{equation}
\begin{equation}
\begin{split}
0=&h''+ \frac{3(K')^2}{16g f_3^2 P^2}-\frac{h (g')^2}{8g^2}
-\frac{9(h')^2}{8h}+ \frac{3h (f_3')^2}{4f_3^2}
+\left(\frac{2 f_3'}{f_3}+\frac{f_2'}{2f_2}-\frac 4\r\right) h'+\frac{h f_2'}{\r f_2}
\\
&+\frac{9K^2}{8f_2 h f_3^4 \r^2}+\left(\frac{4 h}{f_3 \r}+ \frac{h f_2'}{2f_3 f_2}\right) 
f_3'+ \frac{5g P^2}{4f_2 f_3^2 \r^2}+\frac{h (f_2-13 f_3^2-6 f_3)}{f_3^2 \r^2}+9 h^2 H^2\,,
\end{split}
\eqlabel{eq4}
\end{equation}
\begin{equation}
\begin{split}
0=&K''+\left(\frac{f_2'}{2f_2}-\frac {g'}{g}-\frac{h'}{h}-\frac 3\r\right) K'-\frac{2 g K P^2}{h f_2 f_3^2 \r^2}\,,
\end{split}
\eqlabel{eq5}
\end{equation}
\begin{equation}
\begin{split}
0=&g''-\frac{(g')^2}{g}+\left(\frac{2 f_3'}{f_3}+\frac{f_2'}{2f_2}-\frac 3\r\right) g'
+\frac{(K')^2}{4h f_3^2 P^2}-\frac{g^2 P^2}{h f_2 f_3^2 \r^2}\,.
\end{split}
\eqlabel{eq6}
\end{equation}
Additionally we have the first order constraint
\begin{equation}
\begin{split}
0=&(K')^2+\frac{2 h f_3^2 P^2 (g')^2}{g}+\frac{2 f_3^2 P^2 g (h')^2}{h}-12 h P^2 g (f_3')^2
-\frac{8 f_3 h g P^2 (f_3' \r-2 f_3)}{f_2 \r} f_2'\\
&+\frac{16 f_3 g P^2 (4 f_3' h+f_3 h')}{\r}
+\left(96 h f_3-48 h f_3^2-16 h f_2-\frac{4 P^2 g}{f_2}-\frac{2 K^2}{h f_2 f_3^2}\right) \frac{g P^2}{\rho^2}\\
&+48 g P^2 h^2 f_3^2 H^2\,.
\end{split}
\eqlabel{eq7}
\end{equation}
We explicitly verified that the constraint \eqref{eq7} is consistent with 
\eqref{eq2}-\eqref{eq6}.

\subsection{UV asymptotics}\label{uvcond}
The general UV (as $\r\to 0$) asymptotic solution of \eqref{eq2}-\eqref{eq7}  describing the symmetric phase of cascading 
gauge theory takes form
\begin{equation}
\begin{split}
f_2=&1-\a_{1,0}\ (H\r)+\left(-\frac38 P^2 g_0-\frac 14 K_0+\frac14(\a_{1,0})^2+\frac12P^2 g_0\ln\r\right)\ (H\r)^2\\
&+\sum_{n=3}^\infty\sum_{k} a_{n,k}\ (H\r)^{n}\ln^k \r\,,
\end{split}
\eqlabel{f2uv}
\end{equation}
\begin{equation}
\begin{split}
f_3=&1-\a_{1,0}\ (H\r)+\left(-\frac12 P^2 g_0-\frac 14 K_0+\frac14(\a_{1,0})^2+\frac12P^2 g_0\ln\r\right)\ (H\r)^2\\
&+\sum_{n=3}^\infty\sum_{k} b_{n,k}\ (H\r)^{n}\ln^k \r\,,
\end{split}
\eqlabel{f3uv}
\end{equation}
\begin{equation}
\begin{split}
h=&\frac18 P^2 g_0+\frac14 K_0-\frac12 P^2 g_0 \ln \r+\a_{1,0}\left(\frac 12 K_0-P^2 g_0\ln\r\right)
\ (H\r)+\biggl(
\frac{119}{576} P^4 g_0^2\\
&+\frac{31}{96} K_0 P^2 g_0-\frac 14 P^2 g_0\a_{1,0}^2 
+\frac 18 K_0^2+\frac 58 \a_{1,0}^2 K_0-\frac{1}{96} P^2 g_0 (62 P^2 g_0+120 \a_{1,0}^2\\
&+48 K_0) \ln\r
+\frac12 P^4 g_0^2 \ln^2\r
\biggr)\ (H\r)^2
+\sum_{n=3}^\infty\sum_{k} h_{n,k}\ 
(H\r)^{n}\ln^k \r\,,
\end{split}
\eqlabel{huv}
\end{equation}
\begin{equation}
\begin{split}
K=&K_0-2 P^2 g_0 \ln \r-P^2 g_0 \a_{1,0} \ (H\r)
+ \biggl(\frac{1}{16} P^2 g_0 (2K_0+9 P^2 g_0-4\a_{1,0}^2)\\
&-\frac 14 P^4 g_0^2 \ln \r\biggr)\ (H\r)^2
+\sum_{n=3}^\infty\sum_{k} K_{n,k}\ (H\r)^{n}\ln^k \r\,,
\end{split}
\eqlabel{kuv}
\end{equation}
\begin{equation}
g=g_0 \left(1-\frac12 P^2 g_0\ (H\r)^2+\sum_{n=3}^\infty\sum_{k} g_{n,k}\ (H\r)^{n}\ln^k \r\right)\,.
\eqlabel{guv}
\end{equation}
It is characterized by 7 parameters:
\begin{equation}
\{K_0\,,\ H\,,\ g_0\,,\  \a_{1,0}\,,\ a_{4,0}\,,\ a_{6,0}\,,\ a_{8,0}\,,\ g_{4,0}\}\,.
\eqlabel{uvpar}
\end{equation}
In what follows we developed the UV expansion to order $\calo(\r^{12})$
inclusive.

\subsection{IR asymptotics}\label{ircond}
We use a radial coordinate $\r$ that extends to infinity, see \eqref{extend}. Introducing 
\begin{equation}
y\equiv \frac 1\r\,,\qquad h^h\equiv y^{-2}\ h\,,\qquad  f^h_{2,3}\equiv y\ f_{2,3}\,,
\eqlabel{horfunc}
\end{equation}
the general IR (as $y\to 0$) asymptotic solution of  \eqref{eq2}-\eqref{eq7} describing the symmetric phase of cascading 
gauge theory takes form
\begin{equation}
\begin{split}
f_2^h=&f_{2,0}^h-\frac{9 H^2 P^2 (f_{3,0}^h)^2 g_0^h+6 H^4 (K_0^h)^2-17 (f_{2,0}^h)^2 (f_{3,0}^h)^2+6 f_{2,0}^h (f_{3,0}^h)^3} 
{5(f_{3,0}^h)^4}\ y
\\
&+\sum_{n=2} f_{2,n}^h y^{n}\,,
\end{split}
\eqlabel{f2hy}
\end{equation}
\begin{equation}
\begin{split}
f_3^h=&f_{3,0}^h-\frac{H^2 P^2 (f_{3,0}^h)^2 g_0^h+6 H^4 (K_0^h)^2+7 (f_{2,0}^h)^2 (f_{3,0}^h)^2-18 f_{2,0}^h (f_{3,0}^h)^3}
{5f_{2,0}^h (f_{3,0}^h)^3}\ y\\
&+\sum_{n=2} f_{3,n}^h y^{n}\,,
\end{split}
\eqlabel{f3hy}
\end{equation}
\begin{equation}
\begin{split}
h^h=&\frac{1}{4H^2}\biggl(1-\frac25 \frac{(3 H^2 P^2 (f_{3,0}^h)^2 g_0^h+10 H^4 (K_0^h)^2+(f_{2,0}^h)^2 (f_{3,0}^h)^2-6 f_{2,0}^h 
(f_{3,0}^h)^3}
 {(f_{3,0}^h)^4 f_{2,0}^h}\ y\\
&+\sum_{n=2} h_{n}^h y^{n}\biggr)\,,
\end{split}
\eqlabel{hhy}
\end{equation}
\begin{equation}
K=K_0^h+\frac{16 K_0^h g_0^h P^2 H^2 }{5 (f_{3,0}^h)^2 f_{2,0}^h}\ y+\sum_{n=2} K_{n}^h y^{n}\,,
\eqlabel{khy}
\end{equation}
\begin{equation}
g=g_0^h\left (1+\frac{8g_0^h P^2 H^2 }{5 (f_{3,0}^h)^2 f_{2,0}^h}\ y+\sum_{n=2} g_{n}^h y^{n}\right)\,.
\eqlabel{ghy}
\end{equation}
It is characterized by 4 additional parameters:
\begin{equation}
\{K_0^h\,,\ g_0^h\,,\  f_{2,0}^h\,,\ f_{3,0}^h\}\,.
\eqlabel{irpar}
\end{equation}
In what follows we developed the IR expansion to order $\calo(y^{6})$ inclusive.

\subsection{Symmetries}\label{symmetries}
The background geometry \eqref{metricaby}, \eqref{om12} enjoys 4 distinct scaling symmetries. 
We now discuss these symmetries and exhibit their action on the asymptotic parameters \eqref{uvpar}.
\nxt First, we have:
\begin{equation}
\begin{split}
&P\to \l\ P\,,\ g\to \frac 1\l\ g\,,\qquad \{\r,f_i,h,K\}\to \{\r,f_i,h,K\}\,,\qquad 
\{y,f_i^h,h^h\}\to \{y,f_i,h^h\}\,,
\end{split}
\eqlabel{scale1}
\end{equation} 
which acts on the asymptotic parameters as 
\begin{equation}
\begin{split}
&g_0\to \frac 1\l\ g_0\,,\\
&\{K_0\,,
H\,,  \a_{1,0}\,, a_{4,0}\,, a_{6,0}\,, a_{8,0}\,, g_{4,0}\}
\to \{K_0\,, H\,,  \a_{1,0}\,, a_{4,0}\,, a_{6,0}\,, a_{8,0}\,, g_{4,0}\}\,,
\end{split}
\eqlabel{action1}
\end{equation}
and
\begin{equation}
\begin{split}
\{K_0^h\,,\ g_0^h\,,\ f_{2,0}^h\,,\ f_{3,0}^h\}\to \{K_0^h\,,\  \l^{-1} g_0^h\,,\ f_{2,0}^h\,,\ f_{3,0}^h\}\,.
\end{split}
\eqlabel{action1h}
\end{equation}
We can use the exact symmetry \eqref{scale1} to set 
\begin{equation}
g_0=1\,.
\eqlabel{res1}
\end{equation}
\nxt Second, we have:
\begin{equation}
\begin{split}
&P\to \l\ P\,,\ \r\to \frac 1\l\ \r\,,\ h\to \l^2\ h\,,\ K\to \l^2 K\,,\qquad \{H,f_i,g\}\to\{H,f_i, g\}\,,
\\
&\{y,f_2^h,f_3^h,h^h\}\to \{\l y,\l f_2^h,\l f_3^h, h^h\}\,,
\end{split}
\eqlabel{scale2}
\end{equation}
which acts on the asymptotic parameters as 
\begin{equation}
\begin{split}
g_0\to g_0 \,,
\end{split}
\eqlabel{action21}
\end{equation}
\begin{equation}
\begin{split}
\a_{1,0}\to \l\a_{1,0}\,,
\end{split}
\eqlabel{action21h}
\end{equation}
\begin{equation}
\begin{split}
K_0\to \l^2 \biggl(K_0-2P^2 g_0\ \ln\l \biggr)\,,
\end{split}
\eqlabel{action22}
\end{equation}
\begin{equation}
\begin{split}
a_{4,0}&\to \l^4\biggl(a_{4,0}+\frac{1}{48}P^2g_0 (3K_0-P^2g_0)\ \ln\l-\frac{1}{16}P^4g_0^2\ \ln^2\l
\biggr)\,,
\end{split}
\eqlabel{action26}
\end{equation}
\begin{equation}
\begin{split}
g_{4,0}&\to \l^4\biggl(g_{4,0}+\left(-\frac{3}{16} P^2 \a_{1,0}^2 g_0-\frac{5}{64} K_0 P^2 g_0+\frac{37}{96} P^4 g_0^2
+3 a_{4,0}\right)\ \ln\l\\
&+\frac{3}{64} P^2 g_0 (P^2 g_0+2 K_0)\ \ln^2\lambda-\frac{1}{16}  P^4 g_0^2\ \ln^3\l\biggr)\,,
\end{split}
\eqlabel{action25}
\end{equation}
\begin{equation}
\begin{split}
&a_{6,0}\to \l^6\biggl(a_{6,0}+\biggl(\frac{89}{40} P^2 a_{4,0} g_0
-\frac15 P^2 g_0 g_{4,0}+\frac15 K_0 a_{4,0}+\frac{1491}{32000} K_0 P^4 g_0^2\\
&+\frac{689743}{3840000} P^6 g_0^3
+\frac{11}{320} K_0 P^2 \a_{1,0}^2 g_0-\frac{197}{640} P^4 \a_{1,0}^2 g_0^2
+\frac{419}{38400} K_0^2 P^2 g_0\biggr)\ \ln\l\\
&+
\biggl(-\frac{1}{64} P^4 \a_{1,0}^2 g_0^2+\frac{1}{160} K_0^2 P^2 g_0+\frac{171}{3200} K_0 P^4 g_0^2
-\frac12 P^2 a_{4,0} g_0-\frac{1733}{16000} P^6 g_0^3\biggr)\ \ln^2\l
\\
&+\biggl(-\frac{463}{14400} P^6 g_0^3-\frac{3}{160} K_0 P^4 g_0^2\biggr)\ \ln^3\l+\frac{3}{320} P^6 g_0^3\ \ln^4\l
\biggr)\,,
\end{split}
\eqlabel{action27}
\end{equation}
\begin{equation}
\begin{split}
&a_{8,0}\to \l^8\biggl(a_{8,0}+\frac{1}{P^2 g_0 (70 K_0-141 P^2 g_0)} \biggl(
-140 P^4 a_{8,0} g_0^2-\frac{11289869889229}{7468070400000} P^{12} g_0^6\\
&+18 K_0^2 a_{4,0}^2
+\frac{79241}{280} K_0 P^4 \a_{1,0}^2 a_{4,0} g_0^2-\frac{67}{2} K_0 P^4 \a_{1,0}^2 g_0^2 g_{4,0}
+\frac{131}{4} K_0^2 P^2 \a_{1,0}^2 a_{4,0} g_0\\
&-24 K_0 P^2 a_{4,0} g_0 g_{4,0}
-\frac{17122502251}{790272000} K_0 P^8 \a_{1,0}^2 g_0^4-\frac{1264903}{26880} K_0 P^6 \a_{1,0}^4 g_0^3
\\
&+\frac{3642629}{537600} K_0^2 P^6 \a_{1,0}^2 g_0^3-\frac34 K_0^2 P^4 \a_{1,0}^4 g_0^2
-\frac{308363}{560} P^6 \a_{1,0}^2 a_{4,0} g_0^3+\frac{135}{4} P^6 \a_{1,0}^2 g_0^3 g_{4,0}
\\&+\frac{16067}{6720} K_0^3 P^4 \a_{1,0}^2 g_0^2-\frac{53709659}{3087000} K_0 P^6 a_{4,0} g_0^3
-\frac{15332}{1225} K_0 P^6 g_0^3 g_{4,0}-\frac{875}{4} P^4 \a_{1,0}^4 a_{4,0} g_0^2\\
&+\frac{1923781}{33600} K_0^2 P^4 a_{4,0} g_0^2
-\frac{2001}{560} K_0^2 P^4 g_0^2 g_{4,0}+350 P^4 \a_{1,0}^2 a_{6,0} g_0^2
-12 P^4 a_{4,0} g_0^2 g_{4,0}\\
&+\frac{9013}{1120} K_0^3 P^2 a_{4,0} g_0-\frac{5706}{35} K_0 P^2 a_{4,0}^2 g_0
+\frac{17699297459}{592704000} P^{10} \a_{1,0}^2 g_0^5\\
&+\frac{1365178374361}{553190400000} K_0 P^{10} g_0^5
+\frac{4598761}{80640} P^8 \a_{1,0}^4 g_0^4+\frac{2135}{192} P^6 \a_{1,0}^6 g_0^3
\\&+\frac{48152049931}{189665280000} K_0^2 P^8 g_0^4-\frac{33703011407}{148176000} P^8 a_{4,0} g_0^4
+\frac{14708381}{529200} P^8 g_0^4 g_{4,0}\\
&+\frac{402129463}{210739200} K_0^3 P^6 g_0^3
+\frac{3965783}{15052800} K_0^4 P^4 g_0^2+\frac{1315}{6} P^6 a_{6,0} g_0^3
+\frac{49853}{70} P^4 a_{4,0}^2 g_0^2\\
&-8 P^4 g_0^2 g_{4,0}^2\biggr)\ \ln\lambda
+\biggl(-\frac{5436207853}{30732800000} P^8 g_0^4-\frac{35277}{171500} P^6 \a_{1,0}^2 g_0^3
-\frac{1469772959}{31610880000} K_0 P^6 g_0^3\\
&+\frac{489}{8960} P^4 \a_{1,0}^4 g_0^2
+\frac{8889}{89600} K_0 P^4 \a_{1,0}^2 g_0^2+\frac{1953403}{105369600} K_0^2 P^4 g_0^2
+\frac{131}{8960} K_0^2 P^2 \a_{1,0}^2 g_0\\
&-\frac{2780609}{1372000} P^4 a_{4,0} g_0^2
+\frac{859}{9800} P^4 g_0^2 g_{4,0}+\frac{9013}{2508800} K_0^3 P^2 g_0-\frac{157}{140} P^2 \a_{1,0}^2 a_{4,0} g_0
\\
&-\frac{2949}{5600} K_0 P^2 a_{4,0} g_0-\frac{3}{280} K_0 P^2 g_0 g_{4,0}
+\frac{9}{560} K_0^2 a_{4,0}-\frac{36}{35} a_{4,0}^2\biggr)\ \ln^2\lambda\\
&+\biggl(\frac{2671073519}{47416320000} P^8 g_0^4-\frac{180151}{2822400} P^6 \a_{1,0}^2 g_0^3
-\frac{3778787}{56448000} K_0 P^6 g_0^3-\frac{27}{640} K_0 P^4 \a_{1,0}^2 g_0^2
\\&-\frac{4513}{250880} K_0^2 P^4 g_0^2+\frac{8879}{19600} P^4 a_{4,0} g_0^2
+\frac{1}{140} P^4 g_0^2 g_{4,0}+\frac{3}{8960} K_0^3 P^2 g_0-\frac{3}{40} K_0 P^2 a_{4,0} g_0\\
&\biggr)\
\ln^3\lambda+\biggl(\frac{3590117}{112896000} P^8 g_0^4+\frac{93}{4480} P^6 \a_{1,0}^2 g_0^3
+\frac{4537}{179200} K_0 P^6 g_0^3-\frac{3}{1792} K_0^2 P^4 g_0^2\\
&+\frac{3}{70} P^4 a_{4,0} g_0^2\biggr)\ \ln^4\l
+\biggl(-\frac{4617}{448000} P^8 g_0^4+\frac{3}{1600} K_0 P^6 g_0^3\biggr)\ \ln^5\l-\frac{1}{1600} P^8 g_0^4\ \ln^6\l
\biggr)\,,
\end{split}
\eqlabel{action28}
\end{equation}
and 
\begin{equation}
\begin{split}
&\{K_0^h\,,\ g_0^h\,,\ f_{2,0}^h\,,\ f_{3,0}^h\}\to \{\l^2 K_0^h\,,\ g_0^h\,,\ \l f_{2,0}^h\,,\ \l f_{3,0}^h\}\,.
\end{split}
\eqlabel{action2h}
\end{equation}
We can use the exact symmetry \eqref{scale2} to relate different sets of $\{K_0,P\}$.
For the study of perturbative in $P^2/K_0$ expansion we find it convenient to set $K_0=1$ and 
vary $P^2$. To access the infrared properties of the theory we set $P=1$ and vary $K_0$.
Notice that the two approaches connect at $\{K_0=1,P=1\}$.
\nxt Third, we have:
\begin{equation}
\begin{split}
&\r\to \l\ \r\,,\ H\to \frac1\l\ H\,,\qquad \{P\,,\ f_2\,,\ f_3\,, h\,, K\,, g\}\to 
 \{P\,,\ f_2\,,\ f_3\,, h\,, K\,, g\}\,,\\
&\{y,f_2^h,f_3^h,h^h\}\to \{\l^{-1} y,\l^{-1}f_2^h,\l^{-1}f_3^h,\l^2 h^h\}\,,
\end{split}
\eqlabel{scale3} 
\end{equation}
This scaling symmetry acts on the asymptotic parameters as 
\begin{equation}
\begin{split}
\{g_0\,,\ \a_{1,0}\}\to \{g_0\,,\ \a_{1,0} \}\,,
\end{split}
\eqlabel{action31}
\end{equation}
\begin{equation}
\begin{split}
K_{0}\to K_0+2 P^2 g_0 \ln\l\,,
\end{split}
\eqlabel{action312}
\end{equation}
\begin{equation}
\begin{split}
a_{4,0}\to a_{4,0}+\biggl(\frac{1}{48} P^4 g_0^2-\frac{1}{16} K_0 P^2 g_0\biggr)\ \ln\l-\frac{1}{16} P^4 g_0^2\ \ln^2\l\,,
\end{split}
\eqlabel{action34}
\end{equation}
\begin{equation}
\begin{split}
&g_{4,0}\to g_{4,0}+\biggl(\frac{3}{16} P^2 \a_{1,0}^2 g_0+\frac{5}{64} K_0 P^2 g_0-\frac{37}{96} P^4 g_0^2-3 a_{4,0}\biggr)\ \ln\l
+\biggl(\frac{3}{64} P^4 g_0^2\\
&+\frac{3}{32} K_0 P^2 g_0\biggr)\ \ln^2\l+\frac{1}{16} P^4 g_0^2\ \ln^3\l\,,
\end{split}
\eqlabel{action33}
\end{equation}
\begin{equation}
\begin{split}
&a_{6,0}\to a_{6,0}+\biggl(-\frac{89}{40} P^2 a_{4,0} g_0+\frac15 P^2 g_0 g_{4,0}
-\frac15 K_0 a_{4,0}-\frac{1491}{32000} K_0 P^4 g_0^2\\
&-\frac{689743}{3840000} P^6 g_0^3
-\frac{11}{320} K_0 P^2 \a_{1,0}^2 g_0+\frac{197}{640} P^4 \a_{1,0}^2 g_0^2-\frac{419}{38400} 
K_0^2 P^2 g_0\biggr)\ \ln\l
+\biggl(\\
&-\frac{1}{64} P^4 \a_{1,0}^2 g_0^2+\frac{1}{160} K_0^2 P^2 g_0+\frac{171}{3200} K_0 P^4 g_0^2-\frac12 P^2 a_{4,0} g_0
-\frac{1733}{16000} P^6 g_0^3\biggr)\ \ln^2\l\\
&+\biggl(\frac{463}{14400} P^6 g_0^3
+\frac{3}{160} K_0 P^4 g_0^2\biggr)\ \ln^3\l
+\frac{3}{320} P^6 g_0^3\ \ln^4\l\,,
\end{split}
\eqlabel{action35}
\end{equation}
\begin{equation}
\begin{split}
&a_{8,0}\to a_{8,0}+\frac{1}{P^2 g_0 (70 K_0-141 P^2 g_0)} \biggl(\frac{11289869889229}{7468070400000} P^{12} g_0^6
+\biggl(-\frac{17699297459}{592704000} \a_{1,0}^2\\
&-\frac{1365178374361}{553190400000} K_0\biggr) P^{10} g_0^5
+\biggl(\frac{17122502251}{790272000} K_0 \a_{1,0}^2+\frac{33703011407}{148176000} a_{4,0}
\\
&-\frac{14708381}{529200} g_{4,0}-\frac{48152049931}{189665280000} K_0^2-\frac{4598761}{80640} \a_{1,0}^4
\biggr) P^8 g_0^4
+\biggl(\frac{1264903}{26880} K_0 \a_{1,0}^4\\
&+\frac{308363}{560} \a_{1,0}^2 a_{4,0}-\frac{135}{4} \a_{1,0}^2 g_{4,0}
+\frac{53709659}{3087000} K_0 a_{4,0}+\frac{15332}{1225} K_0 g_{4,0}-\frac{402129463}{210739200} K_0^3
\\
&-\frac{2135}{192} \a_{1,0}^6-\frac{1315}{6} a_{6,0}-\frac{3642629}{537600} K_0^2 \a_{1,0}^2\biggr) P^6 g_0^3
+\biggl(\frac 34 K_0^2 \a_{1,0}^4+12 a_{4,0} g_{4,0}\\
&+\frac{2001}{560} K_0^2 g_{4,0}
+\frac{67}{2} K_0 \a_{1,0}^2 g_{4,0}+\frac{875}{4} \a_{1,0}^4 a_{4,0}-\frac{1923781}{33600} K_0^2 a_{4,0}
+140 a_{8,0}-350 \a_{1,0}^2 a_{6,0}\\
&-\frac{79241}{280} K_0 \a_{1,0}^2 a_{4,0}-\frac{16067}{6720} K_0^3 \a_{1,0}^2
+8 g_{4,0}^2-\frac{49853}{70} a_{4,0}^2-\frac{3965783}{15052800} K_0^4\biggr) P^4 g_0^2\\
&+\biggl(-\frac{9013}{1120} K_0^3 a_{4,0}+\frac{5706}{35} K_0 a_{4,0}^2-\frac{131}{4} K_0^2 \a_{1,0}^2 a_{4,0}
+24 K_0 a_{4,0} g_{4,0}\biggr) P^2 g_0\\
&-18 K_0^2 a_{4,0}^2\biggr)\ \ln\l
+\biggl(-\frac{5436207853}{30732800000} P^8 g_0^4+\biggl(-\frac{1469772959}{31610880000} K_0
-\frac{35277}{171500} \a_{1,0}^2\biggr) P^6 g_0^3\\
&+\biggl(\frac{489}{8960} \a_{1,0}^4
+\frac{8889}{89600} K_0 \a_{1,0}^2+\frac{1953403}{105369600} K_0^2-\frac{2780609}{1372000} a_{4,0}
+\frac{859}{9800} g_{4,0}\biggr) P^4 g_0^2\\
&+\biggl(\frac{131}{8960} K_0^2 \a_{1,0}^2
-\frac{2949}{5600} K_0 a_{4,0}-\frac{3}{280} K_0 g_{4,0}+\frac{9013}{2508800} K_0^3
-\frac{157}{140} \a_{1,0}^2 a_{4,0}\biggr) P^2 g_0\\
&+\frac{9}{560} K_0^2 a_{4,0}-\frac{36}{35} a_{4,0}^2\biggr)\ \ln^2\l
+\biggl(-\frac{2671073519}{47416320000} P^8 g_0^4
+\biggl(\frac{180151}{2822400} \a_{1,0}^2\\
&+\frac{3778787}{56448000} K_0\biggr) P^6 g_0^3
+\biggl(-\frac{8879}{19600} a_{4,0}-\frac{1}{140} g_{4,0}+\frac{4513}{250880} K_0^2
+\frac{27}{640} K_0 \a_{1,0}^2\biggr) P^4 g_0^2\\
&+\biggl(\frac{3}{40} K_0 a_{4,0}
-\frac{3}{8960} K_0^3\biggr) P^2 g_0\biggr)\ \ln^3\l
+\biggl(\frac{3590117}{112896000} P^8 g_0^4+\biggl(\frac{93}{4480} \a_{1,0}^2
\\
&+\frac{4537}{179200} K_0\biggr) P^6 g_0^3+\biggl(\frac{3}{70} a_{4,0}-\frac{3}{1792} K_0^2\biggr) 
P^4 g_0^2\biggr)\ \ln^4\l
+\biggl(\frac{4617}{448000} P^8 g_0^4\\
&-\frac{3}{1600} K_0 P^6 g_0^3\biggr)\ \ln^5\l
-\frac{1}{1600} P^8 g_0^4\ \ln^6\l\,,
\end{split}
\eqlabel{action36}
\end{equation}
and  
\begin{equation}
\begin{split}
&\{K_0^h\,,\ g_0^h\,,\ f_{2,0}^h\,,\ f_{3,0}^h\}\to \{K_0^h\,,\ g_0^h\,,\ \l^{-1} f_{2,0}^h\,,\ \l^{-1} f_{3,0}^h\}\,.
\end{split}
\eqlabel{action3h}
\end{equation}
We can use the exact symmetry \eqref{scale3} 
to set
\begin{equation}
H=1\,.
\eqlabel{res3f0}
\end{equation}
\nxt Forth, we have residual diffeomorphisms \eqref{leftover} of the metric parametrization \eqref{metricaby}.
The latter transformations act on asymptotic parameters as 
\begin{equation}
\begin{split}
\{g_0\,,\  H\,,\ K_0\}\to \{g_0\,,\ H\,, K_0 \}\,,
\end{split}
\eqlabel{action41}
\end{equation}
\begin{equation}
\begin{split}
\a_{1,0}\to \a_{1,0}+2\ \frac{\a}{H} \,,
\end{split}
\eqlabel{action412}
\end{equation}
\begin{equation}
\begin{split}
a_{4,0}\to a_{4,0}+\frac 14 P^2 \a_{1,0} g_0\ \frac{\a}{H}+\frac14 P^2 g_0\ \frac{\alpha^2}{H^2}\,,
\end{split}
\eqlabel{action42}
\end{equation}
\begin{equation}
\begin{split}
g_{4,0}\to  g_{4,0}-\frac32 P^2 \a_{1,0} g_0\ \frac{\a}{H}-\frac32 P^2 g_0\ \frac{\alpha^2}{H^2}\,,
\end{split}
\eqlabel{action44}
\end{equation}
\begin{equation}
\begin{split}
&a_{6,0}\to a_{6,0}+\biggl(-\frac{11}{96} P^4 g_0^2 \a_{1,0}-\frac18 P^2 g_0 \a_{1,0}^3
+\frac{5}{32} P^2 g_0 K_0 \a_{1,0}+3 \a_{1,0} a_{4,0}\biggr)\ \frac{\a}{H}\\
&+\biggl(-\frac{11}{96} P^4 g_0^2+\frac{5}{32} K_0 P^2 g_0+3 a_{4,0}\biggr)\ 
\frac{\a^2}{H^2}+\frac14 P^2 \a_{1,0} g_0\ \frac{\a^3}{H^3} +\frac{1}{8}P^2 g_0 \frac{\alpha^4}{H^4}\,,
\end{split}
\eqlabel{action45}
\end{equation}
\begin{equation}
\begin{split}
&a_{8,0}\to a_{8,0}+\biggl(\frac{1791949}{2560000} P^6 \a_{1,0} g_0^3
+\biggl(\frac{16839}{64000} K_0 \a_{1,0}-\frac{10337}{11520} \a_{1,0}^3\biggr) P^4 g_0^2
\\
&+\biggl(-\frac{9}{10} \a_{1,0} g_{4,0}-\frac{473}{1920} K_0 \a_{1,0}^3+\frac{1417}{25600} K_0^2 \a_{1,0}
+\frac14 \a_{1,0}^5+\frac{761}{80} \a_{1,0} a_{4,0}\biggr) P^2 g_0
\\&-5 \a_{1,0}^3 a_{4,0}+\frac{9}{10} K_0 \a_{1,0} a_{4,0}+10 \a_{1,0} a_{6,0}\biggr)\ \frac{\a}{H}
+\biggl(\frac{1791949}{2560000} P^6 g_0^3
+\biggl(-\frac{1793}{1280} \a_{1,0}^2\\
&+\frac{16839}{64000} K_0\biggr) P^4 g_0^2
+\biggl(\frac{761}{80} a_{4,0}-\frac{9}{10} g_{4,0}+\frac{1417}{25600} K_0^2+\frac{99}{640} K_0 \a_{1,0}^2\biggr) P^2 g_0
+10 a_{6,0}\\&+\frac{9}{10} K_0 a_{4,0}\biggr)\ \frac{\a^2}{H^2}
+\biggl(-\frac{145}{144} P^4 g_0^2 \a_{1,0}+\biggl(-\frac{5}{12} \a_{1,0}^3
+\frac{77}{96} K_0 \a_{1,0}\biggr) P^2 g_0\\
&+10 \a_{1,0} a_{4,0}\biggr)\ \frac{\a^3}{H^3}
+\biggl(-\frac{145}{288} P^4 g_0^2+\frac{77}{192} K_0 P^2 g_0+5 a_{4,0}\biggr)\ \frac{\a^4}{H^4}
+\frac14 g_0 P^2 \a_{1,0}\ \frac{\a^5}{H^5}
\\&+\frac{1}{12} P^2 g_0\ \frac{\a^6}{H^6}\,,
\end{split}
\eqlabel{action46}
\end{equation}
and 
\begin{equation}
\begin{split}
\{K_0^h\,,\ g_0^h\,,\ f_{2,0}^h\,,\ f_{3,0}^h\}\to \{K_0^h\,,\ g_0^h\,,\  f_{2,0}^h\,,\  f_{3,0}^h\}\,.
\end{split}
\eqlabel{action4h}
\end{equation}
As mentioned earlier, the diffeomorphisms \eqref{leftover} can be completely fixed requiring that 
\begin{equation}
\lim_{\r\to +\infty} h^{-1/2}\r^{-2}=0\,,
\eqlabel{geocoms}
\end{equation}
\ie in the holographic dual to the symmetric phase of  cascading gauge theory the manifold $\calm_5$ geodesically completes 
in the interior with smooth shrinking of $dS_4$ (see \eqref{metricaby}) as $\r\to +\infty$.

\subsection{Keeping the physical parameters fixed}

Holographic duality between a gauge theory and a supergravity necessitates the dictionary 
relating the parameters of the two. Specifically, the non-zero non-normalizable components 
of the gravitational modes are mapped to parameters of the gauge theory. From 
\eqref{f2uv}-\eqref{guv} these are: $H$ (characterizing the curvature
of the boundary metric  $\del\calm_5$ in \eqref{metricaby}), the asymptotic string coupling 
$g_0$, the number of fractional $D3$ branes $P$, and the asymptotic five-form flux parameter $K_0$. 
It is straightforward to map the former 3 parameters: $H$ is simply the Hubble constant of the 
background geometry on which we formulate the cascading gauge theory; the value of $g_0$ is related to the 
sum of the gauge couplings of the cascading gauge theory in the far UV (see \eqref{defg0}), and the 
parameter $P$ is the rank difference of the cascading gauge theory group factors inducing the 
renormalization group flow. It is a bit more tricky to identify the last gravitational parameter --- $K_0$.
The difficulty arises from the fact that $K_0$ can not be identified in the far $UV$, \ie as 
$\rho\to \infty$ in \eqref{kuv}, and thus it is sensitive to the rescaling of the radial coordinate 
$\rho$. To address this question, the authors of \cite{Bena:2011wh,Dymarsky:2011pm} 
proposed matching the D3-brane Maxwell charge of two cascading geometries (supposedly dual to the same gauge theory)
on a fixed\footnote{Fixing a UV screen requires a careful matching of the radial coordinates.} UV holographic screen.
An alternative (and equivalent) method, first proposed in \cite{aby}, is to notice that $K_0$ must be related to the 
strong coupling scale $\Lambda$ of the cascading gauge theory, see \eqref{diff}. It becomes clear then why rescaling of the 
radial coordinate $\r$ requires modification of $K_0$: holographic radial coordinate serves as an 'energy scale ruler',
and its rescaling necessitates corresponding rescaling of the dimensionful gauge theory parameters 
($H$ and $\Lambda$ in our case). It is also clear that the combination of gravitational parameters dual to the 
ratio of $\frac{H}{\Lambda}$ must be left invariant under the rescaling. Specifically, in our case the 
corresponding combination must be invariant under the gravitational symmetry transformations rescaling the asymptotic radial coordinate $\r$,
\ie the symmetries \eqref{scale2}  and \eqref{scale3}. Turns out that this is  sufficient to unambiguously relate $K_0$ to the strong coupling 
scale of the cascading gauge theory. We point out that this approach was used in \cite{aby} and \cite{abk}, and passed a 
highly nontrivial consistency check of validity of the cascading gauge theory plasma first law of thermodynamics in 
a dual holographic setting. It was also used in \cite{abs3}.

Recall that a symmetry transformation \eqref{scale3} rescales $H$, and a symmetry transformation \eqref{scale2} rescales $P$ 
and affects $K_0$, while 
leaving the combination 
\begin{equation}
\frac{K_0}{P^2g_0}+2 \ln H+
\ln P^2 g_0\ =\ {\rm invariant}\ \equiv -2 \ln\Lambda + 2\ln H = \ln\frac{H^2}{\Lambda^2}
\eqlabel{scalel}
\end{equation}
invariant. The latter invariant defines the strong coupling scale $\Lambda$ of  cascading gauge theory. 
In particular, using the symmetry choices \eqref{res1} and \eqref{res3f0} we identify 
\begin{equation}
\frac{K_0}{P^2}= \ln \frac{1}{\Lambda^2 P^2 }\equiv \frac 1\dd \,.
\eqlabel{physical}
\end{equation}
Notice that \eqref{physical} is not invariant under the symmetry transformation \eqref{scale2}. This is because 
such transformation modifies $P^2 g_0$, and thus changes the theory;  \eqref{physical} is invariant under the residual 
diffeomorphisms \eqref{leftover}.

As defined in \eqref{physical}, a new dimensionless parameter $\dd$ is small when the IR cutoff set by the 
$dS_4$ is much higher than the strong coupling scale $\Lambda$ (and thus  cascading gauge theory is close to be conformal).
In section \ref{pertubative} we develop perturbative expansion in $\dd$.

\subsection{Numerical procedure}\label{numerical}
Although we would like to have an analytic control over the gravitational solution dual to a symmetric 
phase of cascading gauge theory, the relevant equations 
for $\{f_2$, $f_3$, $h$, $K$, $g\}$ \eqref{eq2}-\eqref{eq7} are rather complicated. 
Thus, we have to resort to numerical analysis. Recall that various scaling symmetries of the background equations 
of motion allowed us to set  (see \eqref{res1} and \eqref{res3f0})
\begin{equation}
\lim_{\r\to 0} g\equiv g_0=1\,,\qquad  H=1\,.
\eqlabel{2symm}
\end{equation}
While the metric parametrization \eqref{metricaby} has  residual diffeomorphisms \eqref{leftover}, the latter are fixed 
once we insist on the IR asymptotics at $y\equiv \frac 1\r\to 0$ (see \eqref{geocoms}). Finally, a scaling symmetry 
\eqref{scale2} relates different pairs $\{K_0,P\}$ so that only the ratio $\frac{K_0}{P^2}\equiv \frac 1\dd$ is physically meaningful 
(see \eqref{physical}). In the end, for a fixed $\dd$, the gravitational solution is characterized by 5 parameters in the UV 
and 4 parameters in the IR:
\begin{equation}
\begin{split}
&{\rm UV}:\qquad \{\a_{1,0}\,,\ a_{4,0}\,,\ a_{6,0}\,,\ a_{8,0}\,,\ g_{4,0}\}\,,
\\
&{\rm IR}:\qquad \{K_0^h\,,\ g_0^h\,,\ f_{2,0}^h\,,\ f_{3,0}^h\}\,.
\end{split}
\eqlabel{uvirfinal}
\end{equation} 
Notice that $5+4=9$ is precisely the number of integration constants needed to specify a solution to \eqref{eq2}-\eqref{eq7} ---
we have 5 second order differential equations and a single first order differential constraint: $2\times 5-1=9$.

In practice, we replace the second-order differential equation for $f_2$ \eqref{eq2} with the constraint equation \eqref{eq7},
which we use to  algebraically eliminate $f_2'$ from  \eqref{eq3}-\eqref{eq6}. The solution is found using the 
``shooting'' method as detailed in \cite{abk}. 
 
Finding a ``shooting'' solution in 9-dimensional parameter space  \eqref{uvirfinal} is quite challenging. Thus, we start with 
(leading) analytic results for $\dd\ll 1$ (see section \ref{pertubative}) and construct numerical solution 
for $(K_0=1, P^2)$ slowly incrementing $P^2$ from zero to one. Starting with the solution at $K_0=P^2=1$ 
we slowly decrease $K_0$ while keeping $P^2=1$.

\subsection{Symmetric phase of cascading gauge theory at $\frac{H}{\Lambda}\gg 1$}
\label{pertubative}

In this section we describe perturbative solution in $\dd\ll 1$ \eqref{physical} to 
\eqref{eq2}-\eqref{eq7}.
Such gravitational backgrounds describe cascading gauge theory  on  $dS_4$, which Hubble 
scale $H$  is well above the strong coupling scale $\Lambda$ of cascading gauge theory. 

In the limit $\dd\to 0$ (or equivalently $P\to 0$) 
the gravitational background is simply that of the Klebanov-Witten model \cite{kw} on $dS_4$ \cite{bds4}:
\begin{equation}
\begin{split}
\dd=0: \qquad &f_2^{(0)}=f_3^{(0)}=1+\sqrt{\hK_0}\r\,,
\qquad h^{(0)}=\frac{ \hK_0}{4(1+\sqrt{\hK_0}\r)^2}\,,\\
&K^{(0)}=\hK_0\,,\qquad g^{(0)}=1\,,
\end{split}
\eqlabel{kw}
\end{equation}
where $\hK_0$ is a constant.
Perturbatively, we find
\begin{equation}
\begin{split}
&f_i(\r)=f_i^{(0)}\times\  \sum_{j=0}^\infty \left(\frac{P^2}{\hK_0}\right)^j\ f_{i,j}(\r^2 \hK_0)\,,\qquad 
h(\r)=h^{(0)}\times\ \sum_{j=0}^\infty \left(\frac{P^2}{\hK_0}\right)^j\ h_{j}(\r^2 \hK_0)\,,\\
&K(\r)=\hK^{(0)}\times\ \sum_{j=0}^\infty \left(\frac{P^2}{\hK_0}\right)^j\ K_{j}(\r^2 \hK_0)\,,\qquad
g(\r)=g^{(0)}\times\ \sum_{j=0}^\infty \left(\frac{P^2}{\hK_0}\right)^j\ g_{j}(\r^2 \hK_0)\,.
\end{split}
\eqlabel{pertgen}
\end{equation}
Apart from technical complexity, there is no obstacle of developing perturbative 
solution to any order in $\frac{P^2}{\hK_0}$. For our purposes it is sufficient to do so to order $\calo\left(\frac{P^4}
{\hK_0^2}\right)$.
Notice that explicit $\r$ dependence enters only in combination $\r \sqrt{\hK_0}$, thus, we can set $\hK_0=1$
and reinstall explicit $\hK_0$ dependence when necessary.

Substituting \eqref{pertgen} in \eqref{eq2}-\eqref{eq7} we find to  order $\calo(\dd)$
the following equations
\begin{equation}
\begin{split}
0=&f_{2,1}''- \frac{\r+6}{2\r (\r+1)} f_{2,1}'+\frac{\r+2}{2\r (\r+1)} h_1'-\frac34 (K_1')^2
- \frac{3 \r^2-16 \r-16}{4\r^2 (\r+1)^2} h_1\\
&-\frac{4 K_1+7 f_{2,1}-20 f_{3,1}-3}{(\r+1) \r^2}\,,
\end{split}
\eqlabel{pereq2}
\end{equation}
\begin{equation}
\begin{split}
0=&f_{3,1}''- \frac{\r+6}{2\r (\r+1)} f_{3,1}'+\frac14 (K_1')^2+ \frac{\r+2}{2\r (\r+1)} h_1'
- \frac{3 \r^2-16 \r-16}{4\r^2 (\r+1)^2} h_1\\
&+\frac{5 f_{2,1}+8 f_{3,1}-4 K_1-1}{(\r+1) \r^2}\,,
\end{split}
\eqlabel{pereq3}
\end{equation}
\begin{equation}
\begin{split}
0=&h_1''-\frac{\r+4}{\r (\r+1)} h_1'+\frac 34 (K_1')^2+\frac{(\r+2) (f_{2,1}'+4 f_{3,1}')}{2 \r (\r+1)}
+ \frac{9(\r^2-16 \r-16)}{4\r^2 (\r+1)^2} h_1\\
&-\frac{17 f_{2,1}+68 f_{3,1}-36 K_1-5}{(\r+1) \r^2}\,,
\end{split}
\eqlabel{pereq4}
\end{equation}
\begin{equation}
\begin{split}
0=&K_1''- \frac{\r+6}{2\r (\r+1)} K_1'-\frac{8}{(\r+1) \r^2}\,,
\end{split}
\eqlabel{pereq5}
\end{equation}
\begin{equation}
\begin{split}
0=&g_1''- \frac{\r+6}{2\r (\r+1)} g_1'+(K_1')^2-\frac{4}{(\r+1) \r^2}\,,
\end{split}
\eqlabel{pereq6}
\end{equation}
along with the first order constraint
\begin{equation}
\begin{split}
0=&f_{2,1}'+4 f_{3,1}'+h_1'
+ \frac{(\r+1) \r}{2(\r+2)} (K_1')^2+\frac{(\r+4) (3 \r+4)}{2\r (\r+2) (\r+1)} h_1
\\&+\frac{2 (4 f_{3,1}+f_{2,1}-4 K_1-1)}{(\r+2) \r}\,.
\end{split}
\eqlabel{pereqc}
\end{equation}

Above equations should be solved with $\calo(\dd)$ UV and the IR boundary conditions 
prescribed in sections \ref{uvcond} and \ref{ircond}. 
We solve all the equations numerically. Parameterizing 
the asymptotics as follows
\nxt UV, \ie $\r\to 0$, (the independent coefficients being $\{\a_{1,1,0}$, $k_{1,4,0}$,  
$a_{1,6,0}$, $a_{1,8,0}$, $g_{1,4,0}\}$):
\begin{equation}
\begin{split}
&f_{2,1}= \a_{1,1,0}\ \r+\biggl(-\frac38-\frac12 \a_{1,1,0}+\frac12 \ln\r
\biggr)\ \r^2+ \biggl(\frac18+\frac12 \a_{1,1,0}-\frac12 \ln\r\biggr)\ \r^3\\
&+\biggl(-\frac{5}{24}
-\frac12 \a_{1,1,0}+\frac43 k_{1,4,0}+\frac{9}{16} \ln\r\biggr)\ \r^4+\biggl(\frac{29}{96}+\frac 12 \a_{1,1,0}
-\frac83 k_{1,4,0}-\frac58 \ln\r\biggr)\ \r^5\\
&+\biggl(\frac{1}{160} \ln^2\r+\frac{4}{15} \ln\r k_{1,4,0}
+\frac{25099}{38400} \ln\r+a_{1,6,0}\biggr)\ \r^6+\biggl(-\frac{3}{160} \ln^2\r
-\frac45 \ln\r k_{1,4,0}\\
&-\frac{8379}{12800} \ln\r-\a_{1,1,0}-3 a_{1,6,0}+\frac{98}{15} k_{1,4,0}
-\frac{17513}{25600}\biggr)\ \r^7+\biggl(\frac{3}{8960} \ln^3\r+\frac{3}{140} \ln^2\r k_{1,4,0}
\\
&+\frac{16}{35} \ln\r k_{1,4,0}^2+\frac{89373}{2508800} \ln^2\r+\frac{29791}{19600} \ln\r k_{1,4,0}
+\frac{223043661}{351232000} \ln\r+a_{1,8,0}\biggr)\ \r^8\\
&+\calo(\r^9)\,,
\end{split}
\eqlabel{f21uv}
\end{equation}
\begin{equation}
\begin{split}
&f_{3,1}= \a_{1,1,0}\ \r+ \biggl(-\frac12-\frac12 \a_{1,1,0}+\frac12 \ln\r\biggr)
\ \r^2+ \biggl(\frac14+\frac12 \a_{1,1,0}-\frac12 \ln\r\biggr)\ \r^3+\biggl(
-\frac{41}{192}\\
&-\frac12 \a_{1,1,0}+\frac12 \ln\r\biggr)\ \r^4+ \biggl(\frac{7}{32}+\frac12 \a_{1,1,0}
-\frac12 \ln\r\biggr)\ \r^5+\biggl(-\frac{1}{640} \ln^2\r
-\frac{1}{15} \ln\r k_{1,4,0}\\
&+\frac{19049}{38400} \ln\r
-\frac58 \a_{1,1,0}-\frac14 a_{1,6,0}+\frac{353}{480} k_{1,4,0}
-\frac{381067}{1228800}\biggr)\ \r^6
+\biggl(\frac{3}{640} \ln^2\r+\frac15 \ln\r k_{1,4,0}\\
&-\frac{6229}{12800} \ln\r
+\frac78 \a_{1,1,0}+\frac34 a_{1,6,0}-\frac{1043}{480} k_{1,4,0}+\frac{591377}{1228800}\biggr)\ \r^7
+\biggl(\frac{3}{8960} \ln^3\r\\
&+\frac{3}{140} \ln^2\r k_{1,4,0}
+\frac{16}{35} \ln\r k_{1,4,0}^2-\frac{28227}{2508800} \ln^2\r-\frac{9409}{19600} \ln\r k_{1,4,0}
-\frac45 k_{1,4,0}^2\\
&+\frac{167306161}{351232000} \ln\r
-\frac{231}{64} \a_{1,1,0}-\frac{231}{32} a_{1,6,0}+a_{1,8,0}+\frac{129741}{6400} k_{1,4,0}
-\frac{116879077}{49152000}\biggr)\ \r^8\\
&+\calo(\r^9)\,,
\end{split}
\eqlabel{f31uv}
\end{equation}
\begin{equation}
\begin{split}
&h_{1}=\frac12-2 \ln\r+ \biggl(1-2 \a_{1,1,0}\biggr)\ \r
+ \biggl(\frac{19}{24}+\a_{1,1,0}-\ln\r\biggr)\ \r^2+ \biggl(-\frac{11}{24}-\a_{1,1,0}
\\&+\ln\r\biggr)\ \r^3+ \biggl(\frac{431}{1024}+\a_{1,1,0}+\frac16 k_{1,4,0}
-\frac{127}{128} \ln\r\biggr)\ \r^4+\biggl(-\frac{3359}{7680}-\a_{1,1,0}
-\frac13 k_{1,4,0}\\
&+\frac{63}{64} \ln\r\biggr)\ \r^5+\biggl(\frac{656813}{1536000}+\a_{1,1,0}+\frac{189}{200} 
k_{1,4,0}-\frac{12233}{12800} \ln\r\biggr)\ \r^6+\biggl(-\frac{4213513}{10752000}
\\
&-\a_{1,1,0}-\frac{1201}{600} k_{1,4,0}+\frac{11599}{12800} \ln\r\biggr)\ \r^7
+\biggl(-\frac{9}{7168} \ln^3\r-\frac{9}{112} \ln^2\r k_{1,4,0}-\frac{12}{7} \ln\r k_{1,4,0}^2
\\
&+\frac{12441}{2007040} \ln^2\r+\frac{4147}{15680} \ln\r k_{1,4,0}
+\frac{212}{105} k_{1,4,0}^2-\frac{238628771}{280985600} \ln\r
+\frac{637}{64} \a_{1,1,0}\\
&+\frac{693}{32} a_{1,6,0}
-\frac{15}{4} a_{1,8,0}-\frac{17860741}{313600} k_{1,4,0}+\frac{110837461177}{16859136000}\biggr)\ \r^8+\calo(\r^9)\,;
\end{split}
\eqlabel{h1uv}
\end{equation}
\begin{equation}
\begin{split}
&K_{1}=-2 \ln\r+\r-\frac18 \r^2-\frac{1}{24} \r^3+
\biggl(k_{1,4,0}+\frac{3}{64} \ln\r\biggr)\ \r^4+\biggl(\frac{33}{640}
-2 k_{1,4,0}-\frac{3}{32} \ln\r\biggr)\ \r^5\\
&+\biggl(-\frac{307}{3072}+\frac{35}{12} k_{1,4,0}
+\frac{35}{256} \ln\r\biggr)\ \r^6
+\biggl(\frac{1031}{7168}-\frac{15}{4} k_{1,4,0}-\frac{45}{256} \ln\r\biggr)\ \r^7+\biggl(-\frac{24077}{131072}
\\
&+\frac{1155}{256} k_{1,4,0}+\frac{3465}{16384} \ln\r\biggr)\ \r^8+\calo(\r^9)\,;
\end{split}
\eqlabel{K1uv}
\end{equation}
\begin{equation}
\begin{split}
&g_{1}= -\frac12 \r^2+\frac12 \r^3+\biggl(g_{1,4,0}+\left(-\frac{33}{64}+4 k_{1,4,0}\right) \ln\r
+\frac{3}{32} \ln^2\r\biggr)\ \r^4
+\biggl(-\frac{31}{128}-2 g_{1,4,0}\\
&-2 k_{1,4,0}+\left(\frac{15}{16}-8 k_{1,4,0}\right) \ln\r
-\frac{3}{16} \ln^2\r\biggr)\ \r^5+\biggl(\frac{3671}{9216}+\frac{35}{12} g_{1,4,0}+\frac{161}{36} k_{1,4,0}
\\&+\left(-\frac{497}{384}+\frac{35}{3} k_{1,4,0}\right) \ln\r+\frac{35}{128} \ln^2\r\biggr)\ \r^6
+\biggl(-\frac{533}{1024}-\frac{15}{4} g_{1,4,0}-\frac{83}{12} k_{1,4,0}\\
&+\left(\frac{103}{64}
-15 k_{1,4,0}\right) \ln\r-\frac{45}{128} \ln^2\r\biggr)\ \r^7
+\biggl(\frac{81683}{131072}-\frac12 k_{1,4,0}^2
+\frac{1155}{256} g_{1,4,0}\\
&+\frac{7117}{768} k_{1,4,0}
+\left(-\frac{15499}{8192}+18 k_{1,4,0}\right) \ln\r+\frac{27}{64} \ln^2\r\biggr)\ \r^8+\calo(\r^9)\,;
\end{split}
\eqlabel{g1uv}
\end{equation}
\nxt IR, \ie $y=\frac 1\r\to 0$, (the independent coefficients being $\{a_{1,0}^h, b_{1,0}^h, g_{1,0}^h, k_{1,0}^h\}$: 
\begin{equation}
\begin{split}
&f_{2,1}=a_{1,0}^h+\calo(y)\,,\qquad f_{3,1}=b_{1,0}^h+\calo(y)\,,\qquad  g_{1}=g_{1,0}^h+\calo(y)\,,\\
&K_{1}=k_{1,0}^h+\calo(y)\,,\qquad h_{1}=\left(-\frac65+\frac{18}{5} a_{1,0}^h+\frac{72}{5} b_{1,0}^h-8k_{1,0}^h\right)
\ y +\calo(y^2)\,,
\end{split}
\eqlabel{fhhir}
\end{equation}
we find 
\begin{equation}
\begin{split}
&\a_{1,1,0}=0.43427(8)\,,\qquad k_{1,4,0}=0.04829(9)\,,\qquad a_{1,6,0}=-0.40703(7)\\
&a_{1,8,0}=-0.42707(1)\,,\qquad g_{1,4,0}=-0.26443(7)\,,\qquad a_{1,0}^h=-0.15661(4)\\
&b_{1,0}^h=-0.37883(6)\,,\qquad g_{1,0}^h=-0.72222(2)\,,\qquad k_{1,0}^h=-1.10592(2)
\end{split}
\eqlabel{numres}
\end{equation}

%%%%%SECOND ORDER \delta^2%%%%%%
In an analogous way, it is possible to go to second order in $\delta$ by taking eqs. \eqref{eq2}-\eqref{eq7} and evaluate them with the expansion \eqref{pertgen} to second order in $\delta$. Then, we will get equations for functions ${f_{2,2},f_{3,2},h_2,K_2,g_2}$. As with the first order equations, one uses the UV and IR boundary conditions prescribed in sections \ref{uvcond} and \ref{ircond}. Setting $H=1$, we get that the independent coefficients in the UV are $\{\a_{2,1,0}$, $k_{2,4,0}$, $a_{2,6,0}$, $a_{2,8,0}$, $g_{2,4,0}\}$, while those in the IR are $\{a_{2,0}^h, b_{2,0}^h, g_{2,0}^h, k_{2,0}^h\}$. Solving numerically, we find the values of these constants to be
\begin{equation}
\begin{split}
&\a_{2,1,0}=0.35729(1)\,,\qquad k_{2,4,0}=0.18423(1)\,,\qquad a_{2,6,0}=-0.48877(2)\,,\\
&a_{2,8,0}=-0.60853(7)\,,\qquad g_{2,4,0}=-0.64457(3)\,,\qquad a_{2,0}^h=0.54009(5)\,,\\
&b_{2,0}^h=0.63805(4)\,,\qquad g_{2,0}^h=0.31165(0)\,,\qquad k_{2,0}^h=1.65246(0)\,.
\end{split}
\eqlabel{numres2}
\end{equation}

%%%%%%

We can now identify the leading $\calo(\dd^2)$ 
values of general UV and IR parameters (see \eqref{uvirfinal}):
\begin{equation}
\begin{split}
&\a_{1,0}  = -1-\a_{1,1,0}\ \dd -\a_{2,1,0}\ \dd^2\,, \\
&a_{4,0}  =  \left(-\frac{1}{12}+\frac 43 k_{1,4,0}\right) \dd + \left(-\frac{139}{1152}+\frac{a_{1,1,0}}{24}+\frac{2 g_{1,4,0}}{3}-\frac{22 k_{1,4,0}}{9}+\frac{4 k_{2,4,0}}{3}\right) \dd^2, \\
&g_{4,0}   =   g_{1,4,0}\ \dd + g_{2,4,0}\ \dd^2\,, \\
&a_{6,0}  = \left(a_{1,6,0}+\frac{29}{96}-\frac83 k_{1,4,0}+\frac12 \a_{1,1,0}\right)\ \dd\ + \\
& +  \left(\frac{145}{576}-\frac{5 a_{1,1,0}}{32}-\frac{a_{1,1,0}^2}{4}+\frac{a_{2,1,0}}{2}-\frac{4 g_{1,4,0}}{3}+\frac{44 k_{1,4,0}}{9}-\frac{4 a_{1,1,0} k_{1,4,0}}{3}-\frac{8 k_{2,4,0}}{3}\right. \\
&\left.+a_{2,6,0}\right) \dd^2\,, \\
&a_{8,0} = \left(a_{1,8,0}-3 a_{1,6,0}-\frac{17513}{25600}-\a_{1,1,0}+\frac{98}{15} k_{1,4,0}\right) \dd +\\
& + \left(-\frac{87973}{192000}-\frac{15353 a_{1,1,0}}{25600}-\frac{a_{1,1,0}^2}{2}-2 a_{1,1,0} a_{1,6,0}-a_{2,1,0}-3 a_{2,6,0}+a_{2,8,0} + \right. \\
& \left. + \frac{101 g_{1,4,0}}{30}-\frac{2423 k_{1,4,0}}{180}+\frac{178 a_{1,1,0} k_{1,4,0}}{15}+\frac{98 k_{2,4,0}}{15}\right) \dd^2\,, \\
\end{split}
\eqlabel{matchinga}
\end{equation}
\begin{equation}
\begin{split}
&K_0^h =  1+k_{1,0}^h\ \dd +k_{2,0}^h\ \dd^2 \,,\qquad g_0^h=1+g_{1,0}^h\ \dd +g_{2,0}^h\ \dd^2 \,, \\ 
&f_{2,0}^h  =  1+a_{1,0}^h\ \dd +a_{2,0}^h\ \dd^2\,,\qquad 
 f_{3,0}^h=1+b_{1,0}^h\ \dd +b_{2,0}^h\ \dd^2\,,
\end{split}
\eqlabel{matching}
\end{equation}
where we set $K_0=1$.

\begin{figure}[h!]
\begin{center}
\psfrag{a10}[][][0.7]{{$\a_{1,0}$}}
\psfrag{a40}[][][0.7]{{$a_{4,0}$}}
\psfrag{a60}[][][0.7]{{$a_{6,0}$}}
\psfrag{a80}[][][0.7]{{$a_{8,0}$}}
\psfrag{g40}[][][0.7]{{$g_{4,0}$}}
\psfrag{a20h}[][][0.7]{{$a_{0}^h$}}
\psfrag{b20h}[][][0.7]{{$b_{0}^h$}}
\psfrag{g0h}[][][0.7]{{$g_{0}^h$}}
\psfrag{k0h}[][][0.7]{{$K_{0}^h$}}
\psfrag{dd}[][][0.7]{{$\dd$}}
\includegraphics[width=2in]{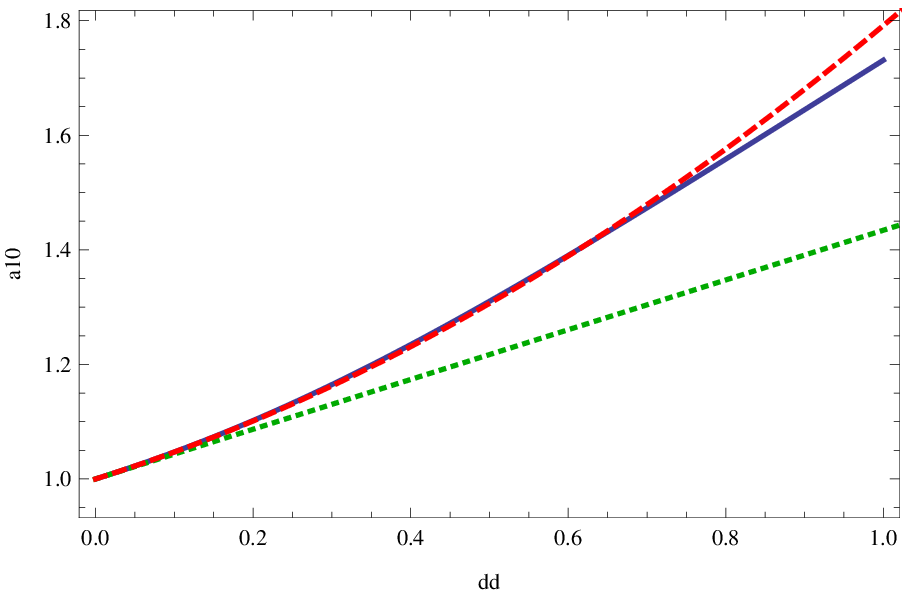}
\includegraphics[width=2in]{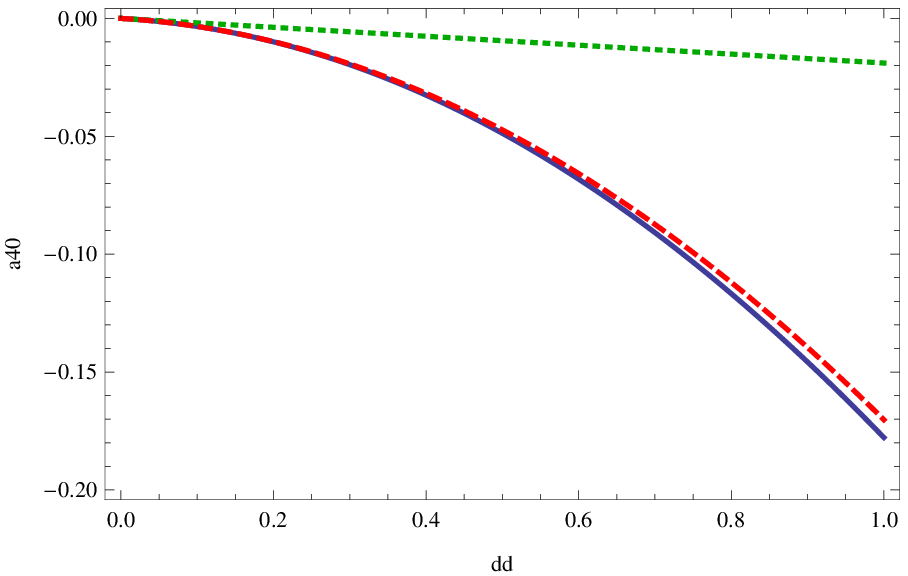}
\includegraphics[width=2in]{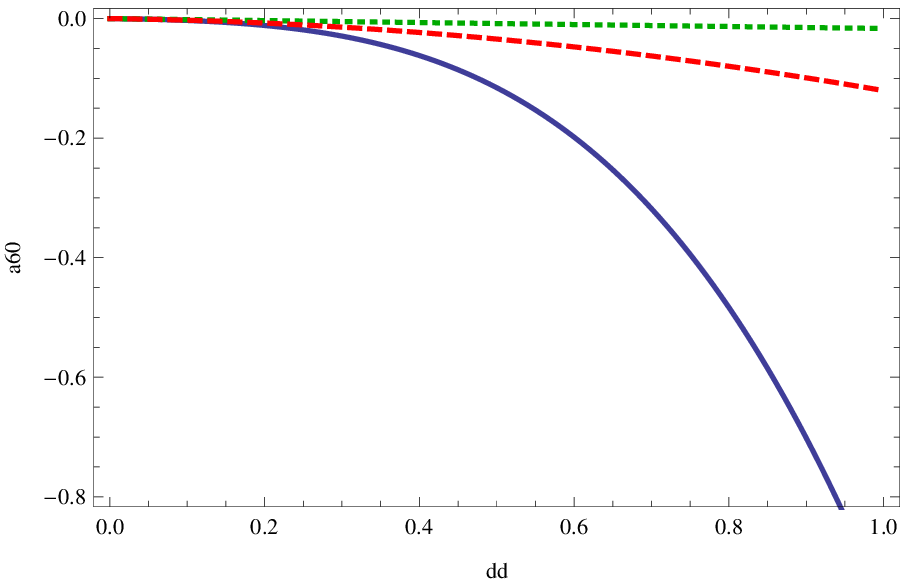}
\includegraphics[width=2in]{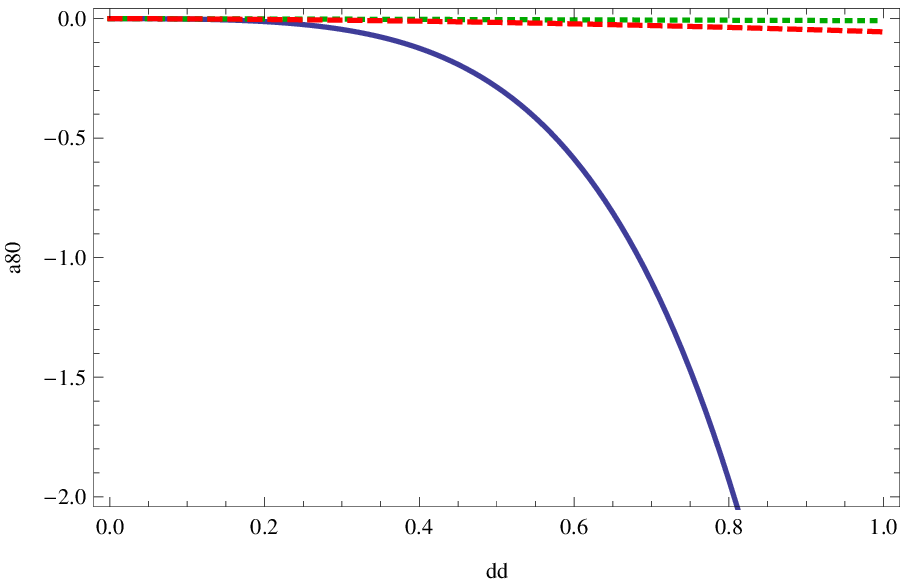}
\includegraphics[width=2in]{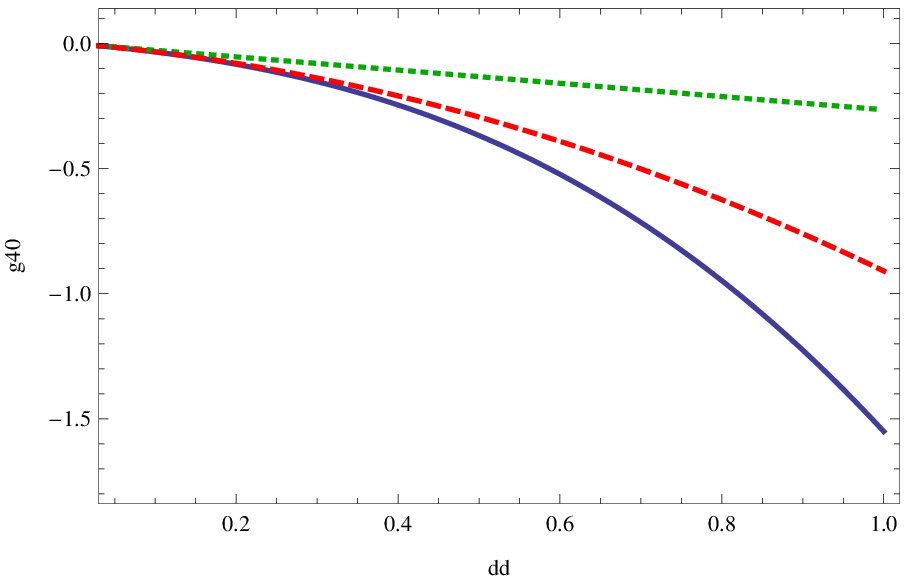}
\includegraphics[width=2in]{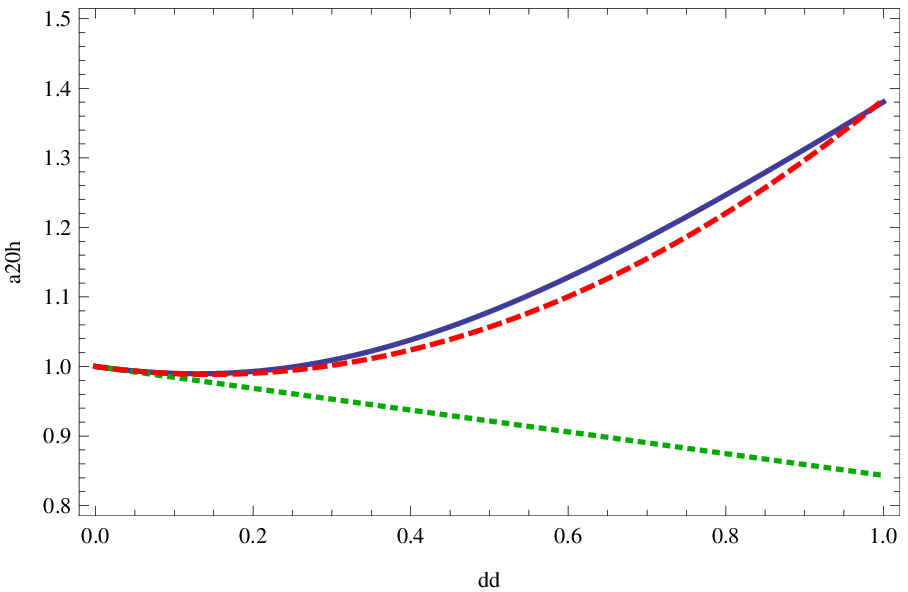}
\includegraphics[width=2in]{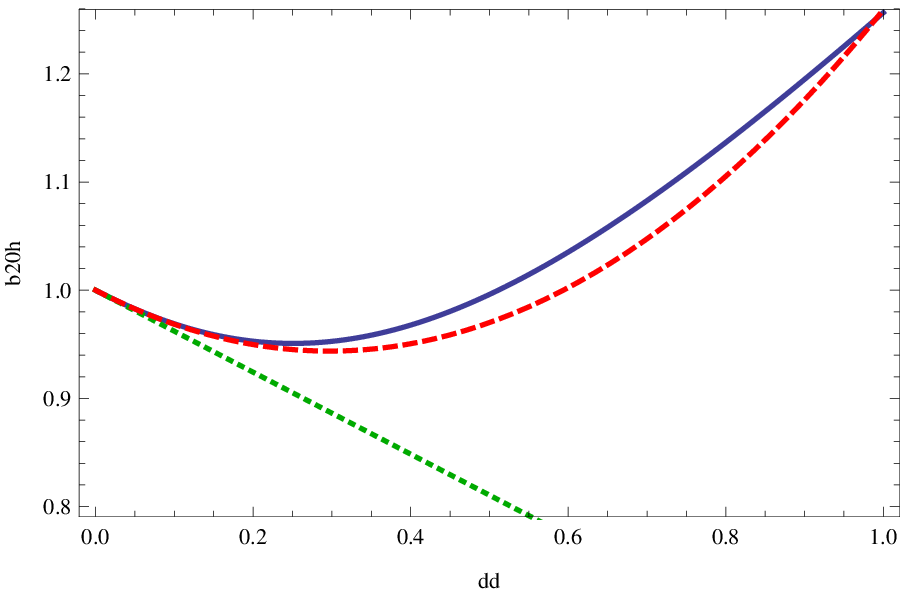}
\includegraphics[width=2in]{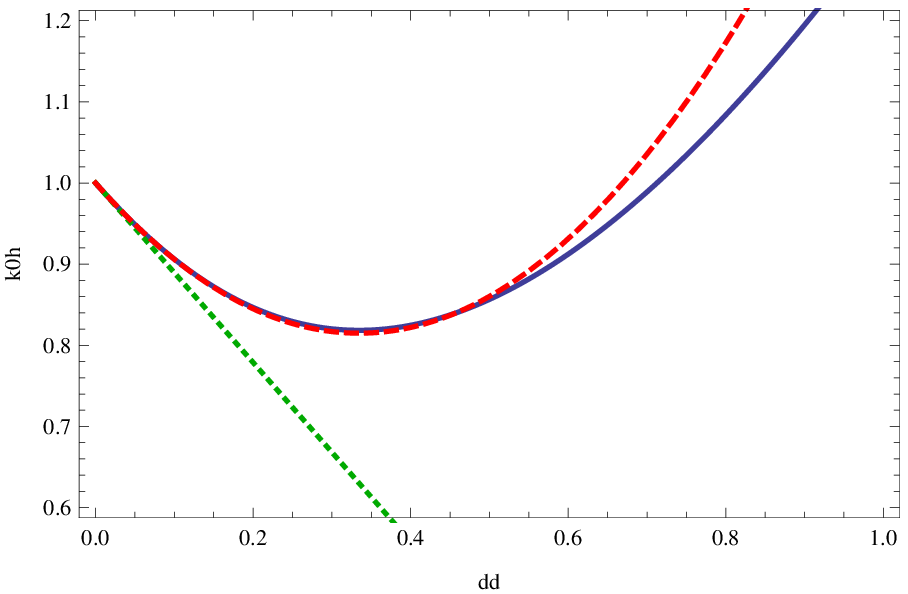}
\includegraphics[width=2in]{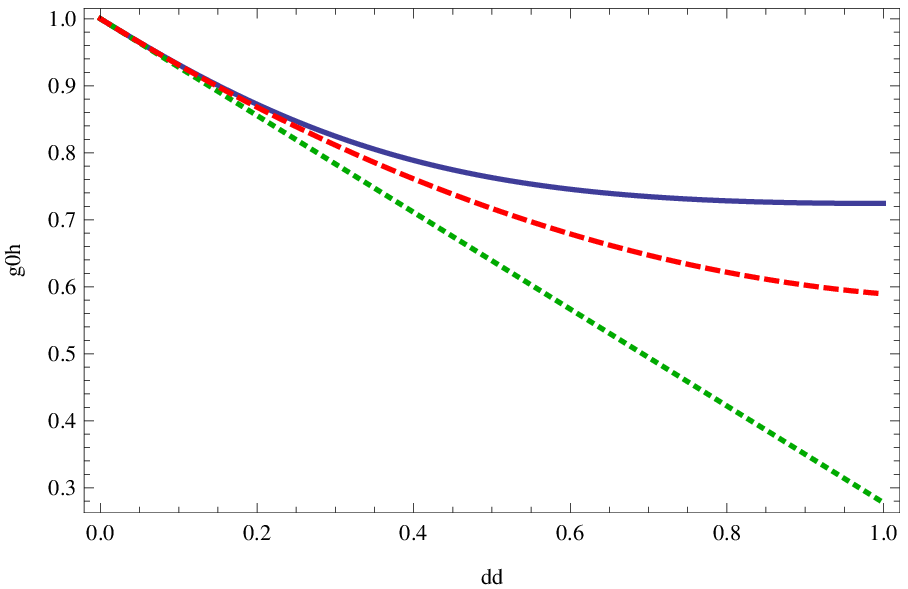}
\end{center}
  \caption{(Colour online) Comparison of values of  UV  parameters $\{\a_{1,0},a_{4,0},a_{6,0},\a_{8,0},g_{4,0}\}$  and IR parameters $\{a_{0}^h, b_{0}^h, K_{0}^h, g_{0}^h \}$
(see \eqref{uvirfinal})
in the range $\dd\in[0,1]$ (blue curves) with their perturbative predictions \eqref{matchinga}-\eqref{matching} at first (green dotted) and second order (red dashed) in $\dd$.
} \label{figure1}
\end{figure}

Figure \ref{figure1} compares the values of general UV and IR parameters $\a_{1,0}$,
$ a_{4,0}$, $a_{6,0}$, $a_{8,0}$, $g_{4,0}$, 
$K_0^h$, $g_{0}^h$, $f_{2,0}^h$, $f_{3,0}^h$  (see \eqref{uvirfinal}), with their perturbative predictions at linear and quadratic order. The results for first and second order will help to correctly initialize the fully non-linear calculation and at the same time provide a verification of the results, at least for small enough $\dd$. 

\section{Cascading gauge theory on $dS_4$ with spontaneously broken chiral symmetry}\label{broken}

\subsection{$R^{1,3}\to dS_{4}$  deformation of Klebanov-Strassler state of cascading gauge theory}\label{kscompact}
$\caln=1$ supersymmetric ground state of cascading gauge theory on $R^{3,1}$ --- referred to as Klebanov-Strassler state --- 
spontaneously breaks chiral symmetry \cite{ks}. 
A natural route to construct a $\csb$ state of the theory on $dS_4$ is to ``deform''  Klebanov-Strassler state:
$R^{1,3}\to dS_4$. We explain now how to achieve this in a ``continuous'' fashion.
  
Consider the five-dimensional metric of the type:
\begin{equation}
ds_5^2=g_{\mu\nu}(y)dy^\mu dy^{\nu}=c_1^2\ \biggl(-dt^2
+\frac{1}{H^2 }\cosh^2(Ht) \left(dS^3\right)^2\biggr)+c_3^2\ (d\r)^2\,,
\eqlabel{5dks}
\end{equation} 
where $c_i=c_i(\r)$.
We will be interested in $\csb$ states of cascading gauge theory on $dS_4$ with a Hubble scale $H$ . 
One can derive equations of motion from \eqref{5action}. Alternatively, we can construct an effective 1-dimensional 
action\footnote{Effectively, in obtaining $S_1$ we perform Kaluza-Klein-like reduction of the effective action $S_5$ on 
$dS_4$.} from \eqref{5action}, by restricting to the metric ansatz \eqref{5dks}, and the $\r$-only dependence 
of the scalar fields $\{\Phi, h_i, \om_i\}$: 
\begin{equation}
S_5\left[g_{\mu\nu},\om_i,h_i,\Phi\right]\ \Longrightarrow\ S_1\left[c_i,\om_i,h_i,\Phi\right]\,.
\eqlabel{1action}
\end{equation}  
It can be verified that equations of motion obtained from $S_1$ coincide with those obtained from 
 \eqref{5action}, provided 
we vary\footnote{This produces the first order constraint similar to \eqref{eq7}.} 
$S_1$ with respect to $c_3$, treating it as an unconstrained field. The 1-dimensional effective action approach 
makes it clear that the only place where the information about $dS_4$ enters is through the evaluation of $R_5$ in \eqref{ric5}:
\begin{equation}
\begin{split}
R_5=-\frac{8 c_1''}{c_3^2 c_1}+\frac{8 c_1' c_3'}{c_3^3 c_1}
-\frac{12 (c_1')^2}{c_3^2 c_1^2}
+\frac{12 \k}{c_1^2}\,,
\end{split}
\eqlabel{r5cuv}
\end{equation}
where derivatives are with respect to $\r$, and $\k=H^2$.

\subsection{Equations of motion}\label{kseoms}
As in \eqref{metricaby} and \eqref{redef} we denote
\begin{equation}
\begin{split}
c_1=&h^{-1/4}\r^{-1}\,,\qquad c_3=h^{1/4}\r^{-1}\,,\qquad \Phi=\ln g\,,\\
h_1=&\frac 1P\left(\frac{K_1}{12}-36\Om_0\right)\,,\qquad h_2=\frac{P}{18}\ K_2\,,\qquad 
h_3=\frac 1P\left(\frac{K_3}{12}-36\Om_0\right)\,,\\
\Om_1=&\frac 13 f_c^{1/2} h^{1/4}\,,\qquad \Om_2=\frac {1}{\sqrt{6}} f_a^{1/2} h^{1/4}\,,\qquad 
\Om_3=\frac {1}{\sqrt{6}} f_b^{1/2} h^{1/4} \,.
\end{split}
\eqlabel{redef2}
\end{equation}
The equations of motion obtained from $S_1\left[c_i,\om_i,h_i,\Phi\right]$ are 
\begin{equation}
\begin{split}
&0=f_c''- \frac{3f_c'}{\r}-3 h f_c \k
-\frac{(f_c')^2}{2f_c}+\frac{5 f_c}{\r^2}+\frac{f_c (g')^2}{8g^2}+\frac{3f'_b f'_c}{4f_b}
+\frac{63 f_a}{16f_b \r^2}+\frac{63 f_b}{16f_a \r^2}+\frac{3 f_c}{f_a \r^2}\\
&-\frac{f_c (f_a')^2}{8f_a^2}
+\frac{3 f_a' f'_c}{4f_a}+\frac{f_c (h')^2}{8h^2}- \frac{f_c (f_b')^2}{8f_b^2}+\frac{3 f_c}{f_b \r^2}-\frac{63}{8 \r^2}
-\frac{K_1^2}{8f_a^2 h^2 f_b^2 \r^2}+ \frac{3g P^2}{2f_a^2 h \r^2}\\
&-\frac{f_c f_a' f'_b}{2f_a f_b}- \frac{27K_1 K_3}{32f_a h f_b g P^2 \r^2}
-\frac{K_2^2 K_1^2}{32f_a^2 h^2 f_b^2 \r^2}+\frac{K_2 K_1^2}{8f_a^2 h^2 f_b^2 \r^2}
- \frac{K_2^2 K_3^2}{32f_a^2 h^2 f_b^2 \r^2}- \frac{3f_c (K_1')^2}{32h f_b^2 g P^2}
\\
&- \frac{3f_c (K_3')^2}{32f_a^2 h g P^2}+\frac{3g P^2 K_2^2}{8h f_b^2 \r^2}+\frac{3 g P^2 K_2^2}{8f_a^2 h \r^2}
-\frac{3 g P^2 K_2}{2f_a^2 h \r^2}-\frac{9 f_c^2}{f_a f_b \r^2}+\frac{f_c h'}{h \r}+\frac{K_2^2 K_1 K_3}{16f_a^2 h^2 f_b^2 \r^2}
\\
&-\frac{K_2 K_1 K_3}{8f_a^2 h^2 f_b^2 \r^2}-\frac{g P^2 f_c (K_2')^2}{12f_a h f_b}
+\frac{27 K_1^2}{64f_a h f_b g P^2 \r^2}+\frac{27 K_3^2}{64f_a h f_b g P^2 \r^2}\,,
\end{split}
\eqlabel{kseq2}
\end{equation}
\begin{equation}
\begin{split}
&0=f_a''-\frac{45 f_a^2}{16f_c f_b \r^2}+\frac{f_a h'}{h \r}
+\frac{g P^2 (K_2')^2}{36h f_b}+\frac{5(K_3')^2}{32f_a h g P^2}
- \frac{f_a f'_b f'_c}{4f_c f_b}-\frac{(f_a')^2}{8f_a}+\frac{5 f_a}{\r^2}
-\frac{3 f_a'}{\r}\\
&-\frac{K_2^2 K_1^2}{32f_c f_a h^2 f_b^2 \r^2}+ \frac{K_2 K_1^2}{8f_c f_a h^2 f_b^2 \r^2}
-\frac{K_2^2 K_3^2}{32f_c f_a h^2 f_b^2 \r^2}-\frac{3 g P^2 K_2}{2f_c f_a h \r^2}+\frac{3 g P^2 K_2^2}{8f_c f_a h \r^2}
\\
&-\frac{9 K_3^2}{64f_c h f_b g P^2 \r^2}-\frac{9 K_1^2}{64f_c h f_b g P^2 \r^2}
+\frac{3 f_a}{f_b \r^2}+\frac{3 f_c}{f_b \r^2}+\frac{9 K_1 K_3}{32f_c h f_b g P^2 \r^2}
+\frac{K_2^2 K_1 K_3}{16f_c f_a h^2 f_b^2 \r^2}
\\&- \frac{K_2 K_1 K_3}{8f_c f_a h^2 f_b^2 \r^2}-\frac{5 f_a g P^2 K_2^2}{8f_c h f_b^2 \r^2}
-\frac{K_1^2}{8f_c f_a h^2 f_b^2 \r^2}+\frac{3g P^2}{2f_c f_a h \r^2}
-\frac{3 f_a (K_1')^2}{32h f_b^2 g P^2}-\frac{9}{\r^2}+\frac{f_a (g')^2}{8g^2}\\
&-3 f_a h \k
+\frac{f_a' f'_b}{2f_b}+\frac{f'_c f_a'}{4f_c}-\frac{f_a (f_b')^2}{8f_b^2}
+\frac{9f_a}{8f_c \r^2}+ \frac{f_a (h')^2}{8h^2}+\frac{27f_b}{16f_c \r^2}\,,
\end{split}
\eqlabel{kseq3}
\end{equation}
\begin{equation}
\begin{split}
&0=f_b''-\frac{3 f'_b}{\r}-\frac{(f_b')^2}{8f_b}
+\frac{5 f_b}{\r^2}-\frac{45 f_b^2}{16f_c f_a \r^2}+\frac{f_b h'}{h \r}
-\frac{K_1^2}{8f_c h^2 f_a^2 f_b \r^2}-\frac{3 f_b (K_3')^2}{32h g f_a^2 P^2}
\\
&-\frac{K_2^2 K_1^2}{32f_c h^2 f_a^2 f_b \r^2}
+ \frac{K_2 K_1^2}{8f_c h^2 f_a^2 f_b \r^2}-\frac{K_2^2 K_3^2}{32f_c h^2 f_a^2 f_b \r^2}
-\frac{9 K_1^2}{64f_c h g f_a P^2 \r^2}+\frac{3 g P^2 K_2^2}{f_c h f_b \r^2}
\\
&-\frac{9K_3^2}{64f_c h g f_a P^2 \r^2}
-\frac{5 g f_b P^2}{2f_c h f_a^2 \r^2}+\frac{3 f_b}{f_a \r^2}+\frac{3 f_c}{f_a \r^2}
-\frac{f_b f'_c f_a'}{4f_c f_a}
+\frac{5 (K_1')^2}{32h g f_b P^2}+ \frac{g P^2 (K_2')^2}{36h f_a}\\
&-\frac{9}{\r^2}
+\frac{27 f_a}{16f_c \r^2}+\frac{9 f_b}{8f_c \r^2}+ \frac{K_2^2 K_1 K_3}{16f_c h^2 f_a^2 f_b \r^2}
-\frac{K_2 K_1 K_3}{8f_c h^2 f_a^2 f_b \r^2}+\frac{5 g f_b P^2 K_2}{2f_c h f_a^2 \r^2}
-\frac{5 g f_b P^2 K_2^2}{8f_c h f_a^2 \r^2}\\
&+\frac{9 K_1 K_3}{32f_c h g f_a P^2 \r^2}
+ \frac{f_b (g')^2}{8g^2}-3 h f_b \k+ \frac{f_a' f'_b}{2f_a}
-\frac{f_b (f_a')^2}{8f_a^2}+\frac{f'_b f'_c}{4f_c}+ \frac{f_b (h')^2}{8h^2}\,,
\end{split}
\eqlabel{kseq4}
\end{equation}
\begin{equation}
\begin{split}
&0=h''+\frac{ K_2^2 K_1^2}{4f_c f_a^2 f_b^2 h \r^2}
-\frac{K_2 K_1^2}{f_c f_a^2 f_b^2 h \r^2}+ \frac{K_2^2 K_3^2}{4f_c f_a^2 f_b^2 h \r^2}
+\frac{9 K_1^2}{16f_c f_a f_b \r^2 g P^2}+\frac{9 K_3^2}{16f_c f_a f_b \r^2 g P^2}
\\
&+\frac{2 h f'_c}{f_c \r}+\frac{4 h f'_b}{f_b \r}+\frac{4 h f_a'}{f_a \r}+ \frac{(K_1')^2}{8f_b^2 g P^2}
+\frac{(K_3')^2}{8f_a^2 g P^2}+\frac{g P^2 K_2^2}{2f_c f_b^2 \r^2}+\frac{ g P^2 K_2^2}{2f_c f_a^2 \r^2}
-\frac{2 g P^2 K_2}{f_c f_a^2 \r^2}+ \frac{f'_c h'}{2f_c}
\\
&+\frac{h' f'_b}{f_b}+\frac{h' f_a'}{f_a}-\frac{16 h}{\r^2}-\frac{(h')^2}{h}+12 h^2 \k
-\frac{K_2^2 K_1 K_3}{2f_c f_a^2 f_b^2 h \r^2}+\frac{K_2 K_1 K_3}{f_c f_a^2 f_b^2 h \r^2}+\frac{K_1^2}{f_c f_a^2 f_b^2 h \r^2}
\\
&+\frac{2 g P^2}{f_c f_a^2 \r^2}+ \frac{g P^2 (K_2')^2}{9f_a f_b)}
-\frac{9 K_1 K_3}{8f_c f_a f_b \r^2 g P^2}-\frac{3 h'}{\r}\,,
\end{split}
\eqlabel{kseq5}
\end{equation}
\begin{equation}
\begin{split}
&0=K_1''-\frac{g K_2^2 K_1 P^2}{f_c f_a^2 h \r^2}+\frac{g K_2^2 K_3 P^2}{f_c f_a^2 h \r^2}
+\frac{4 g K_2 K_1 P^2}{f_c f_a^2 h \r^2}-\frac{2 g K_2 K_3 P^2}{f_c f_a^2 h \r^2}- \frac{9f_b K_1}{2f_c f_a \r^2}
+ \frac{9f_b K_3}{2f_c f_a \r^2}\\
&-\frac{4 g K_1 P^2}{f_c f_a^2 h \r^2}+\frac{K'_1 f'_c}{2f_c}
-\frac{K'_1 g'}{g}-\frac{K'_1 h'}{h}+\frac{f_a' K'_1}{f_a}-\frac{3 K'_1}{\r}-\frac{K'_1 f'_b}{f_b}\,,
\end{split}
\eqlabel{kseq6}
\end{equation}
\begin{equation}
\begin{split}
&0=K_3''+\frac{g K_2^2 K_1 P^2}{f_c f_b^2 h \r^2}-\frac{g K_2^2 K_3 P^2}{f_c f_b^2 h \r^2}
-\frac{2 g K_2 K_1 P^2}{f_c f_b^2 h \r^2}+ \frac{9f_a K_1}{2f_c f_b \r^2}
- \frac{9f_a K_3}{2f_c f_b \r^2}+ \frac{K'_3 f'_c}{2f_c}\\
&-\frac{K'_3 g'}{g}
+\frac{f'_b K'_3}{f_b}-\frac{K'_3 h'}{h}
-\frac{3 K'_3}{\r}-\frac{K'_3 f_a'}{f_a}\,,
\end{split}
\eqlabel{kseq7}
\end{equation}
\begin{equation}
\begin{split}
&0=K_2''-\frac{9f_b K_2}{2f_c f_a \r^2}
- \frac{9f_a K_2}{2f_c f_b \r^2)}+\frac{9 f_b}{f_c f_a \r^2}-\frac{9 K_2 K_1^2}{8f_c g P^2 h f_b f_a \r^2}
+\frac{9 K_2 K_1 K_3}{4f_c g P^2 h f_b f_a \r^2}\\
&-\frac{9 K_2 K_3^2}{8f_c g P^2 h f_b f_a \r^2}
+\frac{9 K_1^2}{4f_c g P^2 h f_b f_a \r^2}- \frac{9K_1 K_3}{4f_c g P^2 h f_b f_a \r^2}
+ \frac{K'_2 f'_c}{2f_c}+\frac{K'_2 g'}{g}-\frac{K'_2 h'}{h}\\
&-\frac{3 K'_2}{\r}\,,
\end{split}
\eqlabel{kseq8}
\end{equation}
\begin{equation}
\begin{split}
&0=g''- \frac{g^2 P^2 K_2^2}{2f_c f_a^2 h \r^2}-\frac{g^2 P^2 K_2^2}{2f_c f_b^2 h \r^2}
+\frac{2 g^2 P^2 K_2}{f_c f_a^2 h \r^2}+\frac{9K_1^2}{16f_c f_a f_b h \r^2 P^2}+\frac{9 K_3^2}{16f_c f_a f_b h \r^2 P^2}
\\&-\frac{(g')^2}{g}- \frac{9K_1 K_3}{8f_c f_a f_b h \r^2 P^2}
+\frac{(K_3')^2}{8f_a^2 h P^2}+\frac{(K_1')^2}{8f_b^2 h P^2}
-\frac{2 g^2 P^2}{f_c f_a^2 h \r^2}-\frac{g^2 P^2 (K_2')^2}{9f_a f_b h}
+ \frac{g' f'_c}{2f_c}\\
&+\frac{g' f_a'}{f_a}+\frac{g' f'_b}{f_b}-\frac{3 g'}{\r}\,.
\end{split}
\eqlabel{kseq9}
\end{equation}
Additionally, we have the first order constraint
\begin{equation}
\begin{split}
&0=\frac89 g^2 (K_2')^2 f_b f_a P^4+(K_3')^2 f_b^2+(K_1')^2 f_a^2-\frac{4 g^2 K_2^2 f_a^2 P^4}{f_c \r^2}
+\frac{4 g f_a^2 f_b^2 P^2 (h')^2}{h}\\
&+\frac{4 h (g')^2 f_a^2 f_b^2 P^2}{g}+\frac{96 h g f_a^2 f_b P^2}{\r^2}+\frac{96 h g f_a f_b^2 P^2}{\r^2}
-\frac{96 h g f_a^2 f_b^2 P^2}{\r^2}-\frac{4 g K_1^2 P^2}{f_c h \r^2}
\\
&+96 h^2 g f_a^2 f_b^2 P^2 \k+\frac{9 K_1 K_3 f_b f_a}{f_c \r^2}
+\frac{32 g f_a^2 f_b^2 P^2 h'}{\r}+\frac{16 g^2 K_2 f_b^2 P^4}{f_c \r^2}
-\frac{4 g^2 K_2^2 f_b^2 P^4}{f_c \r^2}\\
&-\frac{g K_2^2 K_1^2 P^2}{f_c h \r^2}
+\frac{4 g K_2 K_1^2 P^2}{f_c h \r^2}-\frac{g K_2^2 K_3^2 P^2}{f_c h \r^2}
+\frac{64 h g f_a^2 f_b P^2 f'_b}{\r}+\frac{64 h g f_a f_b^2 P^2 f_a'}{\r}
\\&-16 h g f_a f_b P^2 f_a' f'_b-\frac{32 f_c h g f_a f_b P^2}{\r^2}
-\frac{18 h g f_a f_b^3 P^2}{f_c \r^2}-\frac{18 h g f_a^3 f_b P^2}{f_c \r^2}
+\frac{36 h g f_a^2 f_b^2 P^2}{f_c \r^2}\\
&-\frac{9 K_3^2 f_b f_a}{f_c \r^2}
-4 h g f_b^2 P^2 (f_a')^2-4 h g f_a^2 P^2 (f_b')^2-\frac{16 g^2 f_b^2 P^4}{f_c \r^2}
+\frac{2 g K_2^2 K_1 K_3 P^2}{f_c h \r^2}\\
&-\frac{4 g K_2 K_1 K_3 P^2}{f_c h \r^2}
-\frac{8 h g f_b^2 f_a P^2 f'_c f_a'}{f_c}+\frac{32 h g f_a^2 f_b^2 P^2 f'_c}{f_c \r}
-\frac{8 h g f_a^2 f_b P^2 f'_b f'_c}{f_c}-\frac{9 K_1^2 f_b f_a}{2f_c \rho^2}\,.
\end{split}
\eqlabel{kseq10}
\end{equation}
We explicitly verified that for any value $\k$ the constraint \eqref{kseq10} is 
consistent with \eqref{kseq2}-\eqref{kseq9}. Moreover, 
with 
\begin{equation}
f_c=f_2\,,\qquad f_a=f_b=f_3\,,\qquad K_1=K_3=K\,,\qquad K_2=1\,,
\eqlabel{chirallimit}
\end{equation}
equations \eqref{kseq2}-\eqref{kseq10} are equivalent to \eqref{eq2}-\eqref{eq7}.

\subsection{UV asymptotics}\label{ksuv}
The general UV (as $\r\to 0$) asymptotic solution of \eqref{kseq2}-\eqref{kseq10} describing the phase of cascading
 gauge theory with spontaneously broken chiral symmetry takes the form
\begin{equation}
\begin{split}
f_c=&1-\a_{1,0} \r+\biggl(-\frac38 g_0 P^2-\frac14 K_0+\frac14 \a_{1,0}^2+\frac12 P^2 g_0 \ln\r\biggr) \r^2
\\&+\frac14  P^2 \a_{1,0} g_0 \r^3
+\sum_{n=4}^\infty\sum_k f_{c,n,k}\ {\r^n}\ln^k\r\,,
\end{split}
\eqlabel{ksfc}
\end{equation}
\begin{equation}
\begin{split}
f_a=&1-\alpha_{1,0}\r+\biggl(-\frac12 g_0 P^2-\frac 14 K_0+\frac 14 \a_{1,0}^2+\frac12 P^2 g_0 \ln\r
\biggr) \r^2+ f_{a,3,0}\r^3\\
&+\sum_{n=4}^\infty\sum_k f_{a,n,k}\ {\r^n}\ln^k\r\,,
\end{split}
\eqlabel{ksfa}
\end{equation}
\begin{equation}
\begin{split}
f_b=&1- \a_{1,0}\r+\biggl(-\frac12 g_0 P^2-\frac14 K_0+\frac14 \a_{1,0}^2+\frac12 P^2 g_0 \ln\r\biggr)
 \r^2\\&+\biggl(\frac12 P^2 \a_{1,0} g_0-f_{a,3,0}\biggr) \r^3
+\sum_{n=4}^\infty\sum_k f_{b,n,k}\ {\r^n}\ln^k\r\,,
\end{split}
\eqlabel{ksfb}
\end{equation}
\begin{equation}
\begin{split}
h=&\frac18 g_0 P^2+\frac14 K_0-\frac12 P^2 g_0 \ln\r+\biggl(-P^2 g_0 \ln\r+\frac12 K_0\biggr) \a_{1,0} \r
+\biggl(\biggl(
-\frac14 g_0 P^2\\&-\frac54 P^2 g_0 \ln\r
+\frac58 K_0\biggr) \a_{1,0}^2+\frac{119}{576} P^4 g_0^2+\frac{31}{96} P^2 g_0 K_0+\frac18 K_0^2+\frac12 P^4 g_0^2 \ln\r^2
\\&
-\frac{31}{48} P^4 g_0^2 \ln\r-\frac12 \ln\r P^2 g_0 K_0\biggr)\r^2+\biggl(\biggl(-\frac54 P^2 g_0 \ln\r
-\frac{11}{24} g_0 P^2+\frac58 K_0\biggr) \a_{1,0}^3
\\&+\biggl(\frac32 P^4 g_0^2 \ln\r^2-\frac{23}{16} P^4 g_0^2 \ln\r+\frac{19}{64} P^4 g_0^2-\frac32 \ln\r P^2 g_0 K_0
+\frac{23}{32} P^2 g_0 K_0\\
&+\frac38 K_0^2\biggr) \a_{1,0}\biggr)\r^3+\sum_{n=4}^\infty\sum_k h_{n,k}\ {\r^n}\ln^k\r\,,
\end{split}
\eqlabel{ksh}
\end{equation}
\begin{equation}
\begin{split}
K_1=&K_0-2 P^2 g_0 \ln\r-P^2 \a_{1,0} g_0 \r+\biggl(-\frac14 P^2 \a_{1,0}^2 g_0-\frac14 P^4 g_0^2 \ln\r
+\frac{9}{16} P^4 g_0^2\\
&+\frac18 P^2 g_0 K_0\biggr)\r^2
+\biggl(-\frac{1}{12} \a_{1,0}^3 g_0 P^2+\frac{1}{48} g_0 P^2 \biggl(-36 P^2 g_0 \ln\r+13 P^2 g_0\\
&+6 K_0
\biggr) \a_{1,0}+\frac{1}{48} g_0 P^2 \biggl(96 f_{a,3,0} \ln\r+32 f_{a,3,0}+32 k_{2,3,0}
\biggr)\biggr)\r^3\\
&+\sum_{n=4}^\infty\sum_k k_{1,n,k}\ {\r^n}\ln^k\r\,,
\end{split}
\eqlabel{ksK1}
\end{equation}
\begin{equation}
\begin{split}
K_2=&1+\left(k_{2,3,0}-\frac34 \a_{1,0} P^2 g_0 \ln\r+3 f_{a,3,0} \ln\r\right) \r^3
+\sum_{n=4}^\infty\sum_k k_{2,n,k}\ {\r^n}\ln^k\r\,,
\end{split}
\eqlabel{ksK2}
\end{equation}
\begin{equation}
\begin{split}
K_3=&K_0-2 P^2 g_0 \ln\r-P^2 \a_{1,0} g_0 \r+\biggl(-\frac14 P^2 \a_{1,0}^2 g_0-\frac14 P^4 g_0^2 \ln\r+\frac{9}{16}
 P^4 g_0^2\\&+\frac18 P^2 g_0 K_0\biggr)\r^2
+\biggl(-\frac{1}{12} \a_{1,0}^3 g_0 P^2+\frac{1}{48} g_0 P^2 \biggl(12 P^2 g_0 \ln\r+29 P^2 g_0\\
&+6 K_0
\biggr) \a_{1,0}-\frac{1}{48} g_0 P^2 \biggl(96 f_{a,3,0} \ln\r+32 f_{a,3,0}
+32 k_{2,3,0}\biggr)\biggr)\r^3\\&+\sum_{n=4}^\infty\sum_k k_{3,n,k}\ {\r^n}\ln^k\r\,,
\end{split}
\eqlabel{ksK3}
\end{equation}
\begin{equation}
\begin{split}
g=&g_0\biggl(1-\frac12 P^2 g_0\r^2-\frac12 \a_{1,0} P^2 g_0\r^3
+\sum_{n=4}^\infty\sum_k g_{n,k}\ {\r^n}\ln^k\r\biggr)\,.
\end{split}
\eqlabel{ksg}
\end{equation}
It is characterized by 11 parameters:
\begin{equation}
\{K_0\,,\ H\,,\ g_0\,,\  \a_{1,0}\,,\ k_{2,3,0}\,,\ f_{c,4,0}\,,\ f_{a,3,0}\,,\ f_{a,6,0}\,,\ f_{a,7,0}\,,\ f_{a,8,0}\,,\ g_{4,0}\}\,.
\eqlabel{uvparks}
\end{equation}
In what follows we developed the UV expansion to order $\calo(\r^{10})$ inclusive.

\subsection{IR asymptotics}\label{ksir}
As in section \ref{ircond}, we use a radial coordinate $\r$ that extends to infinity, see \eqref{extend}. 
The crucial difference between the IR boundary conditions for a chirally symmetric phase discussed in section \ref{ircond}
and the IR boundary conditions for a $\csb$ phase discussed here is that in the former case the manifold $\calm_5$ geodesically
completes with (a smooth) shrinking to zero size of  $dS_4\subset \calm_5$, while in the latter case, much like in supersymmetric 
Klebanov-Strassler state of cascading gauge theory \cite{ks}, the 10-dimensional uplift of $\calm_5$,
\begin{equation}
\calm_5\ \to\ \calm_{10}=\calm_5\times X_5\,,
\eqlabel{uplift}
\end{equation}
 geodesically completes with (a smooth) shrinking of a 2-cycle in the compact manifold $X_5$ \cite{ks}.        
Introducing 
\begin{equation}
y\equiv \frac 1\r\,,\qquad h^h\equiv y^{-4}\ h\,,\qquad  f^h_{a,b,c}\equiv y^2\ f_{a,b,c}\,,
\eqlabel{kshorfunc}
\end{equation}
the general IR (as $y\to 0$) asymptotic solution of  \eqref{kseq2}-\eqref{kseq10} describing the $\csb$ phase of cascading 
gauge theory takes form
\begin{equation}
\begin{split}
f_c^h=&\frac34 f_{a,0}^h+\biggl(-\frac{19(k_{2,2}^h)^2 P^2 g_0^h}{540h_0^h}-\frac34 f_{a,0}^h h_0^h \k
-\frac{3f_{a,0}^h k_{2,4}^h}{2k_{2,2}^h}-\frac{13P^2 g_0^h}{15(f_{a,0}^h)^2 h_0^h}+\frac65\\
&+\frac{f_{a,0}^h (k_{1,3}^h)^2}{64P^2 g_0^h h_0^h}
-\frac{27}{5 f_{a,0}^h k_{2,2}^h}+ \frac{19(k_{3,1}^h)^2}{320P^2 f_{a,0}^h g_0^h h_0^h}
+\frac{3k_{1,3}^h k_{3,1}^h}{20k_{2,2}^h P^2 f_{a,0}^h g_0^h h_0^h}\biggr)\ y^2\\
&+\sum_{n=2}^\infty f_{c,n}^h\ y^{2n}\,,
\end{split}
\eqlabel{fchks}
\end{equation}
\begin{equation}
\begin{split}
f_a^h=&f_{a,0}^h+\biggl(\frac{17(k_{2,2}^h)^2 P^2 g_0^h}{405h_0^h}+2 f_{a,0}^h h_0^h \k+\frac{f_{a,0}^h k_{2,4}^h}{k_{2,2}^h}
-\frac{4 P^2 g_0^h}{45(f_{a,0}^h)^2 h_0^h}+\frac{11}{5}
+\frac{f_{a,0}^h (k_{1,3}^h)^2}{48P^2 g_0^h h_0^h}\\&+\frac{18}{5 f_{a,0}^h k_{2,2}^h}
- \frac{17(k_{3,1}^h)^2}{240P^2 f_{a,0}^h g_0^h h_0^h}
-\frac{k_{1,3}^h k_{3,1}^h}{10k_{2,2}^h P^2 f_{a,0}^h g_0^h h_0^h}\biggr)\ y^2+\sum_{n=2}^\infty f_{a,n}^h\ y^{2n}\,,
\end{split}
\eqlabel{fahks}
\end{equation}
\begin{equation}
\begin{split}
f_b=&3\ y^2+\sum_{n=2}^\infty f_{b,n}^h\ y^{2n}\,,
\end{split}
\eqlabel{fbhks}
\end{equation}
\begin{equation}
\begin{split}
h^h=&h_0^h+\biggl(-\frac{g_0^h P^2 (k_{2,2}^h)^2}{27f_{a,0}^h}
-2 \k (h_0^h)^2-\frac{4g_0^h P^2}{9(f_{a,0}^h)^3}
- \frac{(k_{1,3}^h)^2}{48g_0^h P^2}- \frac{(k_{3,1}^h)^2}{16g_0^h P^2 (f_{a,0}^h)^2}
\biggr)\ 
y^2\\&+\sum_{n=2}^\infty h_{n}^h\ y^{2n}\,,
\end{split}
\eqlabel{hhks}
\end{equation}
\begin{equation}
\begin{split}
K_1=&k_{1,3}^h\ y^3+\sum_{n=2}^\infty k_{1,n}^h\ y^{2n+1}\,,
\end{split}
\eqlabel{k1hks}
\end{equation}
\begin{equation}
K_2=k_{2,2}^h\ y^2+k_{2,4}^h\ y^4+\sum_{n=3}^\infty k_{2,n}^h\ y^{2n}\,,
\eqlabel{k2hks}
\end{equation}
\begin{equation}
\begin{split}
K_3=&k_{3,1}^h\ y+\biggl(
\frac{41P^2 g_0^h (k_{2,2}^h)^2 k_{3,1}^h}{810f_{a,0}^h h_0^h}
+ \frac{4P^2 g_0^h k_{2,2}^h k_{1,3}^h}{135f_{a,0}^h h_0^h}
+\frac{7}{10} h_0^h k_{3,1}^h \k-\frac15 k_{1,3}^h+\frac{k_{2,4}^h k_{3,1}^h}{k_{2,2}^h}
\\&+ \frac{2P^2 g_0^h k_{3,1}^h}{15(f_{a,0}^h)^3 h_0^h}
+\frac{4k_{3,1}^h}{5f_{a,0}^h}
+\frac{(k_{1,3}^h)^2 k_{3,1}^h}{480P^2 g_0^h h_0^h}
+\frac{18k_{3,1}^h}{5(f_{a,0}^h)^2 k_{2,2}^h}
-\frac{41(k_{3,1}^h)^3}{480P^2 (f_{a,0}^h)^2 g_0^h h_0^h}
\\&-\frac{k_{1,3}^h (k_{3,1}^h)^2}{10P^2 (f_{a,0}^h)^2 g_0^h h_0^h k_{2,2}^h}
\biggr)\ y^3+
+\sum_{n=2}^\infty k_{3,n}^h\ y^{2n+1}\,,
\end{split}
\eqlabel{k3hks}
\end{equation}
\begin{equation}
\begin{split}
g=&g^h_0\biggl(1+\biggl(
\frac{P^2 g_0^h (k_{2,2}^h)^2}{27f_{a,0}^h h_0^h}
+ \frac{4P^2 g_0^h}{9(f_{a,0}^h)^3 h_0^h}
-\frac{(k_{1_3}^h)^2}{48P^2 h_0^h g_0^h}
- \frac{(k_{3,1}^h)^2}{16P^2 (f^h_{a,0})^2 h_0^h g_0^h}
\biggr)\ y^2\\&+\sum_{n=2}^\infty g_{n}^h\ y^{2n}\biggr)\,.
\end{split}
\eqlabel{ghks}
\end{equation}
Notice that the prescribed IR boundary conditions imply 
\begin{equation}
\lim_{y\to 0}\ \om_3^2 = \lim_{y\to 0}\ \frac 16\ f_b\ h^{1/2}=\lim_{y\to 0}\ \frac {y^2}{6}\ f_b\ (h^h)^{1/2}=0\,,
\eqlabel{2cycle}
\end{equation}  
with all the other warp factors in \eqref{10dmetric} being finite. Moreover, see \eqref{10dmetric},
\begin{equation}
\begin{split}
\lim_{y\to 0}\biggl(\om_1^2\ g_5^2+\om_2^2\ [g_3^2+g_4^2]\biggr)=\frac 16 f_{a,0}^h (h_0^h)^{1/2}\left(\frac 12 g_5^2
+g_3^2+g_4^2\right)\,,
\end{split}
\eqlabel{ids3}
\end{equation}
which is the metric of the round $S^3$ which stays of finite size in the deep infrared as the 2-cycle 
fibered over it (smoothly) shrinks to zero size \eqref{2cycle}.
Asymptotic solution \eqref{fchks}-\eqref{ghks} is characterized by 7 additional parameters:
\begin{equation}
\{f_{a,0}^h\,,\ h_{0}^h\,,\ k_{1,3}^h\,,\ k_{2,2}^h\,,\ k_{2,4}^h\,,\ k_{3,1}^h\,, g_0^h\}\,.
\eqlabel{ksirpar}
\end{equation}
In what follows we developed the IR expansion to order $\calo(y^{10})$ inclusive.

\subsection{Symmetries and numerical procedure}
The background geometry \eqref{redef2} dual to a phase of cascading gauge theory with 
spontaneously broken chiral symmetry on $dS_4$ enjoys all the symmetries, 
properly generalized,
discussed in section \ref{symmetries}:
\nxt 
\begin{equation}
P\to \lambda P\,,\ g\to \frac 1\l g\,,\ \{\r,f_{a,b,c},h,K_{1,2,3}\}\to \{\r,f_{a,b,c},h,K_{1,2,3}\}\,,
\eqlabel{kssym1}
\end{equation} 
\nxt 
\begin{equation}
P\to \lambda P\,,\ \r\to \frac 1\l \r\,,\ \{h,K_{1,3}\}\to \l^2\{h,K_{1,3}\}\,,\ 
\{f_{a,b,c},K_{2},g\}\to \{f_{a,b,c},K_{2},g\}\,,
\eqlabel{kssym2}
\end{equation} 
\nxt 
\begin{equation}
\r\to \lambda \r\,,\ H\to \frac1\l H\,,\ 
\{P,f_{a,b,c},h,K_{1,2,3},g\}\to \{P,f_{a,b,c},h,K_{1,2,3},g\}\,,
\eqlabel{kssym3}
\end{equation} 
\nxt \begin{equation}
\left( \begin{array}{c}
P\\
\r  \\
h  \\
f_{a,b,c}\\
K_{1,2,3}\\
g   
\end{array} \right)\
\Longrightarrow \left( \begin{array}{c}
\hat{P}\\
\hr  \\
\hh  \\
\hf_{a,b,c}\\
\hK_{1,2,3}\\
\hat{g}   \end{array} \right)
=
\left( \begin{array}{c}
P\\
{\r}/{(1+\a\ \r)}  \\
(1+\a\ \r)^4\ h \\
(1+\a\ \r)^{-2}\ f_{a,b,c}\\
K_{1,2,3}\\
{g}   \end{array} \right)\,,\qquad \a={\rm const}\,.
\eqlabel{kssym4}
\end{equation}
Thus, much like in section \ref{symmetries}, we can set 
\begin{equation}
g_0=1\,,\qquad H=1\,,\qquad \frac{K_0}{P^2}=\ln\ \frac{1}{\Lambda^2 P^2}\equiv \frac 1\dd\,,
\eqlabel{ksset}
\end{equation}
The residual diffeomorphisms \eqref{kssym4} are actually completely fixed once we insist on the IR asymptotics as 
in \eqref{fchks}-\eqref{ghks}.

The numerical procedure for solving the background equations \eqref{kseq2}-\eqref{kseq10}, subject to the 
boundary conditions \eqref{ksfc}-\eqref{ksg} and \eqref{fchks}-\eqref{ghks}  is identical to the one 
described earlier, see section \ref{numerical}. Given \eqref{ksset}, for a fixed $\dd$,  
the gravitational solution is characterized by 8 parameters in the UV 
and 7 parameters in the IR:
\begin{equation}
\begin{split}
&{\rm UV}:\qquad \{\a_{1,0}\,,\ k_{2,3,0}\,,\ f_{c,4,0}\,,\ f_{a,3,0}\,,\ 
f_{a,6,0}\,,\ f_{a,7,0}\,,\ f_{a,8,0}\,,\ g_{4,0}\}\,,
\\
&{\rm IR}:\qquad 
\{f_{a,0}^h\,,\ h_{0}^h\,,\ k_{1,3}^h\,,\ k_{2,2}^h\,,\ k_{2,4}^h\,,\ k_{3,1}^h\,, g_0^h\}\,.
\end{split}
\eqlabel{ksuvirfinal}
\end{equation} 
Notice that $8+7=15$ is precisely the number of integration constants needed to 
specify a solution to \eqref{kseq2}
-\eqref{kseq10} ---
we have 8 second order differential equations and a 
single first order differential constraint: $2\times 8-1=15$.

In practice, we replace the second-order differential equation for $f_c$ \eqref{kseq2} 
with the constraint equation \eqref{kseq10},
which we use to  algebraically eliminate $f_c'$ from 
\eqref{kseq3}-\eqref{kseq9}. The solution is found using the 
``shooting'' method as detailed in \cite{abk}. 

Ultimately, we are interested in the solution at $\k=H^2=1$. 
Finding such a ``shooting'' solution in 15-dimensional parameter space  
\eqref{ksuvirfinal} is quite challenging. 
Thus, we start with the analytic result for $\k=0$ 
(the Klebanov-Strassler state of cascading gauge theory),
and a fixed value of $\dd$, and  
slowly increase 
$\k$ to $\k=1$.  
We further use the obtained solution as a starting point to explore other 
values of $\dd$.

\subsection{$\k$-deformation of Klebanov-Strassler state}\label{ksk}

We begin with mapping the Klebanov-Strassler solution \cite{ks} to a 
$\k=0$ solution of \eqref{kseq2}-\eqref{kseq10}. 
We set
\begin{equation}
g_0=1\,,\qquad P=1\,.
\eqlabel{setkappa0}
\end{equation}
$\caln=1$ supersymmetric Klebanov-Strassler solution takes form\footnote{See eqs. (2.22) and (2.34) in \cite{ksbh}.}:
\begin{equation}
\begin{split}
&ds_5^2=H_{KS}^{-1/2}\ \left(-dt^2+dx_1^2+dx_2^2+dx_3^2\right)+H_{KS}^{1/2}\ \w_{1,KS}^2\ dr^2\,,\\
&\om_i=\w_{i,KS}\ H^{1/2}_{KS}\,,\qquad h_i=h_{i,KS}\,,
\end{split}
\eqlabel{ks1}
\end{equation}
\begin{equation}
\begin{split}
&h_{1,KS}=\frac{\cosh r-1}{18\sinh r}
\left(\frac{r\cosh r}{\sinh r}-1\right)\,,\qquad 
h_{2,KS}=\frac{1}{18}\left(1-\frac {r}{\sinh r}\right)\,,\\
&h_{3,KS}=\frac{\cosh r+1}{18\sinh r}
\left(\frac{r\cosh r}{\sinh r}-1\right)\,,
\qquad g=1\,,\\
&\w_{1,KS}=\frac{\epsilon^{2/3}}{\sqrt{6}{\hat K_{KS}}}\,,\qquad 
\w_{2,KS}=\frac{\epsilon^{2/3}{\hat K_{KS}}^{1/2}}{\sqrt{2}}\cosh\frac r2\,,\qquad  \w_{3,KS}=\frac{\epsilon^{2/3}
{\hat K_{KS}}^{1/2}}{\sqrt{2}}
\sinh\frac r2\,,
\end{split}
\eqlabel{ksks}
\end{equation}
with 
\begin{equation}
{\hat K_{KS}}=\frac{(\sinh (2r)-2r)^{1/3} }{2^{1/3}\sinh r}\,,\ H'_{KS}=\frac{16((9 h_{2,KS}-1)h_{1,KS}-9 h_{3,KS} h_{2,KS})}
{9\epsilon^{8/3}{\hat K_{KS}}^2\sinh^2 r }\,,\ 
\Om_0=0\,,
\eqlabel{kk}
\end{equation}
where now $r\to \infty$ is the boundary and $r\to 0$ is the IR. Above solution is parametrized by a single 
constant $\e$ which will be mapped to $K_0$, and which in turn will determine all the parameters in \eqref{ksuvirfinal}
once $\k=0$.

Comparing the metric ansatz in \eqref{ks1} and \eqref{5dks}, \eqref{redef2} we identify
\begin{equation}
\frac{(d\r)^2}{\r^4}=(w_{1,KS}(r))^2 (dr)^2\,.
\eqlabel{rrho}
\end{equation}
Introducing 
\begin{equation}
z\equiv e^{-r/3}\,,
\eqlabel{defz}
\end{equation}
we find from \eqref{rrho}
\begin{equation}
\frac1\r=\frac {\sqrt{6}\ (2\e)^{2/3}}{4}\ \int_1^z\ du\  \frac{u^6-1}{u^2(1-u^{12}+12u^6 \ln u)^{1/3}}\,.
\eqlabel{solverho}
\end{equation}
In the UV, $r\to \infty$, $z\to 0$ and $\r\to 0$ we have
\begin{equation}
\begin{split}
&e^{-r/3}\equiv z=
\frac{\sqrt{6}\ (2\e)^{2/3}}{4} \r \biggl(1+\calq \r+\calq^2 \r^2+\calq^3 \r^3+\calq^4 \r^4+\calq^5 \r^5
+\biggl(\frac{27}{80} \e^4 \ln 3+\calq^6\\
&+\frac{27}{800} \e^4-\frac{9}{16} \e^4 \ln 2+\frac{9}{20}
 \e^4 \ln\e+\frac{27}{40} \e^4 \ln\r\biggr) \r^6+\biggl(
-\frac{63}{16} \e^4 \calq \ln 2+\frac{189}{80} \e^4 \calq \ln 3+\calq^7\\
&+\frac{729}{800} \calq \e^4+\frac{63}{20} \e^4 \calq \ln\e
+\frac{189}{40} \calq \e^4 \ln\r\biggr) \r^7+\biggl(\frac{2403}{400} \e^4 \calq^2
-\frac{63}{4} \e^4 \calq^2 \ln 2+\frac{189}{20} \e^4 \calq^2 \ln 3\\
&+\frac{63}{5} \e^4 \calq^2 \ln\e+\calq^8
+\frac{189}{10} \e^4 \calq^2 \ln\r\biggr) \r^8+\biggl(\frac{189}{5} \e^4 \calq^3 \ln\e+\frac{9729}{400}
 \e^4 \calq^3-\frac{189}{4} \e^4 \calq^3 \ln 2\\
&+\frac{567}{20} \e^4 \calq^3 \ln 3+\calq^9+\frac{567}{10} \e^4 \calq^3 
\ln\r\biggr) 
\r^9+\calo(\r^{10}\ln\r)\biggr)\,,
\end{split}
\eqlabel{rrhouv}
\end{equation} 
where 
\begin{equation}
\begin{split}
\calq=&\frac{\sqrt{6}\ (2\e)^{2/3}}{4}\ \biggl\{
\int_0^1\ du\  \biggl(\frac{1-u^6}{u^2(1-u^{12}+12u^6 \ln u)^{1/3}}-\frac{1}{u^2}\biggr)-1\biggr\}\\
=&-\frac{\sqrt{6}\ (2\e)^{2/3}}{4}\ \times\ 0.839917(9)\,.
\end{split}
\eqlabel{qdef}
\end{equation}
In the IR, $r\to 0$, $z\to 1_-$ and $\frac1\r\to 0$ we have
\begin{equation}
\begin{split}
r=\frac{\sqrt 6\ 2^{1/3}}{3^{1/3}\ \e^{2/3}}\ y\ \biggl(1-\frac{2^{2/3}\ 3^{1/3}}{15\ \e^{4/3}}\  y^2+
\frac{71\ 3^{2/3}\ 2^{1/3}}{2625\ \e^{8/3}}\  y^4
+\calo(y^6)\biggr)\,.
\end{split}
\eqlabel{rrhoir}
\end{equation}
Using \eqref{rrhouv} and \eqref{rrhoir}, and the exact analytic solution describing the Klebanov-Strassler state of 
cascading gauge theory \eqref{ksks}, \eqref{kk} we can identify parameters\footnote{We matched the asymptotic
expansions \eqref{ksfc}-\eqref{ksg} and \eqref{fchks}-\eqref{ghks} with the exact solution \eqref{ksks}
to the order we developed them: $\calo(\r^{10})$ and $\calo(y^{10})$ correspondingly.} \eqref{ksuvirfinal} 
\begin{equation}
\begin{split}
&K_0=-\ln 3+\frac53\ \ln2-\frac43\ \ln\e-\frac23\,,\qquad a_{1,0}=2 \calq\,,\\
&k_{2,3,0}=\frac{3 \sqrt{6}}{8} \e^2 (3 \ln3-5 \ln2+4 \ln\e)\,,\qquad f_{c,4,0}=0\,,\qquad f_{a,3,0}=\frac{3\sqrt{6}}{4}\ \e^2\,,\\
&f_{a,6,0}=\biggl(-\frac{27}{16} \ln2
+\frac{81}{50}+\frac{81}{80} \ln3+\frac{27}{20} \ln\e\biggr) \e^4+\frac{3\sqrt{6}}{4} \calq^3  \e^2\,,
\\ &f_{a,7,0}=\frac{3}{800} \calq (2268-1800 \ln2+1440 \ln\e+1080 \ln3) \e^4
+\frac{3\sqrt{6}}{4}  \e^2 \calq^4\,,\\
&f_{a,8,0}=\frac{3}{32} \calq^2 (270-180 \ln2+108 \ln3+144 \ln\e) \e^4
+\frac{3\sqrt{6}}{4} \calq^5  \e^2\,,\qquad g_{4,0}=0\,,
\end{split}
\eqlabel{susyuv}
\end{equation} 
in the UV, and 
\begin{equation}
\begin{split}
&f_{a,0}^h=2^{1/3}\ 3^{2/3}\ \e^{4/3}\,,\qquad h_0^h=\e^{-8/3}\ \times\ 0.056288(0)\,,\\
&k_{1,3}^h=\frac{4\sqrt{6}}{9\ \e^2}\,,\qquad k_{2,2}^h=\frac{2^{2/3}}{3^{2/3}\ \e^{4/3}}\,,\qquad 
k_{2,4}^h=-\frac{11\ 2^{1/3}\ 3^{2/3}}{45\ \e^{8/3}} \,,\\
&k_{3,1}^h=\frac{4\sqrt{6}\ 2^{1/3}\ 3^{2/3}}{27\ \e^{2/3}}\,,\qquad g_0^h=1\,,
\end{split}
\eqlabel{susyir}
\end{equation}
in the IR. Notice that inverting the first identification in \eqref{susyuv}, $\e=\e(K_0)$, we obtain a prediction 
for all the parameters \eqref{ksuvirfinal} as a function of $K_0$. 

\begin{figure}[t]
\begin{center}
\psfrag{fa30}[][][0.7]{{$f_{a,3,0}$}}
\psfrag{fa60}[][][0.7]{{$f_{a,6,0}$}}
\psfrag{k230}[][][0.7]{{$k_{2,3,0}$}}
\psfrag{e}[][][0.7]{{$\epsilon$}}
\includegraphics[width=1.9in]{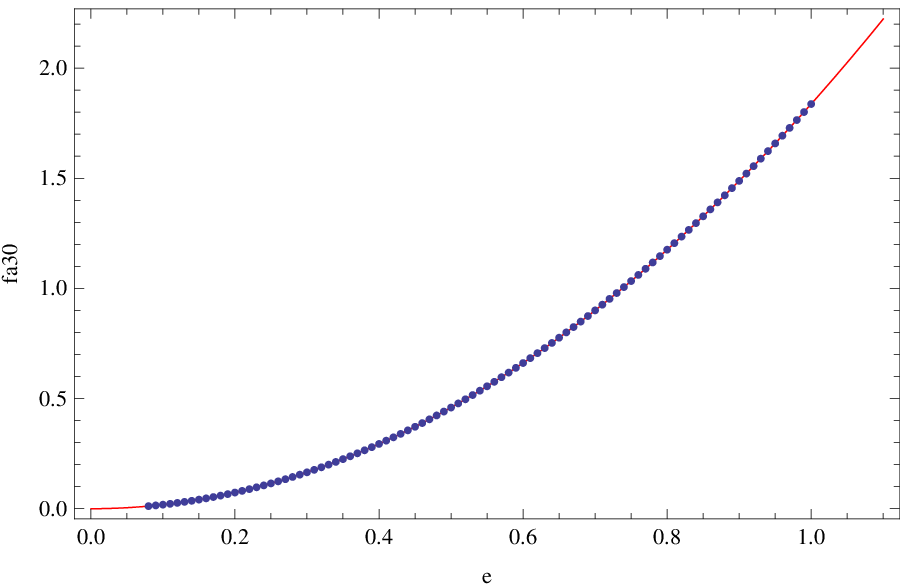}
~
\includegraphics[width=1.9in]{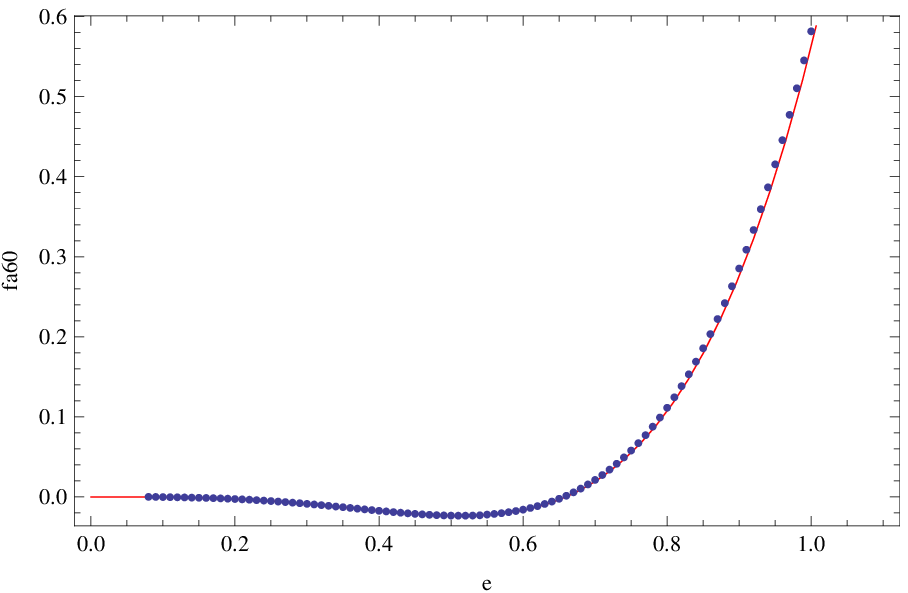}
~
\includegraphics[width=1.9in]{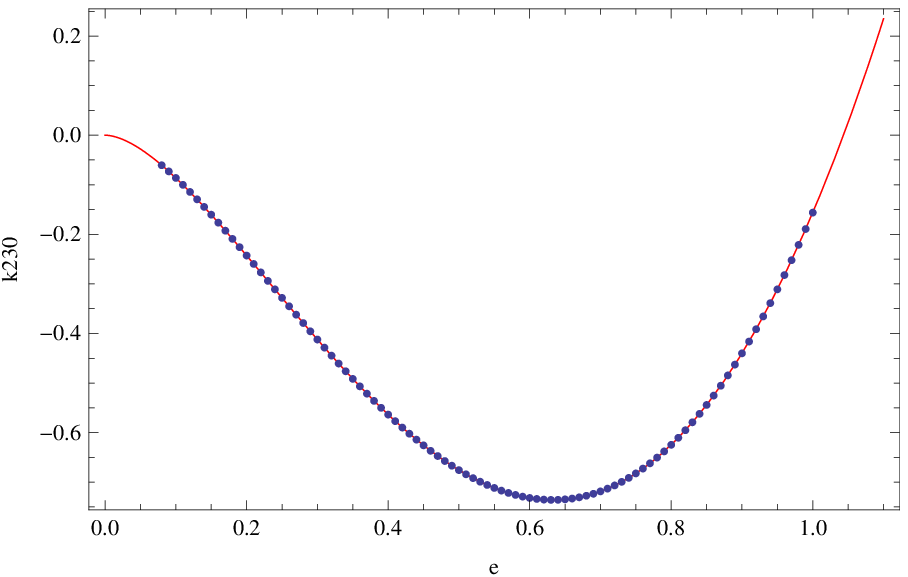}
\end{center}
  \caption{(Colour online) Comparison of values of  select UV  parameters  $\{f_{a,3,0},f_{a,6,0},k_{2,3,0}\}$  
of Klebanov-Strassler state obtained numerically (blue dots) with the analytic prediction (red curves), see \eqref{susyuv}.
} \label{figure7}
\end{figure}

\begin{figure}[t]
\begin{center}
\psfrag{K3h1}[][][0.7]{{$K_{3,1}^h$}}
\psfrag{K2h4}[][][0.7]{{$K_{2,4}^h$}}
\psfrag{K1h3}[][][0.7]{{$K_{1,3}^h$}}
\psfrag{e}[][][0.7]{{$\epsilon$}}
\includegraphics[width=1.9in]{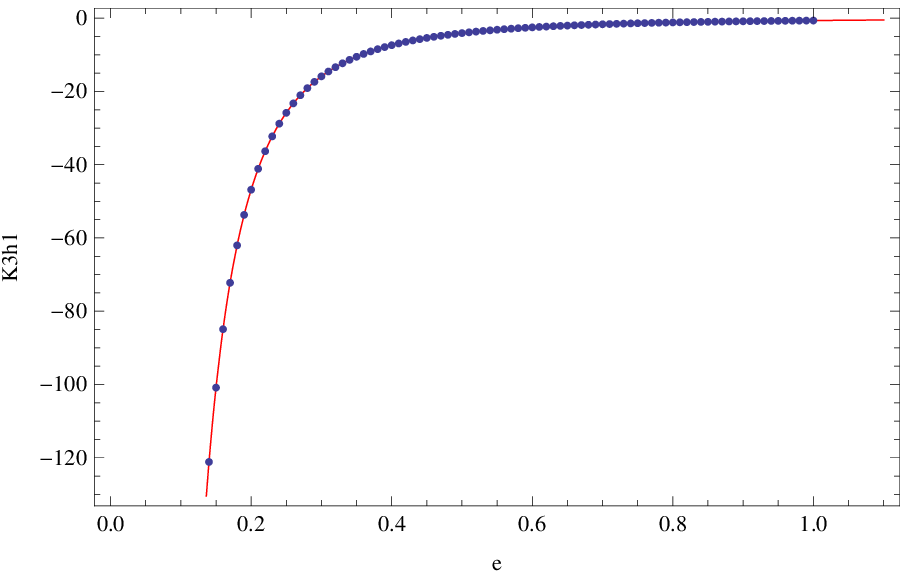}
~
\includegraphics[width=1.9in]{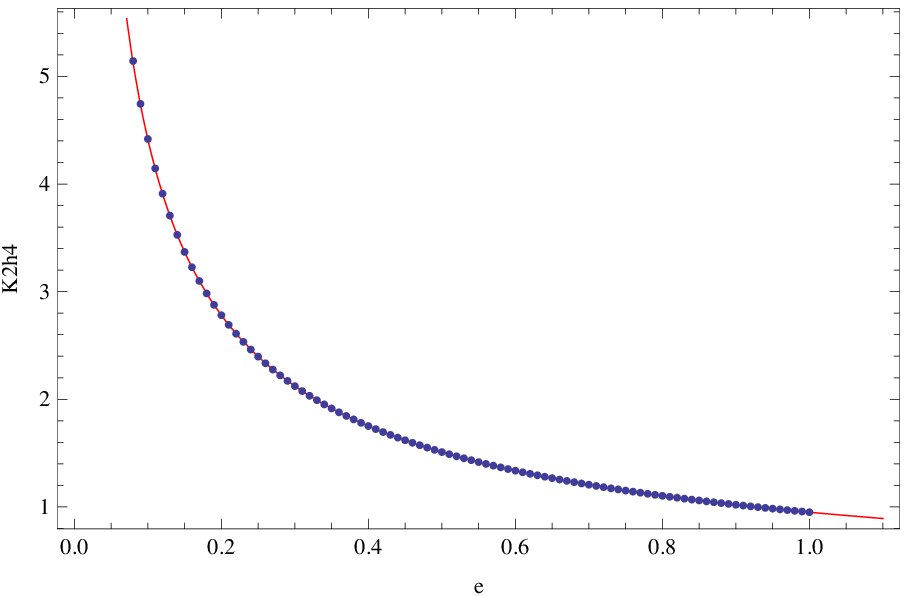}
~
\includegraphics[width=1.9in]{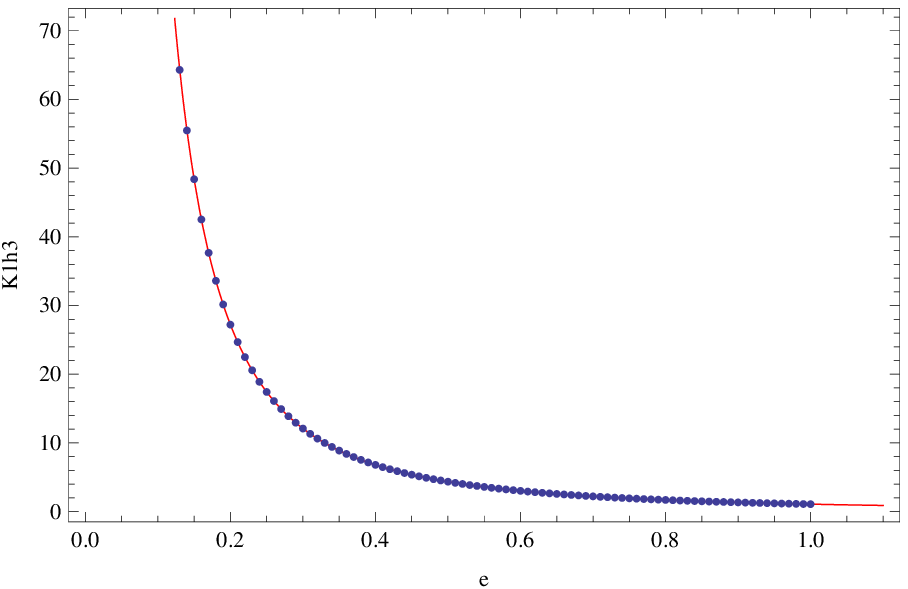}
\end{center}
  \caption{(Colour online) Comparison of values of  select IR  parameters  $\{K_{3,1}^h,K_{2,4}^h, K_{1,3}^h\}$  
of Klebanov-Strassler state obtained numerically (blue dots) with the analytic prediction (red curves), see \eqref{susyir}.
} \label{figure8}
\end{figure}

Figures \ref{figure7} and \ref{figure8} compare the results of select UV and IR parameters in \eqref{ksuvirfinal} 
obtained numerically (blue dots) with analytic predictions (red curves) \eqref{susyuv} and \eqref{susyir} for the 
supersymmetric Klebanov-Strassler state. In this numerical computation we must set $\k=0$. 
Notice that in Klebanov-Strassler state the string coupling is identically constant, \ie $g=1$. The latter in particular
implies that $g_{4,0}= 0$ and $g_0^h=1$. To find our numerical solutions, we set those values as constants and eliminate the second order equation (\ref{kseq9}) for $g$, finding excellent agreement between the expected and the numerical result.

As we mentioned earlier, we are after the states of cascading gauge theory with broken chiral symmetry 
on $dS_4$, \ie the deformations of  Klebanov-Strassler states at $\k=1$. In practice we start with 
numerical Klebanov-Strassler state at $K_0=0.25$ ($P=1$) and increase $\k$ in increments of $\dd\k=10^{-3}$
up to $\k=1$.  
The resulting state is then used as a starting point to explore the states of cascading gauge theory 
on $dS_4$ with $\csb$ for other values of $K_0\ne 0.25$.

\section{Ground state of cascading gauge theory on $dS_4$}\label{transition}

Recall that effective potential $\calv_{eff}$ of a theory on $dS_4$ is defined (by analogy with the free energy density 
in thermodynamics) via 
\begin{equation}
e^{-V_4^E\ \calv_{eff}} ={\calz}_{E}\,,
\eqlabel{defveff}
\end{equation}  
where $\calz_E$ is a Euclidean partition function of the theory on $dS_4$, and $V_4^E$ is a volume of the 
analytically continued de Sitter, $dS_4\to S^4$,
\begin{equation}
V_4^E=\frac {8\pi^2}{3H^4}\,. 
\eqlabel{v4def}
\end{equation}  
For a cascading gauge theory with a dual  gravitational action given by  \eqref{5action},
 the  effective potential is 
\begin{equation}
\calv_{eff}=\int_{\r_{UV}}^{\infty}d\r\ \call_E\,,
\eqlabel{energydef}
\end{equation}    
where $\call_E$ is the Euclidean one-dimensional Lagrangian density corresponding to the state, and $\r_{UV}$ is the 
UV cut-off, regularizing the Euclidean gravitational action in \eqref{energydef}. Briefly, holographic renormalization 
of the theory modifies the effective potential 
\begin{equation}
\int_{\r_{UV}}^{\infty}d\r\ \call_E \to \int_{\r_{UV}}^{\infty}d\r\ \call_E+ S_{GH}^{\r_{UV}}+S_{counterterms}^{\r_{UV}}\,,
\eqlabel{holrenac}
\end{equation}
to include the Gibbons-Hawking and the local counterterms at the cut-off boundary $\r=\r_{UV}$ in a way that would render 
the  renormalized effective potential finite in the limit $\r_{UV}\to 0$. 

Here, we have to distinguish two states of cascading gauge theory: with broken (we use the superscript $ ^b$)
and the unbroken (we use the superscript $ ^s$) chiral symmetry. These states are constructed (numerically) 
in sections \ref{broken} and \ref{symmetric} correspondingly. Given a cascading gauge theory on $dS_4$, 
\ie having fixed its strong coupling scale $\Lambda$, the dilaton asymptotic value $g_0$, the rank offset 
parameter $P$, and the Hubble scale $H$, the true ground state of the theory minimizes the effective potential 
$\calv_{eff}$.

We now present some computational details of $\calv_{eff}^b$ --- 
the effective potential of the state of cascading gauge theory on $dS_4$ with (spontaneously) broken chiral symmetry. 
Using the equations of motion \eqref{kseq2}-\eqref{kseq10}, it is possible to show 
that the on-shell gravitational 
Lagrangian \eqref{5action} takes form
\begin{equation}
\begin{split}
\call_{E}^b=&\frac{108}{16\pi G_5}\times \biggl( \frac{d}{d\r}
\biggl(\frac{2c_1^3c_1'\Omega_1\Omega_2^2\Omega_3^2}{c_3}\biggr)
-6 \k\ c_1^2c_3\Om_1\Om_2^2\Om_3^2
\ 
\biggr)\\
=&-\frac{108}{16\pi G_5}\times \biggl(\
\frac{d}{d\r}\bigg(\frac{f_c^{1/2} f_af_b (\r h'+4 h)}{216 h \r^4}\biggr)
+\frac{\k}{18}\ \frac{hf_af_bf_c^{1/2}}{\r^3}
\ \biggr)
\,,
\end{split}
\eqlabel{lb}
\end{equation}
leading to
\begin{equation}
\begin{split}
\frac{16\pi G_5}{108}\ \calv_{eff}^b=&-\bigg(\frac{f_c^{1/2} f_af_b (\r h'+4 h)}{216 h \r^4}\biggr)\bigg|_{\r=\r_{UV}}^\infty-
\frac{\k}{18}\ \int_{\r_{UV}}^\infty d\r\ \frac{hf_af_bf_c^{1/2}}{\r^3} \\
=&-\bigg(\frac{f_c^{1/2} f_af_b (\r h'+4 h)}{216 h \r^4}\biggr)\bigg|_{\r=\r_{UV}}-
\frac{\k}{18}\ \int_{\r_{UV}}^\infty d\r\ \frac{hf_af_bf_c^{1/2}}{\r^3} \,,
\end{split} 
\eqlabel{seb}
\end{equation}
where we used the fact that (see \eqref{fchks}-\eqref{hhks})
\begin{equation}
\lim_{\r\to \infty}\ \frac{f_c^{1/2} f_af_b (\r h'+4 h)}{216 h \r^4}=-\lim_{y\to 0}
\frac{(f_c^h)^{1/2} f_a^h f_b^h(h^h)'}{216 h^h}=0\,.
\eqlabel{ircontributions}
\end{equation}
Both terms in \eqref{seb} are divergent as $\r_{UV}\to 0$. First, 
using the asymptotic expansion \eqref{ksfc}-\eqref{ksh}, we isolate the 
divergence of the integral in \eqref{seb}:
\begin{equation}
\begin{split}
&\cali^b_{\r_{UV}}\equiv -6 \k\ \int_{\r_{UV}}^1 d\r\ \frac{f_af_bf_c^{1/2}h}{\r^3}\equiv \cali_{finite}^b
+\cali_{\r_{UV},divergent}^b+\calo\left(\r_{UV}\ln^2\r_{UV}\right)\,,
\end{split}
\eqlabel{calidef}
\end{equation}
\begin{equation}
\begin{split}
&\cali_{finite}^b=-6\k \int_0^1 d\r\  \biggl(\frac{f_af_bf_c^{1/2}h}{\r^3}
-\calj_{divergent}^b\biggr)\,,\\
&\calj_{divergent}^b=\frac{1}{\r^3}\biggl(\frac18 g_0 P^2+\frac14 K_0-\frac12 P^2 g_0 \ln\r\biggr)
+\frac{1}{\r^2}\biggl(\frac14 \a_{1,0}^b g_0 P^2 \ln\r\\
&-\frac{1}{16} \a_{1,0}^b (5 g_0 P^2+2 K_0)\biggr)
+\frac{1}{\r}\biggl(-\frac18 P^4 \k g_0^2 \ln^2\r+\biggl(\frac18 K_0 P^2 \k g_0+\frac{5}{48} P^4 \k g_0^2\biggr) \ln\r
\\&+\frac{1}{16} (\a_{1,0}^b)^2 P^2 g_0
+\frac{67}{1152} P^4 \k g_0^2-\frac{5}{96} K_0 P^2 \k g_0-\frac{1}{32} K_0^2 \k\biggr)\,,
\end{split}
\eqlabel{califinite}
\end{equation}
\begin{equation}
\begin{split}
&\cali_{\r_{UV},divergent}^b=-6\k\int_{\r_{UV}}^1 d\r\ \calj_{divergent}^b\\
=&\frac{1}{\r_{UV}^2}\biggl(\frac32 \k g_0 P^2 \ln\r_{UV}-\frac38 \k (-g_0 P^2+2 K_0)\biggr)
\\
&+\frac{1}{\r_{UV}}\biggl(-\frac32 \k \a_{1,0}^b g_0 P^2 \ln\r_{UV}+\frac38 \k \a_{1,0}^b (g_0 P^2+2 K_0)\biggr)
\\
&-\frac14 \k^2 P^4 g_0^2 \ln^3\r_{UV}-\frac{1}{192} \k (-72 K_0 P^2 \k g_0-60 P^4 \k g_0^2) \ln^2\r_{UV}
\\&-\frac{1}{192} \k (-72 (\a_{1,0}^b)^2 P^2 g_0-67 P^4 \k g_0^2+60 K_0 P^2 \k g_0+36 K_0^2 \k) \ln\r_{UV}
\\
&+\biggl\{-\frac{1}{192} \k (-72 \a_{1,0}^b g_0 P^2-144 K_0-144 \a_{1,0}^b K_0+72 g_0 P^2)\biggr\}\,,
\end{split}
\eqlabel{calidiv}
\end{equation}
where in the last line we separated the finite piece coming from the upper limit of integration
in $\cali_{\r_{UV},divergent}^b$. The superscript $\ ^b$ in the UV parameter 
$\a_{1,0}$ is used to indicate that it is computed in the phase with broken chiral symmetry. 
Combining the divergent terms in \eqref{calidiv} with divergences of the boundary term in \eqref{seb}
we find
\begin{equation}
\begin{split}
&\frac{16\pi G_5}{108}\ \calv_{eff}^b=\biggl\{
\calv_{eff,-4}^b\ \frac{1}{\r^4}+\calv_{eff,-3}^b\ \frac{1}{\r^3}+\calv_{eff,-2}^b \frac{1}{\r^2}
+\calv_{eff,-1}^b\ \frac{1}{\r}+\calv_{eff,0}^b \\
&+\calo(\r^0)
\biggr\}\bigg|_{\r=\r_{UV}}\,,
\end{split}
\eqlabel{caleb}
\end{equation} 
with 
\begin{equation}
\begin{split}
\calv_{eff,-4}^b=&\frac{K_0-2 \ln\r}{27(1+2 K_0-4 \ln\r)}\,,
\end{split}
\eqlabel{ebm4}
\end{equation}
\begin{equation}
\begin{split}
\calv_{eff,-3}^b=&\frac{\a_{1,0}^b}{27(1+2 K_0-4 \ln\r)^2} \biggl(16 \ln\r^2-(4 (1+4 K_0)) \ln\r+1+2 K_0+4 K_0^2\biggr)\,,
\end{split}
\eqlabel{ebm3}
\end{equation}
\begin{equation}
\begin{split}
\calv_{eff,-2}^b=&-\frac{1}{3888(1+2 K_0-4 \ln\r)^3} \biggl(6912 \ln\r^4-(192 (37+72 K_0-36 (\a_{1,0}^b)^2)) \ln\r^3\\
&+(32 (43+333 K_0-108 (\a_{1,0}^b)^2+324 K_0^2-324 K_0 (\a_{1,0}^b)^2)) \ln\r^2-(4 (-97\\
&+344 K_0-360 (\a_{1,0}^b)^2+1332 K_0^2-864 K_0 (\a_{1,0}^b)^2+864 K_0^3\\
&-1296 K_0^2 (\a_{1,0}^b)^2)) \ln\r-99-194 K_0+36 (\a_{1,0}^b)^2+344 K_0^2-720 K_0 (\a_{1,0}^b)^2\\
&+888 K_0^3-864 K_0^2 (\a_{1,0}^b)^2+432 K_0^4-864 K_0^3 (\a_{1,0}^b)^2\biggr)\,,
\end{split}
\eqlabel{ebm2}
\end{equation}
\begin{equation}
\begin{split}
\calv_{eff,-1}^b=&\frac{\a_{1,0}^b}{3888(1+2 K_0-4 \ln\r)^4}  \biggl(-27648 \ln\r^5+(1536 (32+45 K_0-6 (\a_{1,0}^b)^2)) \ln\r^4\\
&-(64 (413+1536 K_0-108 (\a_{1,0}^b)^2+1080 K_0^2-288 K_0 (\a_{1,0}^b)^2)) \ln\r^3\\
&+(48 (161+826 K_0-88 (\a_{1,0}^b)^2+1536 K_0^2-216 K_0 (\a_{1,0}^b)^2+720 K_0^3\\
&-288 K_0^2 (\a_{1,0}^b)^2)) \ln\r^2-(16 (134+483 K_0+21 (\a_{1,0}^b)^2+1239 K_0^2\\
&-264 K_0 (\a_{1,0}^b)^2+1536 K_0^3-324 K_0^2 (\a_{1,0}^b)^2+540 K_0^4-288 K_0^3 (\a_{1,0}^b)^2)) \ln\r\\
&+301+1072 K_0-300 (\a_{1,0}^b)^2+1932 K_0^2+168 K_0 (\a_{1,0}^b)^2+3304 K_0^3\\
&-1056 K_0^2 (\a_{1,0}^b)^2+3072 K_0^4-864 K_0^3 (\a_{1,0}^b)^2+864 K_0^5-576 K_0^4 (\a_{1,0}^b)^2\biggr)\,,
\end{split}
\eqlabel{ebm1}
\end{equation}
\begin{equation}
\begin{split}
\calv_{eff,0}^b=&-\frac{1}{432} \ln\r^3+\frac{1}{3456}(13+12 K_0)  \ln\r^2
-\frac{1}{82944} (103+312 K_0-576 (\a_{1,0}^b)^2\\
&+144 K_0^2) \ln\r\,,
\end{split}
\eqlabel{ebm0}
\end{equation}
where we set $P=1$, $g_0=1$, $\k=1$, and used \eqref{ksfc}-\eqref{ksh}. 
 Turns out that all the divergences are removed 
once we include the generalized\footnote{``Generalized" five-dimensional Gibbons-Hawking term is just a dimensional reduction of the 
10-dimensional Gibbons-Hawking term corresponding to \eqref{10dmetric}.} Gibbons-Hawking term, see \cite{aby}, 
\begin{equation}
\begin{split}
S_{GH}^{\r_{UV}}=\frac{108}{8\pi G_5}\ \frac{1}{c_3}\left(c_1^4 \Om_1\Om_2^2\Om_3^2\right)'\bigg|_{\r=\r_{UV}}=
\frac{1}{8\pi G_5}\ \frac{\r}{h^{1/4}}\left(\frac{h^{1/4}f_c^{1/2}f_af_b}{\r^4}\right)'\bigg|_{\r=\r_{UV}}\,,
\end{split}
\eqlabel{ghterm}
\end{equation}
and the local counter-terms obtained in \cite{aby} with the following obvious modifications:
\begin{equation}
\begin{split}
&K^{KT}=\frac 12 K_1+\frac 12 K_3\,,\qquad \Om_1^{KT}=3\Om_1\,,\qquad \Om_2^{KT}=\frac{\sqrt{6}}{2} \left(\Om_2+\Om_3\right) \,.
\end{split}
\eqlabel{modifications}
\end{equation}
We find
\begin{equation}
\begin{split}
&16\pi G_5\ \calv_{eff}^b=3 f_{c,4,0}+\frac{9}{32} (\a_{1,0}^b)^2
+\frac{3}{16} K_0 (\a_{1,0}^b)^2+\frac{59}{48} K_0+\frac{805}{1152}
-\frac34 \a_{1,0}^b K_0\\&-\frac38 \a_{1,0}^b-\frac18 K_0^2
+\cali_{finite}^b+\int_0^1 dy\ (-6 h^h f_a^h f_b^h (f_c^h)^{1/2})+\calv_{ambiguity}^b\,,\\
&\calv_{ambiguity}^b=-36\kappa_1^b K_0^2-36\kappa_2^b K_0-36 \kappa_3^b\,,
\end{split}
\eqlabel{calebfin}
\end{equation}
where $\calv_{ambiguity}^b$ comes from the renormalization scheme ambiguities $\{\k_i^b\}$, see \cite{aby}. Note that 
the ambiguities are completely specified by the gauge theory parameters, \ie $\{K_0,P,g_0\}$ and the Hubble scale $H$, 
(the non-normalizable 
coefficients of the holographic gravitational dual).

Identical analysis for the symmetric phase leads to 
\begin{equation}
\begin{split}
&16\pi G_5\ \calv_{eff}^s=3 a_{4,0}+\frac{805}{1152}-\frac 38
 \a_{1,0}^s+\frac{59}{48} K_0
+\frac{9}{32} (\a_{1,0}^s)^2
-\frac 34 \a_{1,0}^s K_0+\frac{3}{16} (\a_{1,0}^s)^2 K_0\\
&-\frac18 K_0^2
+\cali_{finite}^s+\int_0^1 dy\ (-6 h^h (f_3^h)^2  (y f_2^h)^{1/2})
+\calv_{ambiguity}^s\,,\\
&\cali_{finite}^s=-6 \int_0^1 d\r\  \biggl(\frac{f_3^2f_2^{1/2}h}{\r^3}
-\calj_{divergent}^s\biggr)\,,\\
&\calj_{divergent}^s=\frac{1}{\r^3}\biggl(\frac18+\frac14 K_0-\frac12 \ln\r\biggr)
+\frac{1}{\r^2}\biggl(\frac14 \a_{1,0}^s  \ln\r
-\frac{1}{16} \a_{1,0}^s (5 +2 K_0)\biggr)
\\&+\frac{1}{\r}\biggl(
-\frac18 \ln^2\r+\biggl(\frac18 K_0 +\frac{5}{48} \biggr) \ln\r
+\frac{1}{16} (\a_{1,0}^s)^2 
+\frac{67}{1152} -\frac{5}{96} K_0 -\frac{1}{32} K_0^2 \biggr)\,,
\\
&\calv_{ambiguity}^s=-36\kappa_1^s K_0^2-36\kappa_2^s K_0-36 \kappa_3^s\,.
\end{split}
\eqlabel{calesfin}
\end{equation}

We can now compare the effective potentials of a chirally 
symmetric state and a state spontaneously  breaking chiral symmetry
for a cascading gauge theory on $dS_4$ (we restored the full $\{P,g_0,H\}$ 
dependence)
\begin{equation}
\begin{split}
&16\pi G_5\left(\calv_{eff}^b-\calv_{eff}^s\right)=
3 (f_{c,4,0}- H^4 a_{4,0})+
\frac{3}{16} (-3 P^2 \a_{1,0}^b g_0+2 P^2 g_0-2 K_0 \a_{1,0}^b\\
&+4 K_0)
 H^2 (H \a_{1,0}^s-\a_{1,0}^b)-\frac{3}{32} (3 P^2 g_0+2 K_0) H^2 
(H \a_{1,0}^s-\a_{1,0}^b)^2\\
&+\left(\cali_{finite}^b-\cali_{finite}^s\right)+
H^2\biggl(\int_0^1 dy\ (-6 h^h f_a^h f_b^h (f_c^h)^{1/2})
-\int_0^1 dy\ (-6 h^h (f_3^h)^2  (y f_2^h)^{1/2})\biggr)\,,
\end{split}
\eqlabel{finaldiff}
\end{equation} 
where we used the {\it same} renormalization scheme for computing 
both $\calv_{eff}^b$ and $\calv_{eff}^s$, \ie we set 
\begin{equation}
H^{-4} \k_i\bigg|^b= H^{-4} \k_i\bigg|^s\,,\qquad i=1,2,3\,.
\eqlabel{sameRG}
\end{equation}

\begin{figure}[t]
\begin{center}
\psfrag{V}[][][0.7]{{$\frac{16\pi G_5}{P^4g_0^2}\times\frac{\calv_{eff}^{b,s}}{H^4}$}}
\psfrag{d}[][][0.7]{{$\frac{16\pi G_5}{P^4g_0^2}\times\frac{\calv_{eff}^b
-\calv_{eff}^s}{H^4}$}}
\psfrag{k}[][][0.7]{{$\ln\frac{H^2}{\Lambda^2P^2g_0}$}}
\includegraphics[width=2.8in]{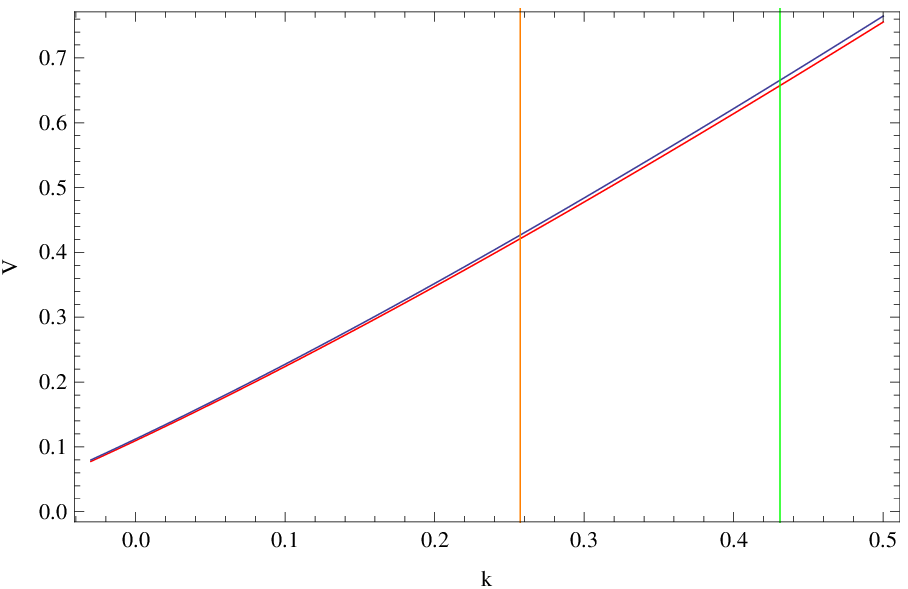}
~
~
\includegraphics[width=2.8in]{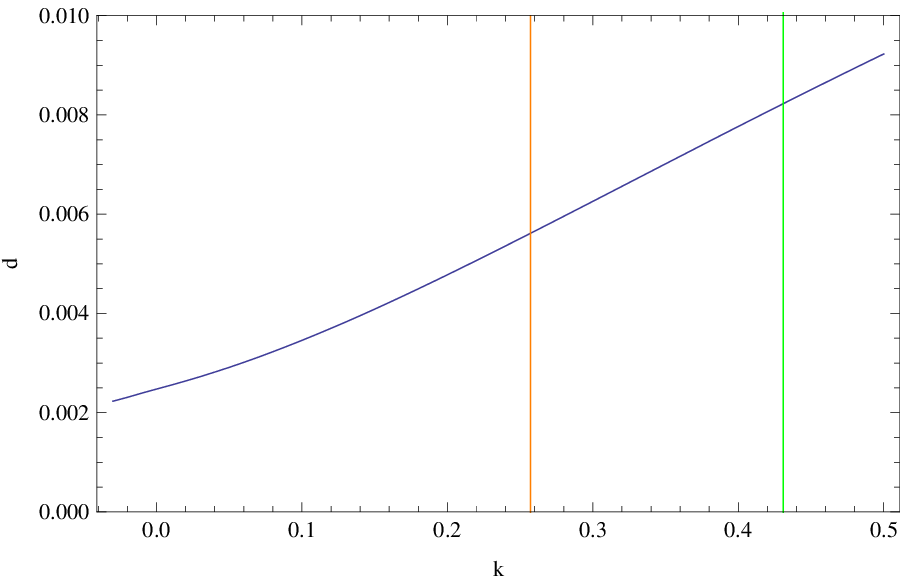}
\end{center}
\caption{(Colour online) Left Panel: effective potentials of the 
chirally symmetric ($V_{eff}^s$, red) and the broken phase ($V_{eff}^b$, blue)
of the cascading gauge theory on $dS_4$. Right Panel: the difference 
$(V_{eff}^b- V_{eff}^s)$.   The vertical  lines 
represent the first order chiral symmetry breaking phase transitions 
of cascading gauge theory on $S^3$ \cite{abs3} (green line) 
and at finite temperature \cite{abk}
(orange line). }
\label{figure9}
\end{figure}

Figure \ref{figure9} presents effective potentials 
(and their difference) between the  state with spontaneously broken chiral 
symmetry, $\calv_{eff}^b$, and the chirally symmetric state, $\calv_{eff}^s$, 
of cascading gauge theory on
$dS_4$ as a function of $\ln\frac{H^2}{\Lambda^2}$. 
Over the range of $\frac H\Lambda$ 
studied\footnote{It is difficult to keep our current numerical 
procedure stable for smaller values of $\frac H\Lambda$.},
\begin{equation}
\frac{16\pi G_5}{P^4g_0^2}\times\frac{\calv_{eff}^{b}-\calv_{eff}^{s}}{H^4}>0\,,\qquad 
\ln\frac{H^2}{\Lambda^2}\ge -0.03\,,
\eqlabel{difffin}
\end{equation}
implying that chirally symmetric phase is a true ground state of 
cascading gauge theory on $dS_4$. For comparison, 
the vertical green and orange lines 
indicate the first order chiral symmetry 
breaking phase transitions of cascading gauge theory on 
$S^3$ \cite{abs3} and at finite temperature \cite{abk}.  

\section{Properties of $dS_4$ deformed KT/KS geometries}\label{propertiesnew}
Given numerical constructions of $dS_4$ deformed  KT/KS geometries as in section \ref{symmetric}, 
we can compute the D3 brane charge at the tip of the conifold. Following \cite{landscape}, 
we find (see \eqref{khy})
\begin{equation}
Q^{D3,s}=\frac{1}{27\pi}\lim_{y\to 0} K(y) = \frac{K_0^h}{27\pi} \,, 
\eqlabel{chargektds}
\end{equation} 
and (see \eqref{k1hks}-\eqref{k3hks})
\begin{equation}
Q^{D3,b}=\frac{1}{54\pi}\lim_{y\to 0} \biggl(K_1(y) (2-K_2(y))+K_2(y) K_3(y)\biggr)=0\,,
\eqlabel{chargeksds}
\end{equation} 
where we use superscripts $ ^b$ and $ ^s$ to denote chiral symmetry broken (deformed KS) and 
chiral symmetry unbroken (deformed KT) phases.  

\begin{figure}[t]
\begin{center}
\psfrag{q}[][][0.7]{{$\frac{27\pi }{P^2g_0}Q^{D3,s}$}}
\psfrag{l}[][][0.7]{{$\ln\left(\frac{27\pi}{P^2g_0}Q^{D3,s}\right)$}}
\psfrag{k}[][][0.7]{{$\ln\frac{H^2}{\Lambda^2P^2g_0}$}}
\includegraphics[width=2.5in]{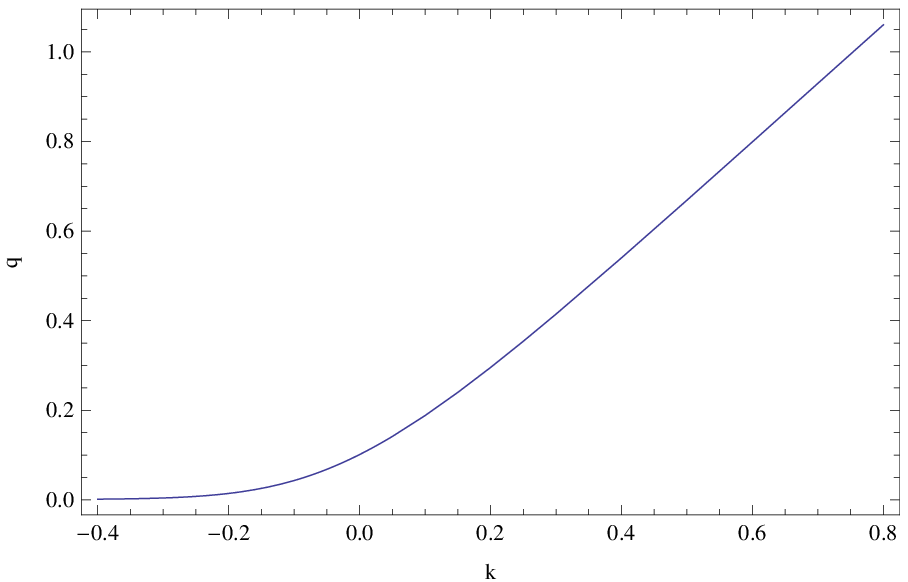}
~
~
\includegraphics[width=2.5in]{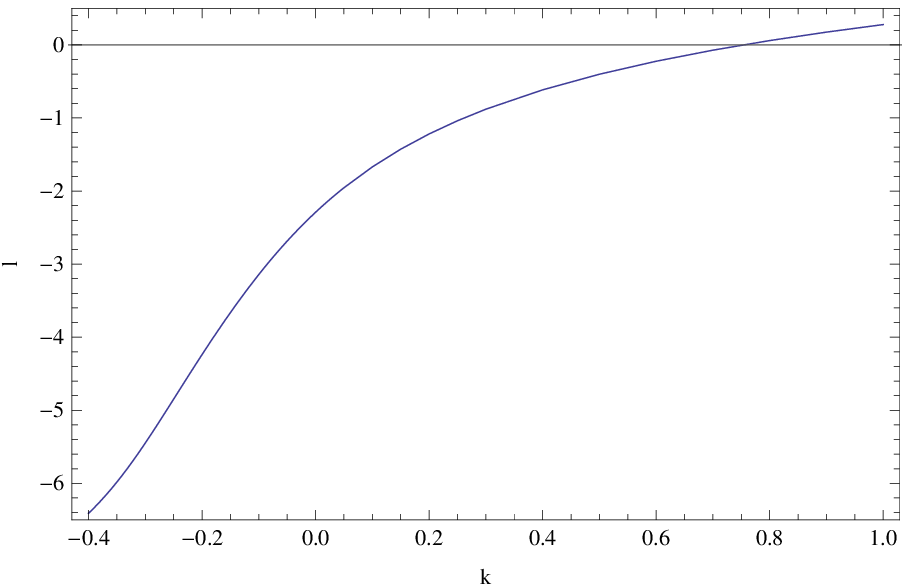}
\end{center}
\caption{(Colour online) Left Panel: D3 brane charge at the tip of the conifold of the $dS_4$ deformed KT throat geometry, $Q^{D3,s}$, 
as a function of $\frac H\Lambda$. Right Panel: logarithm of D3 brane charge at the tip of the conifold 
of the $dS_4$ deformed KT throat geometry, $Q^{D3,s}$, as a function of $\frac H\Lambda$. }
\label{figure10}
\end{figure}

Figure \ref{figure10} presents  D3 brane charge at the tip of the conifold of the $dS_4$ deformed KT throat geometry, $Q^{D3,s}$, 
as a function of $\frac H\Lambda$. Note that over all the range of parameters accessible with our numerical code
$Q^{D3,s}>0$.

\section{Properties of $S^3$ deformed KT/KS geometries}\label{propertiesold}
Using numerical constructions of $S^3$ deformed  KT/KS geometries presented in \cite{abs3}, 
we can compute the D3 brane charge at the tip of the conifold. Following \cite{landscape}, 
we find (see eq.(3.24) of \cite{abs3})
\begin{equation}
Q^{D3,s}=\frac{1}{27\pi}\lim_{y\to 0} K(y) = \frac{K_0^h}{27\pi} \,, 
\eqlabel{chargekts}
\end{equation} 
and (see eqs.(5.34)-(5.36) of \cite{abs3})
\begin{equation}
Q^{D3,b}=\frac{1}{54\pi}\lim_{y\to 0} \biggl(K_1(y) (2-K_2(y))+K_2(y) K_3(y)\biggr)=0\,.
\eqlabel{chargekss}
\end{equation} 
where we use superscripts $ ^b$ and $ ^s$ to denote chiral symmetry broken (deformed KS) and 
chiral symmetry unbroken (deformed KT) phases.

\begin{figure}[t]
\begin{center}
\psfrag{q}[][][0.7]{{$\frac{27\pi }{P^2g_0}Q^{D3,s}$}}
\psfrag{k}[][][0.7]{{$\ln\frac{\mu_3^2}{\Lambda^2P^2g_0}$}}
\psfrag{w}[][][0.7]{{$\mu_{3,\csb}$}}
\psfrag{e}[][][0.7]{{$\mu_{3,tachyon}$}}
\psfrag{r}[][][0.7]{{$\mu_{3,negative}$}}
\includegraphics[width=2.5in]{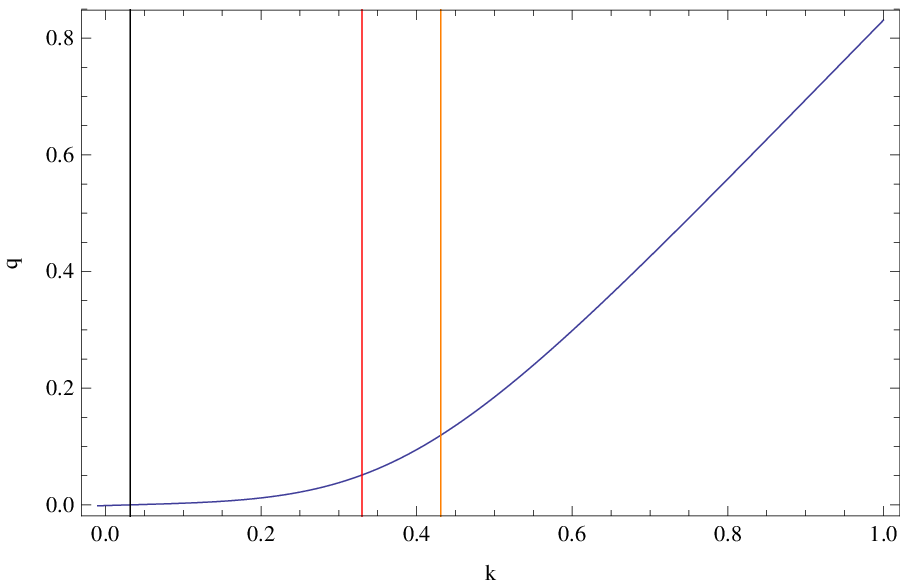}
~
\includegraphics[width=2.65in]{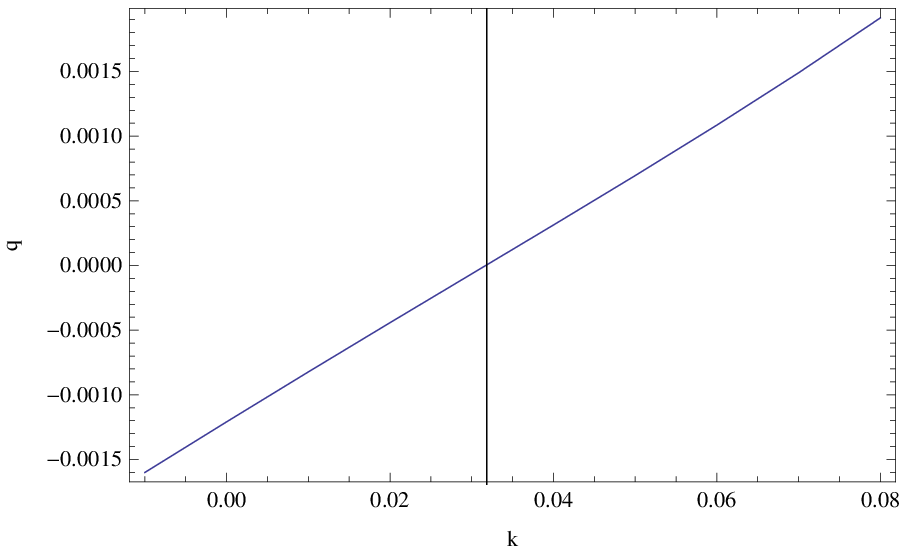}
\end{center}
\caption{(Colour online) D3 brane charge at the tip of the conifold of the $S^3$ deformed KT throat geometry, $Q^{D3,s}$, 
as a function of $\frac{\mu_3}{\Lambda}$.  The vertical orange line represents the value of the compactification 
scale $\mu_{3,\csb}$ below which it becomes energetically favourable to tunnel to $S^3$ deformed KS throat geometry,
with spontaneous breaking of chiral symmetry. The vertical red line represents the value of the compactification 
scale $\mu_{3,tachyon}$ below which some of the linearized fluctuations (spontaneously breaking the chiral symmetry)
become tachyonic. The vertical black lines denote  the value of the compactification 
scale $\mu_{3,negative}$ below which $Q^{D3,s}<0$. 
}
\label{figure11}
\end{figure}

Figure \ref{figure11} presents  D3 brane charge at the tip of the conifold of the $S^3$ deformed KT throat geometry, $Q^{D3,s}$, 
as a function of $\frac {\mu_3}{\Lambda}$. Here, unlike the $dS_4$ deformed KT throat geometry, we find that  $Q^{D3,s}$
can become negative! This happens whenever 
\begin{equation}
\mu_3< \mu_{3,negative}\,,\qquad \ln  \frac{\mu_{3,negative}^2}{\Lambda^2 P^2 g_0}=0.0318(3)\,,
\eqlabel{negativekts3}
\end{equation}
which is represented by black vertical lines in figure \ref{figure11}. However, these negative values of  $Q^{D3,s}$ are not physical.
The issue is that prior we reach the compactification scale $\mu_{3,negative}$, namely at $\mu_{3,\csb}$ \cite{abs3} 
\begin{equation}
\mu_{3,\csb}> \mu_{3,negative}\,,\qquad \ln  \frac{\mu_{3,\csb}^2}{\Lambda^2 P^2 g_0}=0.4309(8)\,,
\eqlabel{defmucsb}
\end{equation}
chirally symmetric phase of cascading gauge theory  on $S^3$ undergoes a first order phase transition 
to a symmetry broken phase (deformed KS geometry), where $Q^{D3,b}=0$, see \eqref{chargekss}. 
This first order transition is further enhanced by perturbative tachyonic instabilities in chirally symmetric 
phase which arise at a slightly lower value of 
$\mu_3$, namely at $\mu_{3,tachyon}$ \cite{abs3}
\begin{equation}
\mu_{\csb}> \mu_{3,tachyon}> \mu_{3,negative}\,,\qquad \ln  \frac{\mu_{3,tachyon}^2}{\Lambda^2 P^2 g_0}=0.3297(3)\,.
\eqlabel{defmutachyon}
\end{equation}
Thus, a correct behaviour of the D3 charge 
at the tip of the conifold in $S^3$ deformed throat geometries is 
\begin{equation}
Q^{D3}=\begin{cases}
&Q^{D3,s}>0\,,\qquad \mu_3> \mu_{3,\csb}\,;\\
&Q^{D3,b}=0\,,\qquad \mu_3\le \mu_{3,\csb}\,.\\
\end{cases}
\eqlabel{D3final}
\end{equation}  
Once again, the D3 charge at the tip of the conifold is never negative. 

%%%%%%%%%%%%%%%%%%%%%%%%%%%%%%%%%%%%%%%%
\section*{Acknowledgments}
AB thanks IPMU and 
Isaac Newton Institute for Mathematical Sciences
for hospitality where parts of this work were completed.
Research at Perimeter
Institute is supported by the Government of Canada through Industry
Canada and by the Province of Ontario through the Ministry of
Research \& Innovation. We gratefully acknowledge further support by an
NSERC Discovery grant.

\end{document}